\documentclass[11pt,a4paper]{article}
\pdfoutput=1

\usepackage[colorlinks=true, linkcolor=black!50!blue, urlcolor=blue, citecolor=blue, anchorcolor=blue]{hyperref}
\usepackage[font=small,labelfont=bf,margin=0mm,labelsep=period,tableposition=top]{caption}
\usepackage[a4paper,top=1.5cm,bottom=2cm,left=1.5cm,right=1.5cm,bindingoffset=0mm]{geometry}

\usepackage{placeins}
\usepackage{cite}

\usepackage{graphicx}
\usepackage{amsmath}
\usepackage{xcolor}
\usepackage{booktabs}
\usepackage{tabularx}
\usepackage{xspace}
\graphicspath{{../}{./figures/}}

\numberwithin{equation}{section}
\numberwithin{figure}{section}
\numberwithin{table}{section}

\newcommand{\be}{\begin{equation}}
\newcommand{\ee}{\end{equation}}
\newcommand{\la}{\left\langle}
\newcommand{\ra}{\right\rangle}
\newcommand{\lp}{\left(}
\newcommand{\rp}{\right)}
\newcommand{\tr}{\mathrm{tr}}
\def\lsim{\mathrel{\rlap{\lower4pt\hbox{\hskip1pt$\sim$}}
    \raise1pt\hbox{$<$}}}

\newcommand{\pythia}{{\sc\small Pythia8}\xspace}
\newcommand{\sherpa}{{\sc\small Sherpa}\xspace}
\newcommand{\herwig}{{\sc\small Herwig7}\xspace}
\newcommand{\powheg}{{\sc\small POWHEG}\xspace}
\newcommand{\powhegres}{{\sc\small POWHEG-RES}\xspace}
\newcommand{\powhegtwo}{{\sc\small POWHEG-V2}\xspace}
\newcommand{\powhegtwomc}{{\sc\small POWHEG-V2mc}\xspace}
\newcommand{\mgamc}{{\sc\small MG5\_aMC}\xspace}
\newcommand{\genie}{{\sc\small GENIE}\xspace}
\newcommand{\yadism}{{\sc\small YADISM}\xspace}
\newcommand{\lhapdf}{{\sc\small LHAPDF}\xspace}
\newcommand{\apfel}{{\sc\small APFEL}\xspace}
\newcommand{\nnpdf}{{\sc\small NNPDF4.0}\xspace}

\begin{document}
\newgeometry{top=1.5cm,bottom=1.5cm,left=1.5cm,right=1.5cm,bindingoffset=0mm}

$\qquad$
\vspace{1.6cm}

\begin{center}
  {\Large \bf Event Generation for Neutrino and Muon DIS at the LHC}\\
  \vspace{1.1cm}
  {\small
  Paulina Hernandez-Sainz$^{1}$, Felix Kling$^{3}$, Jelle Koorn$^{1,2}$, and Juan Rojo$^{1,2}$
  }\\

\vspace{0.7cm}

{\it \small
  ~$^{1}$Nikhef Theory Group, Science Park 105, 1098 XG Amsterdam, The Netherlands\\[0.1cm]
  ~$^{2}$Department of Physics and Astronomy, VU Amsterdam, De Boelelaan 1081, \\[0.1cm] 1081 HV Amsterdam, The Netherlands\\[0.1cm]
  ~$^{3}$Universit\"at Bonn, Regina-Pacis-Weg 3, D-53113 Bonn, Germany\\[0.1cm]
 }

\vspace{1.0cm}

{\bf \large Abstract}

\end{center}

Neutrinos and muons produced in the forward direction of LHC collisions and detected by the FASER and SND@LHC experiments enable a comprehensive
deep-inelastic scattering (DIS) programme, covering both neutral and charged currents, at TeV energies.
This programme aims to provide stringent constraints on proton and nuclear structure as well as on hadron fragmentation, and to offer novel insight into neutrino interactions at the highest energies achieved in a laboratory.
Realising this potential
requires accurate and precise simulations of neutral- and charged-current DIS.
We systematically compare the predictions of NLO general-purpose Monte Carlo (MC) event generators for neutrino- and muon-initiated DIS at the LHC (\sherpa, \herwig, and \powheg) against analytical structure-function calculations (\yadism) and dedicated neutrino simulation packages (\genie, {\sc\small NuWro}, {\sc\small GiBUU}).
Within this framework, we quantify the role of higher-order QCD corrections, (nuclear) parton distribution functions (PDFs), parton shower and hadronisation modelling, and QED radiation both for inclusive observables and for the FASER fiducial selections.
We also present representative applications to the FASER neutrino programme: comparisons with its most recent data from the emulsion and electronic sub-detectors, a study of strange PDF sensitivity in the dimuon final state, and predictions for charged-hadron production in semi-inclusive DIS.
Our analysis highlights the agreement of modern event generators for inclusive and differential DIS observables once common settings are adopted, while differences remain for observables sensitive to the hadronic final state.

\clearpage
\tableofcontents

\section{Introduction}
\label{sec:intro}

The Large Hadron Collider (LHC) produces an intense, highly energetic flux of neutrinos (and muons) in the forward direction~\cite{Kling:2021gos,Bai:2020ukz,Bai:2021ira,Bai:2022xad,Maciula:2022lzk,Bhattacharya:2023zei,Buonocore:2023kna,Kling:2023tgr,John:2025qlm}, and since 2022 two dedicated far-forward detectors, FASER~\cite{Feng:2017uoz, FASER:2019dxq,FASER:2020gpr,FASER:2022hcn} and SND@LHC~\cite{SNDLHC:2022ihg}, have been recording and studying them~\cite{FASER:2021mtu,FASER:2023zcr,SNDLHC:2023pun,SNDLHC:2023mib,SNDLHC:2024qqb,Ariga:2025lhcnu,SNDLHC:2026oxu,SNDLHC:2026mip,SNDLHC:2026yqa,FASER:2024hoe,FASER:2025qaf,FASER:2025sge,FASER:2026hzm,FASER:2026jhh,FASERnu:2026conf,Boyd:2026xuf,FASER:2026charm}.
These are the highest-energy neutrinos ever produced in a laboratory, and both experiments observe them through the deep-inelastic scattering (DIS) process.
The study of neutrino and muon DIS at these LHC far-forward experiments has a unique potential to improve current determinations of proton and nuclear structure~\cite{Cruz-Martinez:2023sdv,Francener:2025pnr} as encoded in the (nuclear) parton distribution functions (PDFs)~\cite{Gao:2017yyd,Klasen:2023uqj,Kovarik:2019xvh,Ethier:2020way} as well as of light and heavy hadron fragmentation~\cite{Albino:2008gy}.
For instance, charm production in muon DIS could provide conclusive evidence for intrinsic charm in the proton~\cite{Brodsky:1980pb,Ball:2022qks,Guzzi:2022rca,Brodsky:2015fna,Ball:2016neh}, while the analogous measurement in neutrino DIS would represent a sensitive probe of proton strangeness~\cite{Faura:2020oom}.
This potential impact further improves with the upgraded detectors and increased luminosities of the High-Luminosity LHC (HL-LHC)~\cite{FASER:2025upgrade,SNDLHC:2025eppsu,SNDLHC:2026why}, as well as with the proposed Forward Physics Facility (FPF)~\cite{FPF:2025bor,FPFWorkingGroups:2025rsc,Feng:2022inv,Anchordoqui:2021ghd,Adhikary:2024nlv} and with hypothetical neutrino detectors placed at the surface exit points of the LHC neutrino beam~\cite{Ariga:2020tnq, Kamp:2025phs}.

The interpretation of current and future FASER, SND@LHC, and FPF measurements in terms of proton and nuclear structure as well as of hadron fragmentation is only possible with accurate and precise simulations of neutral-current (NC) and charged-current (CC) DIS for both the leptonic and hadronic final states.
Indeed, fiducial signal selections, for instance those associated with vertex identification, often involve exclusive cuts on quantities such as the number and momenta of charged particles, the angle between the outgoing charged lepton and the hadronic system, or the presence of a secondary muon from charm decay.
Therefore, realising the precision QCD physics potential of the LHC far-forward experiments demands state-of-the-art Monte Carlo (MC) simulations~\cite{Buckley:2011ms,Campbell:2022qmc} that account for higher-order QCD and QED corrections, modern PDFs and parton shower algorithms, and tailored hadronisation models~\cite{Skands:2014pea,Fieg:2023kld}.
These criteria are not met by some of the neutrino event generators used to interpret the FASER and SND@LHC data, notably \genie~\cite{Andreopoulos:2009rq,Andreopoulos:2015wxa,NuSTEC:2017hzk}.

With this motivation, in this work we present a systematic benchmark of predictions for neutrino- and muon-initiated DIS at the LHC, obtained at next-to-leading order (NLO) accuracy with three independent general-purpose MC event generators: \sherpa~\cite{Sherpa:2019gpd,Bothmann:2024}, \powheg~\cite{Nason:2004rx,Frixione:2007vw,Alioli:2010xd,Banfi:2023mhz,Buonocore:2024pdv,vanBeekveld:2024ziz,FerrarioRavasio:2024kem}, and \herwig~\cite{Bellm:2015jjp,Bewick:2023tfi}.
 These predictions are compared with the analytical structure-function calculations provided by \yadism~\cite{Candido:2024rkr,Candido:2023utz} up to next-to-next-to-leading order (NNLO) in the QCD expansion, as well as with the predictions of dedicated neutrino simulation packages, in particular \genie.
 Furthermore, we also study other neutrino generators, such as
 {\sc\small GiBUU}~\cite{Buss:2011mx} and {\sc\small NuWro}~\cite{Golan:2012rfa,Juszczak:2009qa}, frequently used for the interpretation of oscillation experiments. 
 In these benchmark comparisons, identical inputs (such as PDFs, electroweak parameters, heavy-quark mass scheme, and  beam energy) are adopted whenever possible to disentangle the effects of the different physical mechanisms entering the theory predictions.
 
Equipped with this framework, we quantify in detail
the role of higher-order QCD corrections, PDF choice, parton shower and hadronisation modelling, charm quark mass effects, and QED radiation both for inclusive observables and for FASER fiducial selections. 
We also consider nuclear PDF uncertainties arising from the tungsten target of FASER and SND@LHC, though nuclear modifications to shower and hadronisation due to rescattering and absorption are left for future work. 
Our analysis highlights the agreement of modern event generators at the inclusive level, while significant differences emerge for fiducial observables sensitive to the hadronic final state. 
While we find that \genie agrees with NLO generators for inclusive rates, we show that, first, this agreement arises from an accidental cancellation of independent effects and that, second, larger differences arise for exclusive observables and final-state distributions. 

We also present three representative applications of the pipeline developed in this work to the FASER neutrino programme.
First, we carry out comparisons of theoretical predictions from different event generators with the most recent FASER neutrino measurements~\cite{FASER:2024hoe,FASERnu:2026conf,FASER:2026nu680,FASER:2024ref,FASER:2026enu2d}, obtained both with the emulsion and with the electronic sub-detectors.
Second, we evaluate event yields for the dimuon signal at FASER, a sensitive probe of the proton strangeness content~\cite{Faura:2020oom,Helenius:2026uuz}, which can be accessed through the tracking spectrometer, complementing direct measurements of charm production in the FASER$\nu$ emulsion detector~\cite{FASER:2026charm}.  
Third, we estimate the expected event yields for charged-hadron production in semi-inclusive DIS (SIDIS), a process for which NNLO QCD corrections have recently become available~\cite{Goyal:2023zdi,Bonino:2024qbh,Bonino:2025ident,Bonino:2025qta,Goyal:2024tmo,Bonino:2024wgg,Bonino:2025sidis}  and which provides unparalleled sensitivity to the flavour separation of hadron fragmentation functions~\cite{Bertone:2017tyb,Bertone:2018ecm,Borsa:2021ran,deFlorian:2017lwf,Moffat:2021dji}. 

The outline of this paper is as follows.
First, Sect.~\ref{sec:generators} presents the event generators and analytic tools considered in this benchmark.
Section~\ref{sec:settings} describes the common settings, fiducial regions, FASER selections, and the associated physical observables.
Then Sect.~\ref{sec:inclusive} compares the generators with the structure-function calculations for inclusive scattering and differential distributions in the DIS variables.
Section~\ref{sec:charm} is dedicated to charm production, with emphasis on the different heavy-quark schemes and the sensitivity of the charm-hadron composition to the hadronisation model.
Section~\ref{sec:hadronic} extends the comparison to observables sensitive to the modelling of the hadronic final state in the FASER$\nu$ selection, and studies the impact of the parton shower algorithm and QED radiation.
Section~\ref{sec:faser-app} presents representative applications to FASER: the comparison with the emulsion and electronic-detector measurements, a study of the PDF dependence of the dimuon signal, and predictions for charged-hadron production in neutrino SIDIS. 
We summarise our findings and outline possible future developments in Sect.~\ref{sec:conclusions}.

Additional studies are collected in the appendices.
Appendix~\ref{app:implementation} discusses the implementation of these event generator settings and how our results can be reproduced and extended.
Appendix~\ref{app:nondis} quantifies the contribution of non-DIS processes to fiducial differential distributions as estimated with \genie.
Appendix~\ref{app:npdf} estimates the effects of nuclear PDF systematic uncertainties for muon and neutrino DIS at TeV energies.
Finally, App.~\ref{app:nugen} compares \genie with two other dedicated neutrino generators, {\sc\small GiBUU} and {\sc\small NuWro}.

\section{Monte Carlo event generators}
\label{sec:generators}

In this section we present the MC event generators that will be considered in this benchmark.
We focus on three NLO-accurate general-purpose MC generators, namely \sherpa, \herwig, and \powheg. 
For inclusive observables and charm production, their predictions are compared with the analytical structure-function calculations provided by \yadism.
For completeness, we have also studied predictions provided by the leading-order (LO) MCs \pythia and \mgamc, finding excellent agreement with the other MCs when run at LO.\footnote{Results for \pythia and \mgamc are not shown in this work but are available from the project repository, see App.~\ref{app:implementation}.}
Furthermore, we consider event generators tailored to neutrino physics, with emphasis on \genie, the current default MC in FASER and SND@LHC, as well as {\sc\small NuWro} and {\sc\small GiBUU} (the latter two in App.~\ref{app:nugen}).
Table~\ref{tab:generators} indicates the main features of these generators.
For each generator we indicate the version used, the source of the hard process, the QCD perturbative order, and the shower and hadronisation algorithms.
We now provide more details on each of these generators in turn.

\begin{table}[!tbp]
  \centering
  \small
  \renewcommand{\arraystretch}{1.60}
  \setlength{\tabcolsep}{4pt}
  \resizebox{\textwidth}{!}{%
  \begin{tabular}{llllll}
    \toprule
    Generator & Version & Hard process & Order & Shower & Hadronisation \\
    \midrule
    \pythia      & 8.311  & internal DIS ME        & LO            & $p_T$-ordered, dipole recoil & Lund string \\
                 &        &                        &               & (also {\sc Vincia}~\cite{Fischer:2016vfv}, {\sc Dire}~\cite{Hoche:2015sya}) & \\
    \mgamc       & 3.7.2  & tree-level ME          & LO            & {\sc\small Pythia}~8.311 & Lund string \\
                 &        &                        &               & (also parton level) & \\
    \genie       & 3.06.02 & {\sc EMDIS} / {\sc DIS-CC} (+charm)  & LO (Bodek--Yang) & none & {\sc AGKY} + {\sc\small Pythia}~8.312 \\
                 &        & tune {\tt G18\_02a}    &               &  & \\
    \midrule 
    \sherpa      & 3.0.5  & {\sc Comix} / {\sc Amegic}~\cite{Krauss:2001iv} & NLO ({\sc MC@NLO}) & CS dipole (also {\sc Dire}) & {\sc AHADIC} cluster \\
    \herwig      & 7.3.0  & {\sc MEDISNC} / {\sc MEDISCC} & NLO ({\sc POWHEG}) & angular-ordered & cluster \\
    \powhegres   & {\tt DIS\_v} & POWHEG-BOX-RES   & NLO           & {\sc\small Pythia}~8.311 & Lund string \\
                 &        &                        &               & or {\sc\small Herwig}~7.3.0 & or cluster \\
    \powhegtwo   & {\tt nu-DIS} & POWHEG-BOX-V2    & NLO           & {\sc\small Pythia}~8.311 & Lund string \\
                 &        &                        &               & or {\sc\small Herwig}~7.3.0 & or cluster \\
    \genie       & 3.06.02 & {\sc HEDIS}, tune {\tt GHE19\_00a} & NLO (\apfel)  & none & {\sc LEPTO}-type + {\sc\small Pythia}~8.312 \\
    \midrule
    \yadism      & 0.13.11 & DIS structure functions & LO, NLO, NNLO & --- & --- \\
    \bottomrule
  \end{tabular}}
  \vspace{0.5cm}
  \caption{LO and NLO event generators considered in this benchmark.
    For each generator we indicate the version, the source of the hard process (ME: matrix element), the QCD perturbative order, and the shower and hadronisation algorithms (CS: Catani--Seymour).
  \sherpa and \herwig can also be run at LO through the same matrix elements.
  In this work both \genie configurations perform hadronisation and decays with \pythia rather than with {\sc Pythia6}, the \genie default.
  The last row corresponds to the analytical structure-function calculation of \yadism, interfaced to {\sc EKO}~0.15.5~\cite{Candido:2022tld} for the DGLAP evolution.
  Appendix~\ref{app:nugen}
  presents results for two other neutrino generators,  {\sc\small NuWro} and {\sc\small GiBUU}.  
  }
  \label{tab:generators}
\end{table}

\paragraph{\pythia.}  {\sc\small Pythia}~8.311~\cite{Bierlich:2022pfr} simulates DIS processes
through its internal LO neutral- and charged-current matrix elements, showered
with the $p_T$-ordered shower and hadronised with the Lund string model~\cite{Andersson:1983ia}.  
The baseline shower uses the dipole-recoil scheme~\cite{Cabouat:2017rzi}, in which initial-state radiation recoils against the colour dipole rather than the scattered lepton.
In this work, \pythia is considered both as a stand-alone generator and as the provider of shower and hadronisation algorithms for some of the NLO generators.
Unless otherwise specified, here \pythia is run with the Monash tune~\cite{Skands:2014pea} for non-perturbative physics. Other tunes, such as the forward physics tune of~\cite{Fieg:2023kld}, can be easily implemented. 

\paragraph{\sherpa.}  \sherpa~v3.0.5~\cite{Sherpa:2019gpd,Bothmann:2024} provides either an LO sample from {\sc Comix}~\cite{Gleisberg:2008fv} or an NLO sample using the MC@NLO~\cite{Frixione:2002ik} scheme for the matching to the Catani--Seymour (CS) dipole
shower~\cite{Schumann:2007mg}.
Hadronisation in \sherpa is based on the {\sc AHADIC} cluster model~\cite{Winter:2003tt,Chahal:2022rid}.  

\paragraph{\herwig.}  {\sc\small Herwig}~7.3.0~\cite{Bellm:2015jjp,Bewick:2023tfi} provides DIS simulations at LO through the {\sc MEDISNC} and {\sc MEDISCC} modules, and at NLO through its POWHEG implementation matched to the angular-ordered shower, and uses a cluster hadronisation model~\cite{Webber:1983if}.
\herwig admits two independent NLO implementations.  The first matches {\sc
Matchbox}~\cite{Platzer:2011bc} one-loop amplitudes to the shower in the MC@NLO scheme; the second is
\herwig's native POWHEG implementation, built on the same matrix-element classes as
the LO calculation, and is the one adopted here. 
The value of the intrinsic $k_T$ used in the \herwig simulations is adjusted to match the DIS kinematics. 

\paragraph{\powheg.}  Two independent POWHEG~\cite{Nason:2004rx,Frixione:2007vw,Alioli:2010xd}
implementations are used in this work.
First,
\powhegres is the
{\tt DIS\_v} process~\cite{Banfi:2023mhz} of POWHEG-BOX-RES~\cite{Jezo:2015aia}, covering inclusive and charm
production in both currents with mass effects neglected; it supplies the NLO
entry for NC muon scattering. 
Second, \powhegtwo is the {\tt nu-DIS}
process of POWHEG-BOX-V2~\cite{Buonocore:2024pdv}, developed for neutrino DIS with a massive
final-state charm quark, and supplies the NLO entry for charged-current
neutrino scattering.  
By default both are showered and hadronised with \pythia, and hence with
Lund-string hadronisation; showering and hadronisation with \herwig is studied as a variant.
Mass effects in \powhegtwo are only available for the generation of charm-tagged events, while for inclusive DIS production charm-quark mass effects are neglected. 
For inclusive generation, both \powheg variants agree with each other and with the ZM-VFNS \yadism calculation at the per-cent level at NLO. 
See also~\cite{Castro:2026xyr} for a recent POWHEG-BOX-RES calculation of heavy-quark pair production in NC DIS.

\paragraph{\mgamc.}  \mgamc~3.7.2~\cite{Alwall:2014hca} is used to generate events at LO matched to \pythia.
In its current implementation, this generator cannot be used to simulate DIS events at NLO accuracy.

\paragraph{\genie.}   
As opposed to the other generators considered here, the \genie~\cite{Andreopoulos:2009rq,Andreopoulos:2015wxa} settings such as PDF choice or matrix-element (ME) model cannot be straightforwardly modified and are only available for a fixed set of tunes. 
Here LO simulations for neutrino scattering are produced with \genie using 
the {\tt G18\_02a\_00\_000} tune~\cite{GENIE:2021zuu}, which is the same configuration that FASER uses~\cite{FASER:2024ykc}.
This tune is built on the Bodek--Yang model~\cite{Bodek:2002vp,Bodek:2003wd} with
GRV98LO~\cite{Gluck:1998xa}.  
For NC muon scattering we use the
{\sc EMDIS} channel of that tune.
For CC scattering we also consider the {\sc HEDIS} module~\cite{Soto:2021vdc} of {\tt GHE19\_00a}, built on BGR18 structure functions~\cite{Bertone:2018dse} at NLO through \apfel~\cite{Bertone:2013vaa}, achieving NLO accuracy for inclusive distributions.
This tune evaluates its structure functions with the {\tt NNPDF31sx\_nlo\_as\_0118\_LHCb\_nf\_6} PDF set~\cite{Ball:2017otu,Bertone:2018dse,NNPDF:2017mvq} rather than with \nnpdf.
To separate PDF effects from other model parameters, we also generate LO events where GRV98
is replaced by \nnpdf.

\genie does not simulate a parton shower, either initial- or final-state, for the configurations considered here: the struck quark and the colour-connected target remnant are passed directly to string fragmentation, so \genie final states contain no perturbative QCD radiation beyond the LO partonic process.
In the default tune the hadronic system is handled by the {\sc AGKY} model~\cite{Yang:2009zx}, which takes hadron multiplicities and momenta from the KNO parametrisation~\cite{Koba:1972ng} below a hadronic invariant mass ($W$) of $2.3$~GeV and from \pythia string fragmentation above $3$~GeV, interpolating in between; {\sc HEDIS} instead uses a {\sc LEPTO}-derived~\cite{Ingelman:1996mq} interface to \pythia at all $W$.
In {\sc HEDIS} the NLO corrections enter only as a reweighting of the LO cross-section, so its final states are LO in structure. 
Furthermore, \genie treats charm production\footnote{In the course of this work we found that the charm-production DIS cross-section of \genie multiplies by the $W$-boson propagator factor $(1+Q^2/M_W^2)$ instead of dividing by it. 
This overestimates the charm cross-section for $\nu_\mu$ scattering on a proton by about $4\%$ at $E_\nu=400$~GeV and $8\%$ at 1~TeV, growing rapidly at higher energies, while the fiducial inclusive cross-section changes by less than $0.2\%$. All \genie predictions in this work are obtained with the corrected cross-section.}  as a separate process~\cite{Aivazis:1993pi} and subtracts it from the inclusive DIS cross-section internally.  
The charm channel must therefore always be generated alongside the
inclusive one.
Concerning NC scattering, the {\sc EMDIS} channel consists of a three-flavour calculation where charm is absent from the initial and final states. 

An advantage of \genie as compared to most general-purpose generators is that it can account also for non-DIS processes such as quasi-elastic, resonance, and diffractive scattering which are then combined with the DIS contribution.
It can also cover the shallow-inelastic scattering (SIS) region at low $Q^2$, which provides a non-negligible contribution to total event yields~\cite{Candido:2023utz}.
In App.~\ref{app:nondis} we quantify the importance of non-DIS contributions for the FASER neutrino kinematics, finding that these effects can in general be neglected for all relevant energies provided one restricts the calculation to the DIS region {\it e.g.} by applying $Q^2>4$~GeV$^2$ and $W>3$~GeV cuts. 
Whenever necessary, predictions simulated with NLO generators can be complemented with the contributions from outside the DIS region (the SIS and resonance regions) as estimated by \genie using the pipeline presented in this work.
This can be done, for instance, by adding the \genie prediction for the kinematic region excluded by the DIS cuts, evaluated for the same selection, with a smooth interpolation across the boundary where differential distributions require it.
An example of this is presented in Sect.~\ref{sec:faser-app} for the comparison with the FASER electronic-detector muon-neutrino data, where the $Q^2<4$~GeV$^2$ contribution from the SIS region is taken from \genie.

\paragraph{\yadism.} Analytic calculations of the DIS structure functions up to NNLO and approximate N$^3$LO (aN$^3$LO)~\cite{NNPDF:2024nan} and accounting for heavy-quark mass effects are provided by
\yadism~\cite{Candido:2024rkr}, using the same PDFs, renormalisation and factorisation scales and electroweak inputs as the MC generators. 
Cross-sections are assembled from $F_2$, $F_L$ and $xF_3$ and then integrated over the
fiducial region.
\yadism further allows the heavy-quark scheme to be varied, comparing massless calculations in the zero-mass variable-flavour-number scheme (ZM-VFNS) with the FONLL-matched ones~\cite{Forte:2010ta} which account for heavy-quark mass effects.
\yadism has been benchmarked with other widely used DIS tools such as \apfel~\cite{Bertone:2013vaa} and {\sc\small HOPPET}~\cite{Salam:2008qg,Karlberg:2025hxk}.
\section{Benchmark settings}
\label{sec:settings}

In this section we summarise the settings and inputs used for this benchmark comparison of MC event generators for muon and neutrino DIS at the LHC. 
We specify  the kinematics, the fiducial regions, the common physics inputs and the observables.
Technical implementation details, including the per-generator cards, 
normalisation conventions, and the validation procedure, are collected in full in the public
repository of this project~\cite{benchmarkrepo}, see also App.~\ref{app:implementation}.

\paragraph{Kinematics and fiducial region.}
We consider a lepton (neutrino or muon) with energy $E_\ell$ incident on a tungsten (W) nucleus at rest in two channels:
neutral-current scattering 
\be 
\mu^\pm  + A \to \mu^\pm + X \, ,
\ee 
via $\gamma^*/Z$ exchange, and
charged-current scattering
\be
\nu_\ell~(\bar{\nu}_\ell) + A \to \ell^-~(\ell^+) +  X \, , \quad \ell=e,\mu\, ,
\ee
via $W^+~(W^-)$ exchange.  
We consider a range of energies between $E_\ell=400$~GeV and $E_\ell=4$~TeV, with representative results for differential distributions shown for $E_\ell=1$ TeV. 
Cross-sections are given per nucleon: each generator is run separately on a proton and on a neutron target, and the two are weighted with the corresponding $(Z,A)$ values of the tungsten target of FASER neglecting nuclear effects (which are studied separately in App.~\ref{app:npdf}).
Event generation is performed either in the centre-of-mass frame or in the target rest frame, depending on which mode is more computationally efficient.
All observables and selection requirements are evaluated in the target rest frame.

The fiducial region of the benchmark is defined by
\begin{equation}
  Q^2 > 4~{\rm GeV}^2 , \qquad W > 3~{\rm GeV},
  \label{eq:fiducial}
\end{equation}
with $Q^2$ and $W^2$ being the standard DIS invariants, $Q^2$ built from
the beam and scattered-lepton momenta and $W^2 = M_N^2 + 2P\cdot q - Q^2$ also
from the momentum $P$ of the target nucleon at rest, with $M_N$ the nucleon mass. 
The generators are run with generation cuts looser than, or at most equal to, those of Eq.~\eqref{eq:fiducial}, and the fiducial region is then imposed on the generated final-state events.
The lower limit on $Q^2$ keeps the calculation in the region of validity of perturbative QCD, while the lower bound on the invariant mass of the hadronic final state $W$ avoids the region where  hadronisation models are unreliable.
For the specific case of charm-production observables the lower limit on $W$ is raised to 5~GeV.

\paragraph{FASER selections.}
\label{sec:faser-selections}
The fiducial region of Eq.~\eqref{eq:fiducial} enables a clean comparison of the predictions of event generators, both inclusive and differential, with the corresponding analytic structure-function calculations.
Additional selection requirements need to be imposed to match the acceptance of the FASER experiment.
For this, we define three further
selections that emulate FASER analyses, all nested 
inside Eq.~\eqref{eq:fiducial}.
Different selections are needed since FASER can perform measurements with different detector subcomponents, each with their own set of requirements on the detected final-state particles.  

\paragraph{\it Spectrometer selection.}  The FASER electronic-detector analyses~\cite{FASER:2023zcr,FASER:2024ref,FASER:2026enu2d} select $\nu_\mu$ and $\bar{\nu}_\mu$ charged-current interactions in a $200$~mm diameter fiducial volume of the tungsten target by requiring a single track passing through all three spectrometer stations with momentum $p_\mu > 100$~GeV, lying within $95$~mm of the detector axis and at an angle $\theta < 25$~mrad to it. 
Therefore our truth-level proxy for the FASER spectrometer selection is given by
\begin{equation}
  E_{\mu'} > 100~{\rm GeV} , \qquad \theta_{\mu'} < 25~{\rm mrad} \, ,
  \label{eq:tierS}
\end{equation}
where with $\mu'$ we indicate the final-state muon. 
Since $\theta_{\mu'} \simeq \sqrt{Q^2(1-y)}/E_{\mu'}$ in this regime, the additional angular requirement on the outgoing muon removes part of the  high-$Q^2$, high-$y$ corner of the fiducial region.

\paragraph{\it Emulsion selection.}  
The FASER$\nu$ emulsion neutrino analyses~\cite{FASER:2024hoe,FASER:2026nu680} are vertex-based.
The current FASER$\nu$ event selection therefore requires more than four tracks converging on the vertex, of which more than three have $\tan\theta \le 0.1$, and identifies the CC muon by a track penetrating more than one hundred tungsten plates with $p_\mu > 200$~GeV and $\tan\theta > 0.005$, back-to-back with the remaining vertex tracks in azimuth.
Taking into account these considerations, 
our truth-level proxy for the FASER$\nu$ emulsion selection is 
\begin{equation}
  E_{\mu'} > 200~{\rm GeV} , \quad
  \theta_{\mu'} > 5~{\rm mrad} , \quad
  N_{0.5} \ge 5 , \quad N_{0.1} \ge 4 , \quad
  \Delta\phi(\mu', X) > \pi/2 ,
  \label{eq:tierE}
\end{equation}
where $N_{0.5}$ and $N_{0.1}$ count charged particles satisfying $\tan\theta < 0.5$ and $<0.1$ respectively, defined at hadron level under the same $c\tau > 10$~mm stability convention as every other observable in this work,\footnote{In the specific case of charmed mesons, the emulsion detectors can reconstruct $\gamma c\tau \sim 3$~mm tracks, and hence a $D^\pm$ appears as one track and a $D^0$ as none. } and $\Delta\phi(\mu', X)$ is the azimuthal separation between the scattered lepton and the vector sum of all stable charged hadrons ($X=\sum h^\pm$).  
In addition, we impose that reconstructed charged particles have energies $E_{\rm ch}\ge 1$~GeV, given that tracks need to pass several tungsten plates in order to be reconstructed.
This cut also improves the infrared robustness of the FASER observables, since soft particles, such as the MeV-energy fragments from the breakup of the nuclear remnant in real events, are excluded.

The efficiency of the emulsion selection depends sensitively 
on the charged-particle multiplicity, one of the observables for which, as we will show in Sect.~\ref{sec:hadronic}, event generators differ the most. 
The same sensitivity propagates into the
experimental estimate for the incoming neutrino energy $E_\nu$.
For instance,  the $2026$ emulsion analysis reconstructs $E_\nu$ with a multivariate estimator whose inputs include the summed
charged-hadron momenta~\cite{FASER:2026nu680}, and hence the modelling of hadronic activity directly affects
the unfolded cross-section estimate. 
Another interesting observable in this context is the fraction of energy going into (unobserved) neutral hadrons, which affects the estimate of the hadronic energy resolution.

\paragraph{\it Dimuon selection.} 
We also consider an extension of the spectrometer selection where we request a secondary  muon of charge opposite to the primary, the so-called dimuon selection:
\begin{equation}
  \textrm{Spectrometer~selection} \quad \textrm{and} \quad \exists\, \mu^{\pm} :\;
  E_{\mu^{\pm}} > 100~{\rm GeV} , \quad \theta_{\mu^{\pm}} < 25~{\rm mrad} \, ,
  \label{eq:tierD}
\end{equation}
where a secondary $\mu^+$ is associated with a primary $\mu^-$ and vice versa.
This selection is not a further acceptance requirement but rather a charm tag, given that an
opposite-sign secondary muon with $E_\mu > 100$~GeV comes overwhelmingly from the
semileptonic decay of a charm hadron.
Its rate is set by the charm fraction, a semileptonic branching ratio of $\mathcal{O}(10\%)$, and the acceptance for the decay muon, and it is of order $10^{-3}$ of the inclusive DIS cross-section.
Being a charm tag, the dimuon selection is evaluated with $W > 5$~GeV, as all charm-production observables. 

\paragraph{Physics inputs.}

The common physics specifications to which all generators are configured for the benchmark study, unless otherwise specified, are collected in Table~\ref{tab:settings}.
Whenever variations with respect to the parameters listed in this table are implemented, these are mentioned specifically.
For instance, as mentioned in 
Sect.~\ref{sec:generators}, not all of these options can be implemented in \genie.
Concerning proton structure, we adopt the {\tt NNPDF40\_nnlo\_as\_01180} PDF set~\cite{NNPDF:2021njg} via the \lhapdf interface~\cite{Buckley:2014ana} throughout. 
As one of the applications to be discussed in Sect.~\ref{sec:faser-app}, predictions for alternative PDF sets are also provided. 
Since the target is a tungsten nucleus, we apply isospin symmetry to construct the neutron PDFs from the proton ones. 
Nuclear PDF effects are not part of the benchmark settings and are studied separately in App.~\ref{app:npdf}.
The analytic structure-function calculations  of \yadism include target-mass corrections (TMCs)~\cite{Schienbein:2007gr}.

\begin{table}[!tbp]
  \centering
  \small
  \renewcommand{\arraystretch}{1.5}
  \begin{tabularx}{\textwidth}{Xl}
    \toprule
    Setting & Value \\
    \midrule
    Beams                     & Fixed-$E_\ell$ $\mu^-$ (NC) or $\nu_\mu$ (CC) on a tungsten nucleus at rest \\
    Fiducial region           & $Q^2 > 4$~GeV$^2$, $W > 3$~GeV ($W > 5$~GeV for charm production) \\
    Parton distributions      & {\tt NNPDF40\_nnlo\_as\_01180}~\cite{NNPDF:2021njg} via \lhapdf~\cite{Buckley:2014ana} \\
    Renormalisation and factorisation scales                    & $\mu_R = \mu_F = Q$ \\
    Electroweak scheme        & $G_\mu$-matched: $\alpha_{\rm em} = 1/132.119$, $\sin^2\theta_W = 0.22291$ \\
    QED radiation             & switched off \\
    Heavy-quark scheme        & massless charm and bottom (zero-mass, five-flavour scheme) \\
    Target-mass corrections   & included in the structure-function calculations \\
    Stable-particle criterion & $c\tau > 10$~mm \\
    Multiple interactions     & switched off \\
    Final-state interactions     & switched off \\
    \bottomrule
  \end{tabularx}
  \vspace{0.2cm}
  \caption{The common specification to which the event generators are configured for the benchmark study.
    }
  \label{tab:settings}
\end{table}

The renormalisation and factorisation scales are set dynamically to $\mu_F=\mu_R=Q$, and their variation using the seven-point prescription~\cite{NNPDF:2019ubu} is used to estimate missing higher-order uncertainties (MHOUs), whose size we quote as half the width of the resulting band.
We use the $G_\mu$-matched electroweak scheme and switch off QED radiation in the simulations; the impact of the latter is studied separately in Sect.~\ref{sec:hadronic}.
Multiple parton interactions play no role, since only one of the two colliding particles carries partons, while final-state interactions with the cold nuclear medium are neglected and their study is beyond the scope of this paper.

The heavy-quark scheme is strictly massless, both for charm and bottom quarks, so that generators can be compared against zero-mass structure functions without a scheme ambiguity. 
We estimate the impact of charm mass effects both from the analytical structure-function side, using the FONLL general-mass scheme~\cite{Forte:2010ta}, up to NNLO for neutral-current scattering and up to NLO for charged-current scattering, for which the massive coefficient functions, although known at NNLO~\cite{Berger:2016inr,Gao:2017kkx,Caola:2026kvl}, are not implemented in \yadism, and from the event generator side using \powhegtwo for charm production in neutrino DIS~\cite{Buonocore:2024pdv}, which accounts for charm mass effects in the $s\to c$ and $d \to c$ transitions.
Whenever mass effects are included in the simulations, we indicate so explicitly. 
The massive CC calculation of \powhegtwo is benchmarked with its \yadism counterpart in the fixed-flavour-number scheme with $n_f=3$ (FFNS $n_f=3$), finding excellent agreement. 

The stability criterion which defines a final-state hadron in the detector is $c\tau>10$~mm.
The chosen criterion implies that we assume that charmed hadrons, whose $c\tau$ is between about $45$ and $310$~$\mu$m, never appear as final-state particles and therefore  they must be counted at the point of production.
While at FASER$\nu$ this is not necessarily true ($D$-mesons may be detected before decay), this distinction is immaterial from the point of view of the present study.

\paragraph{Observables.}
Three main groups of physical observables are considered in the present work:

\begin{enumerate}
  \item Observables constructed from the DIS kinematics, namely from the incoming and outgoing lepton momenta and the total energy  of the hadronic final state.
  These are the momentum transfer, $Q^2$, the momentum fraction, $x_{\rm Bj}$, the inelasticity, $y$, and the invariant mass of the hadronic final state, $W$ (all Lorentz invariants) as well as the scattered-lepton  energy $E_{\ell'}$ and angle $\theta_{\ell'}$  and the energy transfer $\nu$ (or equivalently the total hadronic energy $E_h$) in the target rest frame.
  These observables can be evaluated both with analytical structure functions as well as at the event generator level, and hence are used to validate the implementation of the latter. 

  \item Observables depending on the final-state hadronic activity, such as the charged multiplicity $N_{\rm ch}$, the energy of the leading charged hadron $E_{\rm lead}$, and $\Delta\phi$, the azimuthal separation between the scattered lepton and the vector sum of the charged hadrons.  
  These observables probe the shower structure and the hadronisation model, and are directly relevant to event selection at the LHC far-forward detectors. 

  \item Observables constructed from a charm tag (see below), such as DIS observables restricted to charm-tagged events and the 
   multiplicities and energy spectra of $D^\pm$ and $D^0/\bar{D}^0$ mesons, as well as the ratio between charged and neutral $D$-mesons.  
      
\end{enumerate}

\paragraph{Charm tagging.}
\label{sec:charmdef}
To connect predictions from the analytical structure-function framework for charm production with those provided by event generators we need a functional definition of charm tagging. 
From the event generator side, a charm event is tagged by the presence of at least one charmed hadron produced in the event after parton showering and hadronisation.
This requirement hence applies to open charm of any species, its excited
states, charmed baryons, and charmonium.
Furthermore, this charmed hadron must be prompt. 
We note that this is a hadron-level definition, chosen to mimic charm tagging in experimental FASER analyses and is different from a charm tag 
on the hard process.
For instance, in NC DIS, in addition to charm-initiated scattering, $\gamma^*/Z + c \to c$ and photon--gluon fusion, $\gamma^* g \to c\bar{c}$, this definition also adds charm produced by $g \to c\bar{c}$ splitting inside the shower and the charm carried by the beam remnant, which are absent from the corresponding structure function analytical calculations (though the former are partially captured by the NNLO corrections). 

On the structure function side, neutral-current charm production is defined as the charm-flavour contribution to $F_2$, $F_L$ and $xF_3$ following~\cite{Forte:2010ta}, namely the contribution from those diagrams where the charm quark couples to the virtual photon.
Diagrams which have charm in the final state but where the light quark couples to the virtual photon are included in the light component of the NC structure functions. 
Up to NLO this definition coincides with a tag on the charm quark in the hard process, and from NNLO onwards the two differ.
The hadron-level tag used for the event generators differs from both definitions since it also counts the charm produced in shower splittings and during fragmentation, as well as charm carried by the beam remnant. 
Nevertheless, as we will show in Sect.~\ref{sec:charm}, good agreement is found for inclusive and differential charm-tagged cross-sections between the generator and analytical predictions for the FASER kinematics. 

Concerning charged-current DIS, charm production is defined~\cite{Ball:2011mu} as all contributions proportional to the CKM matrix elements $V_{cd}$ or $V_{cs}$, that is, by the diagrams where the charm quark couples to the $W^\pm$ boson.
In this case, the partonic definition differs from a tag on the charm quark in the hard-process final state already at LO. 
For instance, the channel $\nu \bar{c} \to \mu^- \bar{s}$ carries $|V_{cs}|^2$ and so belongs to the charm structure function definition, but there the charm quark is in the initial state and is absent from the hard-process final state. 
Nevertheless, charm tagging may still arise via the $c$ present in the beam remnant as required by the vanishing charm number of the nucleon. 

We note that \sherpa's parton shower treats the charm quark as massless and produces $g \to c\bar c$ at
a rate an order of magnitude higher than the massive showers of \pythia and
\herwig, unphysically inflating the $D$-meson rates.
To correct for this, charm tags in \sherpa are obtained by demanding both a charm quark in the event hard record and a charm hadron produced in
the event.

Charm-production observables are evaluated in a fiducial region with a tighter requirement on the hadronic invariant mass, $W > 5$~GeV, applied to both currents, to the dimuon selection, and also to the inclusive cross-sections that normalise charm fractions.
In NC scattering, the purely massless ZM calculation populates charm-initiated events down to the $W > 3$~GeV boundary, whereas in a showered event the anticharm quark is left in the beam remnant, so that a hadronic final state can only be formed above a threshold of $W \simeq 4.4$~GeV.
Below this threshold \pythia and \herwig cannot hadronise such events and discard them, depleting the charm-tagged cross-section and degrading the agreement with the analytical charm structure-function calculation.

\paragraph{Theoretical uncertainties.}
As mentioned above, MHOUs associated with the truncation of the QCD expansion are estimated through the standard seven-point variation of $\mu_R$ and $\mu_F$ about the central choice $\mu_R = \mu_F = Q$.
This is done both in \yadism~\cite{Candido:2024rkr,Candido:2023utz} and in \powheg, in the former case exploiting that the scale dependence of the coefficient functions is known analytically 
from the scale independence of the structure functions, and in the latter case using its built-in event reweighting feature.
We have verified that MHOUs estimated using these two independent approaches are fully consistent. 

We also estimate the theoretical uncertainties associated with the nucleon PDF through dedicated studies where again both \yadism and \powheg calculations are repurposed for different PDF sets as well as each set's own
uncertainty, which requires no additional calculation.
Specifically, in addition to NNPDF4.0, we consider predictions from CT18~\cite{Hou:2019efy}, MSHT20~\cite{Bailey:2020ooq}, ABMP16~\cite{Alekhin:2017kpj}, and ATLASpdf21~\cite{ATLAS:2021vod}, in all cases at NNLO accuracy and taking the strong coupling to be $\alpha_s(m_Z)=0.118$.
Furthermore, results are also presented with GRV98LO~\cite{Gluck:1998xa}, which is the PDF set underlying \genie's baseline tune.

\section{Event generation for inclusive DIS}
\label{sec:inclusive}
\label{sec:results}

We consider first the results of event generator benchmarking for inclusive DIS, where only the fiducial cuts of Eq.~(\ref{eq:fiducial}) are applied, and compare predictions both for total cross-sections and for differential distributions in the DIS kinematic variables.
Subsequently we consider charm-tagged cross-sections (Sect.~\ref{sec:charm}), observables dependent on the final-state hadronic system for the FASER$\nu$ acceptance (Sect.~\ref{sec:hadronic}), and applications to FASER where cross-sections are weighted by the LHC neutrino fluxes (Sect.~\ref{sec:faser-app}). 

\begin{figure}[!tbp]
  \centering
  \includegraphics[width=0.98\textwidth]{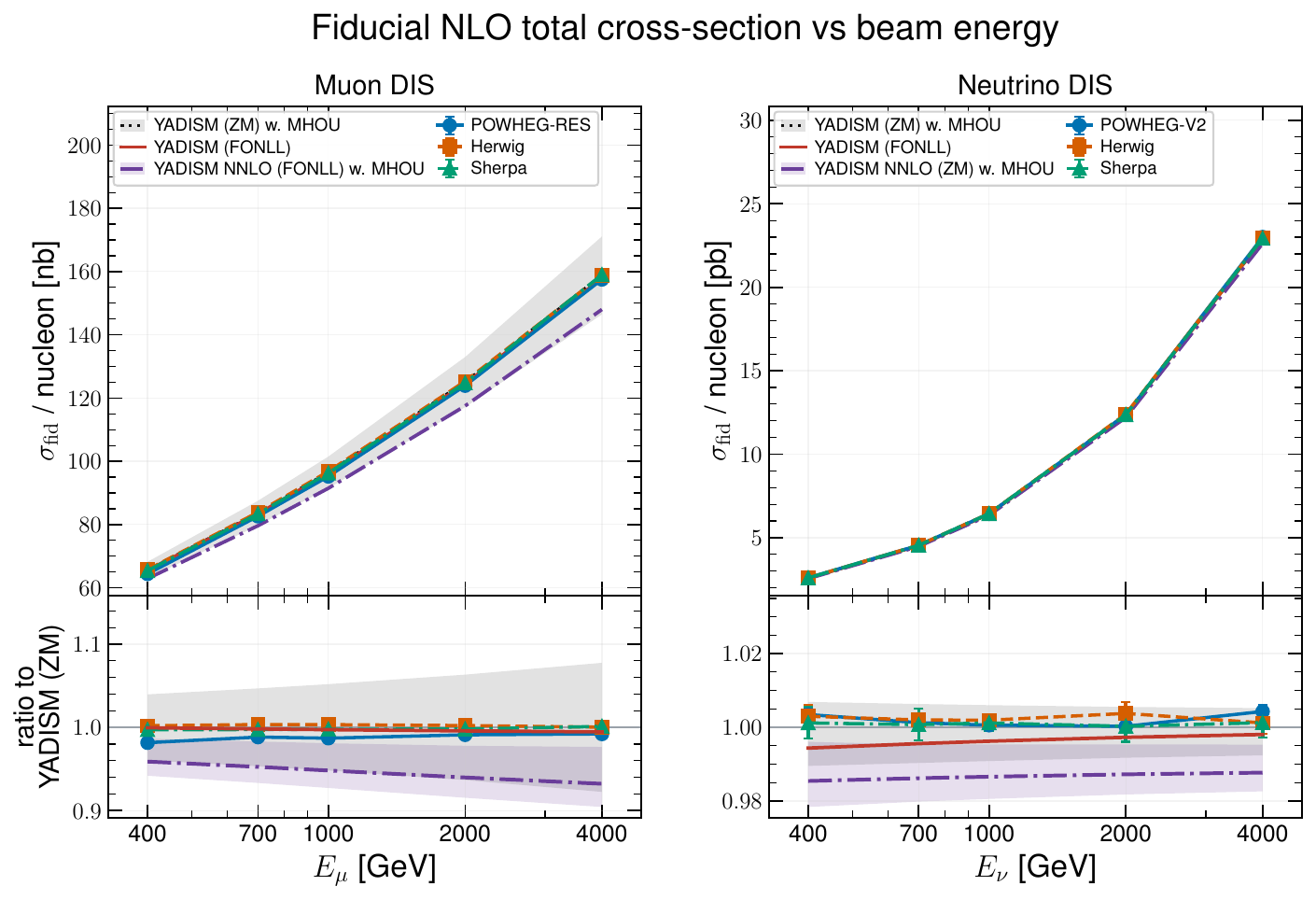}
  \caption{Inclusive fiducial cross-section per nucleon on a tungsten target as a function of the beam energy for muon $(\mu^-)$ neutral-current (left) and muon-neutrino ($\nu_\mu$) charged-current (right panel) scattering in the fiducial region defined by Eq.~(\ref{eq:fiducial}).
We provide results for three NLO-matched event generators (\sherpa, \herwig, and \powheg) as well as for the analytical calculations from \yadism at NLO (ZM-VFNS and FONLL) and at NNLO (FONLL for NC and ZM-VFNS for CC).
The bottom panels display the ratio to the massless NLO \yadism calculation.
The error bands in the generator predictions correspond to the MC statistical uncertainties.
    }
  \label{fig:sigma-vs-energy}
\end{figure}

\paragraph{Total fiducial cross-sections.}
\label{sec:incl-nlo}
\label{sec:incl-scales}
Fig.~\ref{fig:sigma-vs-energy} displays the inclusive per-nucleon cross-section in the fiducial region  Eq.~(\ref{eq:fiducial}) evaluated  on a tungsten target as a function of the beam energy for muon  neutral-current and muon-neutrino charged-current DIS.
We provide results for three NLO-matched event generators, \sherpa, \herwig, and \powheg, as well as for the analytical structure-function calculations from \yadism at NLO and NNLO.
MHOUs are evaluated from the \yadism prediction at each order, in the same scheme as the corresponding central curve.
At NLO, \yadism predictions are provided in the ZM-VFNS and FONLL schemes, so that their difference estimates the impact of charm mass effects in the total cross-section.
At NNLO, we use FONLL for neutral-current and ZM-VFNS for charged-current scattering, since in the latter case the massive coefficient functions, although known at NNLO~\cite{Berger:2016inr,Gao:2017kkx,Caola:2026kvl}, are not yet implemented in \yadism. 
The bottom panels display the ratio to the massless NLO \yadism calculation.

The total fiducial rates predicted by the three NLO-matched generators agree among themselves and with the \yadism reference at the $\lsim 2\%$ level.
Residual differences are covered by the MC statistics, except for \powhegres, which lies $1$--$2\%$ below the reference for muon DIS.
NNLO corrections range between $-4\%$ and $-7\%$ for muon DIS and are around $-1.5\%$ for neutrino DIS; for both currents the NLO and NNLO MHOU bands overlap.
MHOUs being larger for NC as compared to CC is partially explained by the fact that the former probe smaller values of $Q^2$ due to the $Q^{-4}$ enhancement of the virtual photon propagator. 
Raising the momentum-transfer cut to $Q^2 > 11$~GeV$^2$ roughly halves the MHOUs in NC DIS, although they remain larger than in CC.
Charm mass effects, as estimated by the \yadism FONLL calculation, are negligible at the level of inclusive fiducial cross-sections. 
The agreement between the three generators and the \yadism reference shown in Fig.~\ref{fig:sigma-vs-energy}, despite the use of different parton shower algorithms and hadronisation models, confirms that fiducial cross-sections are fully determined by the DIS matrix elements, which are identical in all the massless NLO calculations considered.

\begin{figure}[!tbp]
  \centering
  \includegraphics[width=0.96\textwidth]{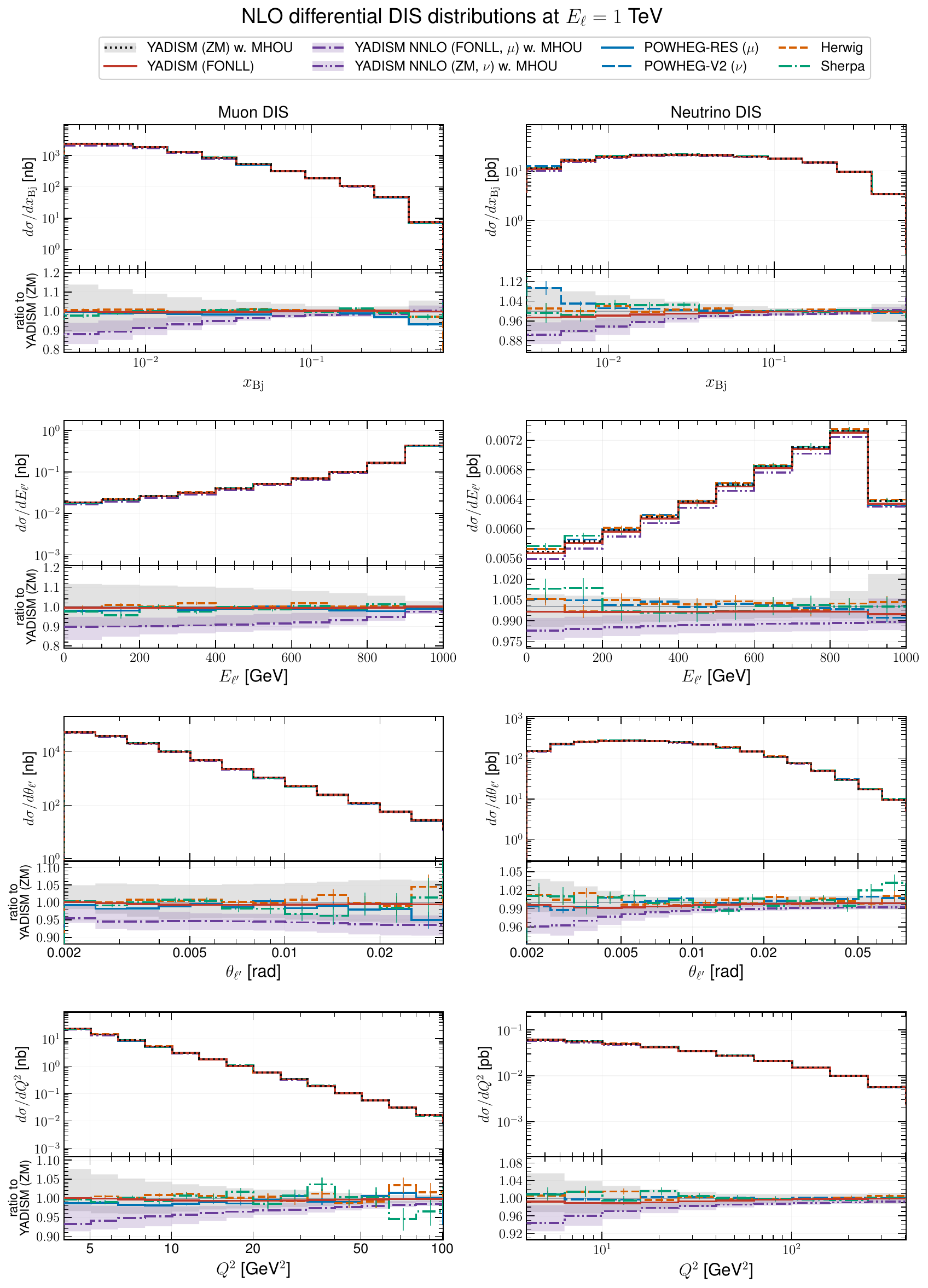}
  \caption{Same as Fig.~\ref{fig:sigma-vs-energy} for differential distributions in the fiducial region at $E_\ell=1$ TeV.
From top to bottom, we show the distributions in Bjorken $x_{\rm Bj}$, the scattered-lepton
    energy $E_{\ell'}$ and polar angle $\theta_{\ell'}$, and the momentum transfer $Q^2$, evaluated in the target rest frame.
    Error bars indicate  the  Monte Carlo integration statistics.}
  \label{fig:dis-distributions}
\end{figure}

\paragraph{Differential distributions.}
\label{sec:incl-diff}
Fig.~\ref{fig:dis-distributions} displays the same benchmark comparison between the predictions of NLO-matched event generators and analytical calculations as in Fig.~\ref{fig:sigma-vs-energy} for differential distributions in the same fiducial region.
For illustration purposes we show $E_\ell=1$ TeV as a
representative energy; we have verified that the results are qualitatively similar at other incoming lepton energies.
As in Fig.~\ref{fig:sigma-vs-energy}, we quote the differential cross-sections per nucleon on a tungsten target, whose integral reproduces the values of Fig.~\ref{fig:sigma-vs-energy}.
Error bars in the generator predictions are the  Monte Carlo integration statistics.
The bottom panels show the ratio to the NLO massless \yadism reference calculation.

From top to bottom, Fig.~\ref{fig:dis-distributions} displays the distributions in Bjorken $x_{\rm Bj}$, the scattered-lepton energy $E_{\ell'}$ and polar angle $\theta_{\ell'}$, and the momentum transfer $Q^2$, in all cases in the target rest frame.
Note that these distributions are not all independent, since for a fixed beam energy the scattering kinematics only has two degrees of freedom.
For instance, $Q^2 = 2 E_\ell E_{\ell'} (1 - \cos\theta_{\ell'})$.
For this reason, the hadronic energy $E_h$ and inelasticity $y$ distributions are not shown, since for fixed $E_\ell$ they are both linear in $E_{\ell'}$, hence comparing the different calculations for the latter is sufficient. 
These are observables that the  analytic \yadism calculation can predict, since they are built from the leptonic kinematics without reference to the hadronic final state.

The main result of Fig.~\ref{fig:dis-distributions} is the excellent agreement between the NLO generators and the analytical \yadism ZM reference for both currents, echoing that of the  integrated cross-sections.
Residual differences are at the per-cent level, are consistent with the MC statistical fluctuations, and overwhelmingly covered by the NLO MHOU band. 
As for the fiducial cross-sections, this agreement is reassuring since DIS distributions should not be sensitive to the modelling of the hadronic final state and depend only on the NLO partonic matrix element, all other physical inputs kept the same.

From Fig.~\ref{fig:dis-distributions} we also find that, as was the case for inclusive cross-sections, NNLO corrections can be sizeable for NC scattering across the whole phase space.
For CC scattering, NNLO corrections can be considerably larger than at the inclusive level, reaching for example $-10\%$ at small $x_{\rm Bj}$ and $-6\%$ at small $Q^2$.
This result strengthens the need for DIS event generators reaching NNLO accuracy to ensure the full scientific exploitation of the FASER and SND@LHC far-forward physics programme.
Charm-mass effects, on the other hand, remain negligible also at the level of differential distributions, and in particular are always smaller than the spread between generators.
We conclude that for inclusive and differential DIS distributions, purely massless calculations should be reliable for the FASER kinematics.

\begin{figure}[!tbp]
  \centering
  \includegraphics[width=0.98\textwidth]{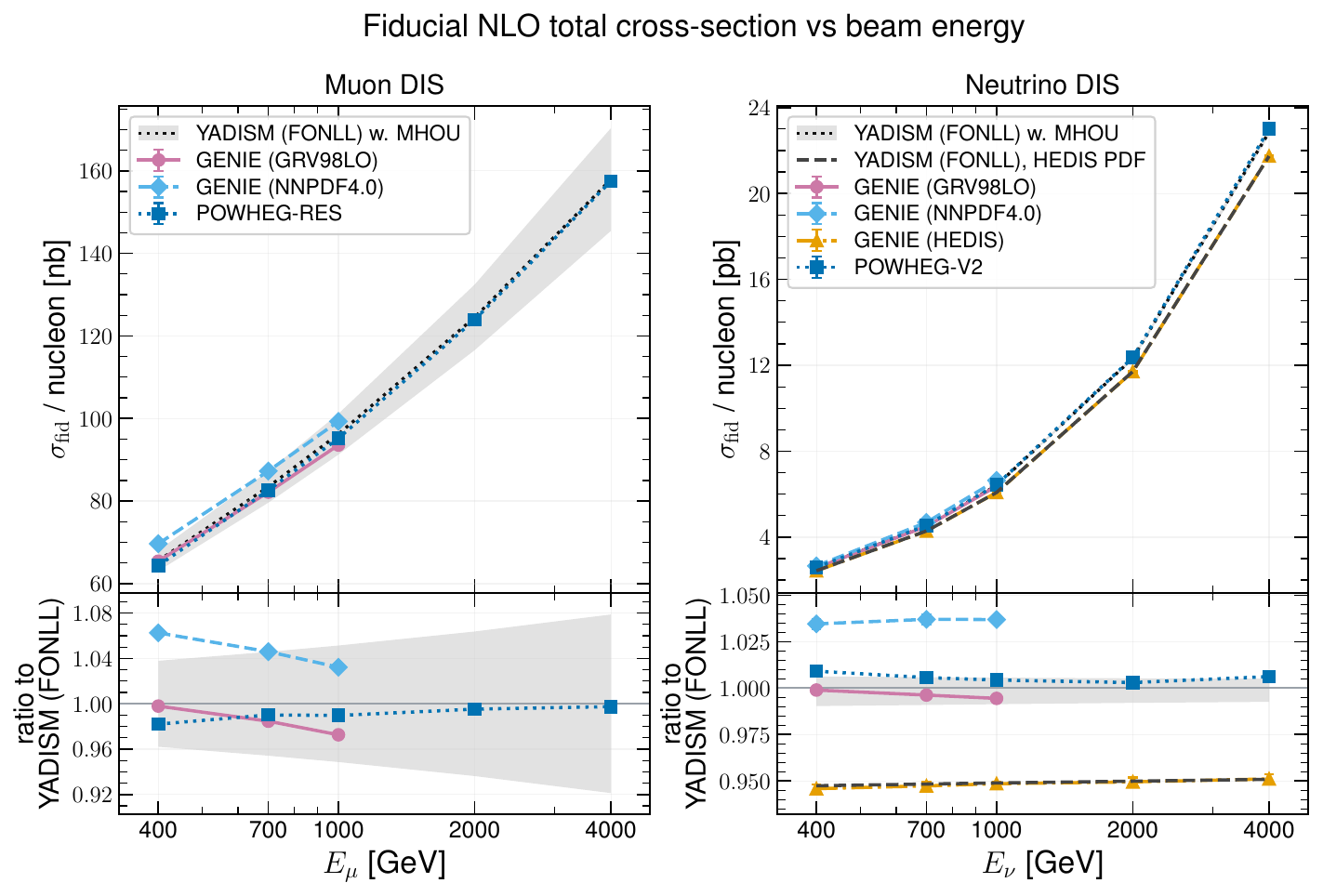}
  \caption{Same as Fig.~\ref{fig:sigma-vs-energy}
 comparing the total inclusive cross-sections predicted by \yadism NLO in the FONLL scheme with \powhegres (muon DIS) and \powhegtwo (neutrino DIS) and different \genie configurations: GRV98LO (default), same with NNPDF4.0, and the {\sc\small HEDIS} tune. 
  In the right panel we also display the  \yadism FONLL prediction evaluated with the same PDF set as {\sc\small HEDIS} uses, namely {\tt NNPDF31sx\_nlo\_as\_0118\_LHCb\_nf\_6}.
  The lower panels display the ratio to the \yadism FONLL reference.
  The default \genie tune stops at $E_\ell=1$ TeV, its declared limit of validity. 
 }
  \label{fig:genie-vs-energy}
\end{figure}

\paragraph{Comparison with \genie.}
Having established that three NLO-matched event generators agree with the structure-function calculation for a common set of physics settings, we move to discuss the comparison with \genie, the baseline generator used in FASER.
Fig.~\ref{fig:genie-vs-energy} displays, in the same format as Fig.~\ref{fig:sigma-vs-energy}, a comparison of the total inclusive cross-sections predicted by \yadism NLO in the FONLL general-mass scheme with \powheg and different \genie tunes.
 In the left panel, we show predictions with the \genie~{\tt G18\_02a} tune (LO matrix elements in the Bodek--Yang model) and GRV98LO and NNPDF4.0 NNLO as PDF sets.
 In the right panel, we show the same LO tune and in addition the NLO {\tt GHE19\_00a} tune with {\sc\small HEDIS}, which evaluates NLO structure functions with \apfel. 
  The lower panels display the ratio to the \yadism FONLL reference.
  In all panels, the \yadism MHOUs are displayed.
  For the case of the {\tt G18\_02a} tune we stop the energy scan at $E_\ell=1$ TeV, its declared limit of validity.\footnote{Within our pipeline it is possible to generate \genie splines extending to higher values of $E_\ell$, should the user require it.} 
The {\tt G18\_02a} tune of \genie with GRV98LO as input PDF is the current baseline calculation used in FASER data analysis. 
We use \yadism FONLL as reference since this \genie cross-section model also accounts for the charm mass effects. 

From Fig.~\ref{fig:genie-vs-energy} we observe that the default \genie tune agrees with both NLO calculations (\yadism and \powheg) due to an accidental cancellation of competing effects, namely the PDF and the matrix-element model.
For muon DIS, using in \genie the same PDF set as the NLO calculations (NNPDF4.0) results in a cross-section up to 6\% higher than the \yadism NLO reference.
For neutrino DIS a similar effect is found, and there the NLO tune based on {\sc\small HEDIS} predicts a suppression of the inclusive rates by 5\% in the considered $E_\nu$ range.
This last offset is entirely a PDF effect.
Indeed, {\tt GHE19\_00a} evaluates its structure functions with {\tt NNPDF31sx\_nlo\_as\_0118\_LHCb\_nf\_6}, whose light-quark densities lie below those of \nnpdf over the $x$ range that dominates the rate.\footnote{This PDF set was tailored for applications to ultra-high-energy (UHE) neutrino astronomy, with $D$-meson data from the LHCb experiment constraining the small-$x$ gluon PDF~\cite{Gauld:2016kpd}.} 
When the \yadism FONLL calculation is evaluated with the same PDF set, the agreement with the  {\sc\small HEDIS} predictions is excellent. 
From the assessment of Fig.~\ref{fig:genie-vs-energy} one can infer that, while the default tune of \genie reproduces the predictions of NLO QCD calculations with modern PDFs, this agreement will not extend to differential observables due to its accidental nature.

\begin{figure}[!tbp]
  \centering
  \includegraphics[width=0.96\textwidth]{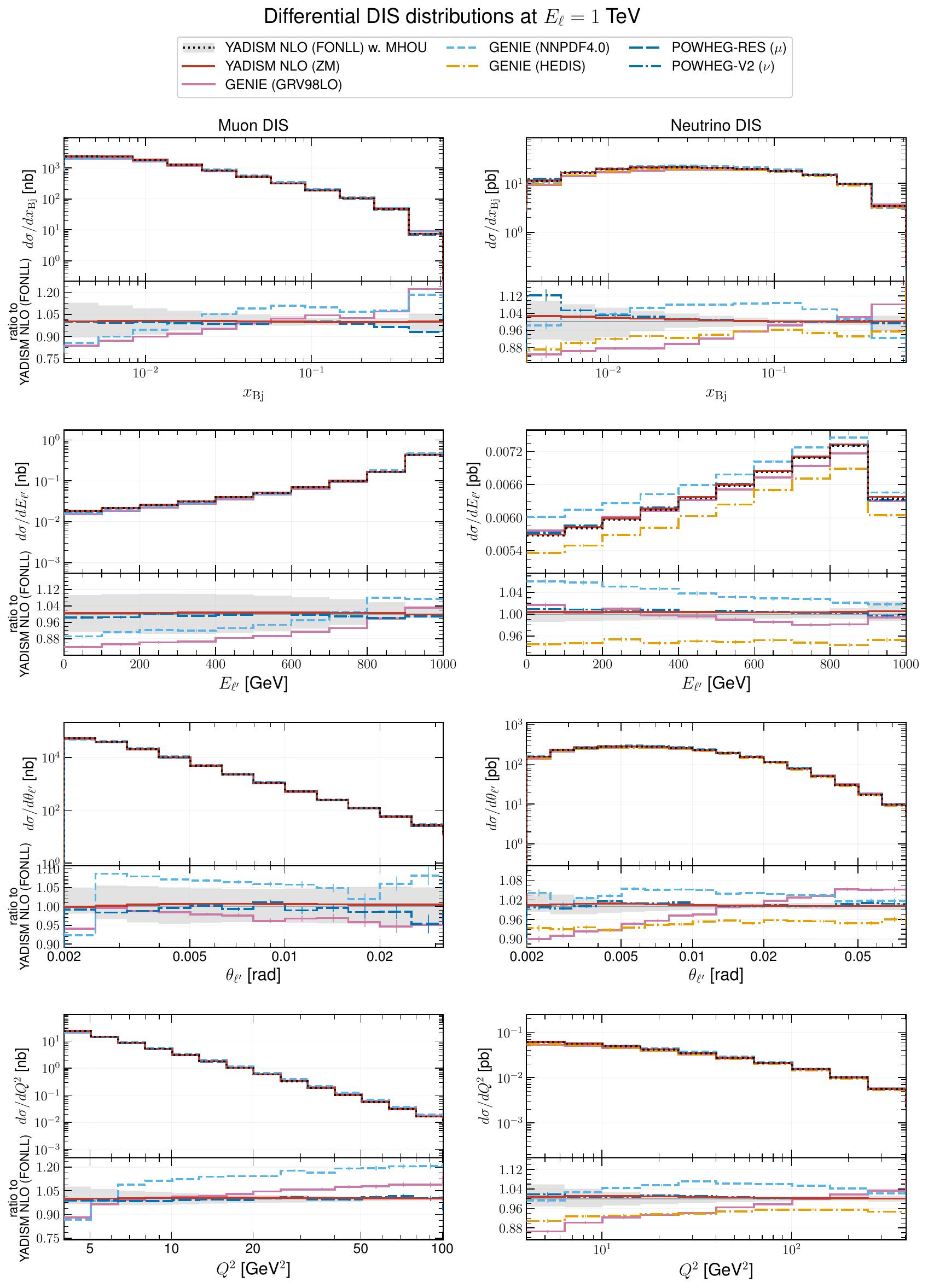}
  \caption{Same as Fig.~\ref{fig:dis-distributions}
  now comparing the same muon and neutrino DIS calculations of Fig.~\ref{fig:genie-vs-energy} including the \genie predictions considered in this work (two tunes and a PDF variant).
}
  \label{fig:genie-dis-distributions}
\end{figure}

To illustrate this,  Fig.~\ref{fig:genie-dis-distributions} compares the same differential distributions of Fig.~\ref{fig:dis-distributions} between \yadism, \powheg, and the \genie tunes as shown in Fig.~\ref{fig:genie-vs-energy}.
As opposed to the inclusive case, differences between the default \genie tune (with GRV98LO as input PDF) and the NLO QCD calculations (\yadism and \powheg, which agree with each other for all observables) can be significant.
For example, these differences range between $-16\%$ and $+22\%$ ($-15\%$ and $+8\%$) for the $x_{\rm Bj}$ distribution in muon (neutrino) DIS, and up to $-6\%$ ($-10\%$) for the corresponding $\theta_{\ell'}$ distributions.
Furthermore, these differences are in general well outside the MHOU band of the NLO calculation, and are removed neither by changing the PDF to NNPDF4.0 in \genie nor by changing the tune to {\sc\small HEDIS}. 
One concludes that using \genie instead of NLO QCD calculations could introduce a systematic modelling bias in the analysis and interpretation of muon and neutrino DIS at FASER and SND@LHC.

In Sect.~\ref{sec:hadronic} we revisit the comparison for observables at the level of the FASER$\nu$ selection, and in Sect.~\ref{sec:faser-app} with the cross-sections weighted by the energy-dependent LHC neutrino fluxes.

\section{Charm production}
\label{sec:charm}

Following the discussion of inclusive DIS observables in the previous section, we move to study charm production; see Sect.~\ref{sec:settings} for the charm-tagging definition adopted.
For this charm-production analysis we use somewhat tighter fiducial cuts, namely $Q^2>4$~GeV$^2$ and $W>5$~GeV  instead of Eq.~(\ref{eq:fiducial}) to account for the larger typical invariant mass of the hadronic final state in this subprocess. 

\begin{figure}[!tbp]
  \centering
  \includegraphics[width=0.99\textwidth]{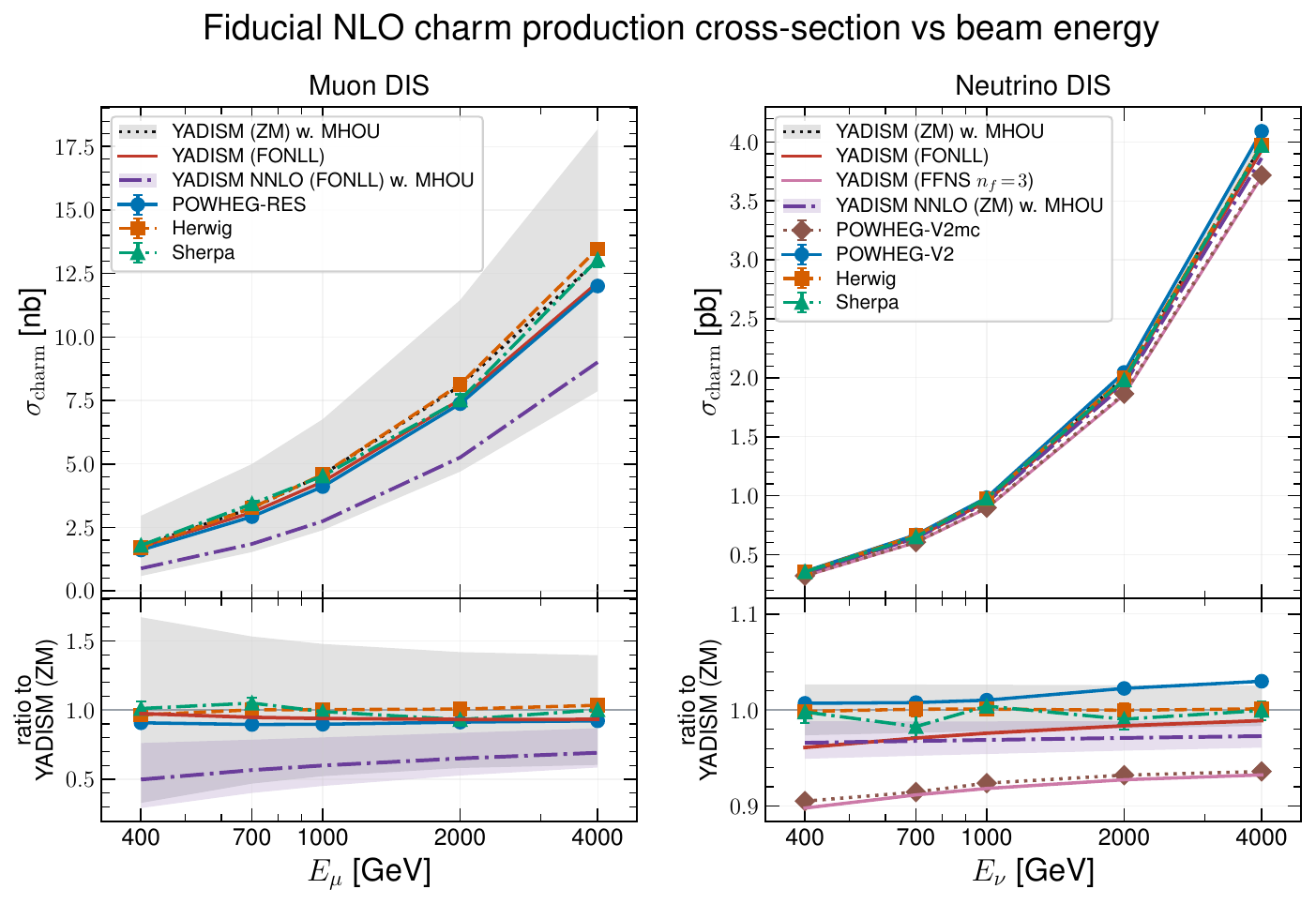}
  \caption{Same as Fig.~\ref{fig:sigma-vs-energy} now for inclusive charm production in the fiducial region with $W>5$~GeV; see Sect.~\ref{sec:settings} for the charm-tagging definition.
For neutrino DIS, we also display the predictions of \powhegtwo for which $m_c\ne 0$ (\powhegtwomc), together with the \yadism calculation in the FFNS $n_f=3$ scheme.
}
  \label{fig:charm-vs-energy}
\end{figure}

\paragraph{Fiducial cross-sections.}
Fig.~\ref{fig:charm-vs-energy} displays the same comparison as Fig.~\ref{fig:sigma-vs-energy} now for inclusive charm production.
For neutrino DIS, we also display the predictions of \powhegtwo for which we allow a non-zero value of the charm mass, $m_c\ne 0$, which are denoted by \powhegtwomc in the following.
To benchmark the latter, we also consider here the \yadism calculation in the FFNS $n_f=3$ scheme, with no charm in the proton, which corresponds to the same heavy-quark mass scheme as that adopted in the \powhegtwomc predictions.

From Fig.~\ref{fig:charm-vs-energy} we observe that the three NLO-matched generators agree with the \yadism ZM NLO reference within a few per cent for neutrino DIS and within about $10\%$ for muon DIS, with residual differences largely covered by the MHOU band of the NLO calculation.
Charm quark mass effects in the matched calculation can be estimated from the shift between the ZM-VFNS and FONLL 
\yadism predictions and amount to $3$--$7\%$ for muon DIS and to $1$--$4\%$ for neutrino DIS, in the latter case comparable to the magnitude of the NLO MHOU band.

Charm mass effects at the event generator level in neutrino DIS are quantified by comparing \powhegtwomc with \powhegtwo, with the former lying $8$--$10\%$ below the latter depending on the value of $E_\nu$.
\powhegtwomc does not resum potentially large logarithms $\ln(Q^2/m_c^2)$ encoded in the charm PDF as the FONLL calculation does, and hence it may underestimate the charm-production cross-section.
The \powhegtwomc prediction agrees closely with the corresponding \yadism calculation in the same mass scheme (FFNS $n_f=3$).
The spread between purely massive, massless, and matched calculations in the right panel of Fig.~\ref{fig:charm-vs-energy} emphasises the importance of carefully accounting for heavy-quark mass effects in predictions for charm production in DIS event generation.  

Fig.~\ref{fig:charm-vs-energy} also indicates that the perturbative behaviour of NC and CC predictions is quite different.
For neutrino DIS, the NNLO correction to the charm cross-section is at most $-3\%$ at every energy, and the size of the MHOUs is reduced from about $2.5\%$ at NLO to $1.5$--$2\%$ at NNLO.
We note however that this comparison is carried out at the purely massless level, and accounting for mass effects in the FONLL matched calculation extended to NNLO may modify this conclusion.
For charm production in muon DIS instead, the perturbative series converges slowly, with NNLO corrections being as large as $-50\%$, though a marked reduction of the associated MHOU band is also observed at the highest energies. 
This slow convergence follows from the low-$Q^2$ behaviour of the charm coefficient functions near threshold, driven by the $Q^{-4}$ enhancement of the photon propagator which pins the neutral-current rate against the $Q^2$ cut at every beam energy.
As for the inclusive DIS case discussed in Sect.~\ref{sec:inclusive}, a significant improvement of the perturbative convergence of charm production is observed by raising the $Q^2$ cut above $4$~GeV$^2$.
For instance, for $Q^2>11$~GeV$^2$ the NNLO/NLO ratio of the muon charm cross-section increases from $0.51$--$0.74$ to $0.78$--$0.86$ in the considered energy range, while the size of the MHOUs, at most $47\%$ at both NLO and NNLO for $Q^2>4$~GeV$^2$, is reduced to $24\%$ at NLO and to $16\%$ at NNLO, the largest values being found at $E_\mu=400$~GeV.

\begin{figure}[!tbp]
  \centering
  \includegraphics[width=0.7\textwidth]{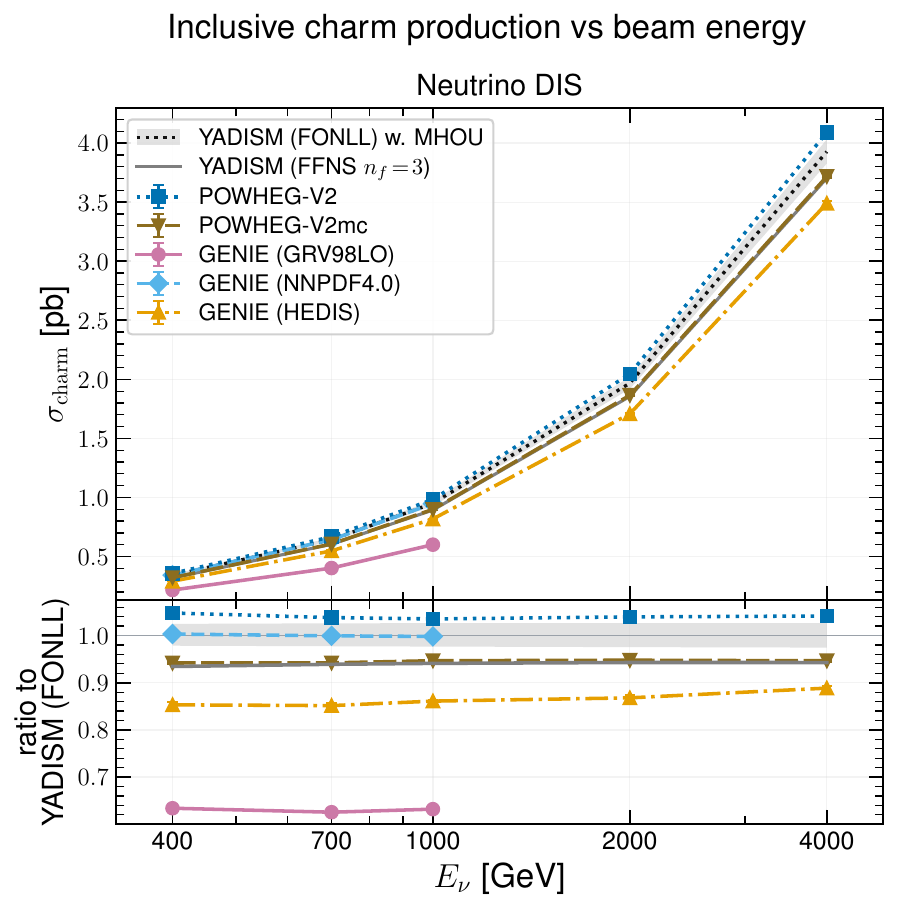}
  \caption{Same as Fig.~\ref{fig:charm-vs-energy}, comparing the NLO \yadism (FONLL and FFNS $n_f=3$) and \powheg calculations of charm production in neutrino-induced DIS with those from \genie for the same configurations considered  in  Fig.~\ref{fig:genie-vs-energy}.
    The bottom panels display the ratio to the \yadism FONLL prediction.
    }
  \label{fig:genie-charm-vs-energy}
\end{figure}

Fig.~\ref{fig:genie-charm-vs-energy} compares the NLO \yadism (FONLL and FFNS $n_f=3$) and \powheg calculations of charm production in neutrino DIS shown in Fig.~\ref{fig:charm-vs-energy} with those from \genie for the same configurations considered  in  Fig.~\ref{fig:genie-vs-energy}.
The default \genie tune (GRV98LO) lies about $40\%$ below the FONLL reference, a deficit which is mostly related to the choice of PDFs and in particular to the shape and magnitude of the strange sea PDF. 
Indeed, substituting \nnpdf for GRV98 within the same \genie tune results in a prediction in agreement with the \yadism FONLL reference. 
The NLO {\sc\small HEDIS} tune also predicts a suppression of up to 15\% as compared to the reference, also explained by PDF differences as was the case for inclusive cross-sections in Sect.~\ref{sec:inclusive}.
This PDF sensitivity emphasises that FASER neutrino DIS data provide a powerful probe of the strange content of the nucleon, as we further quantify in Sect.~\ref{sec:faser-app}.

\begin{figure}[!tbp]
  \centering
  \includegraphics[width=0.94\textwidth]{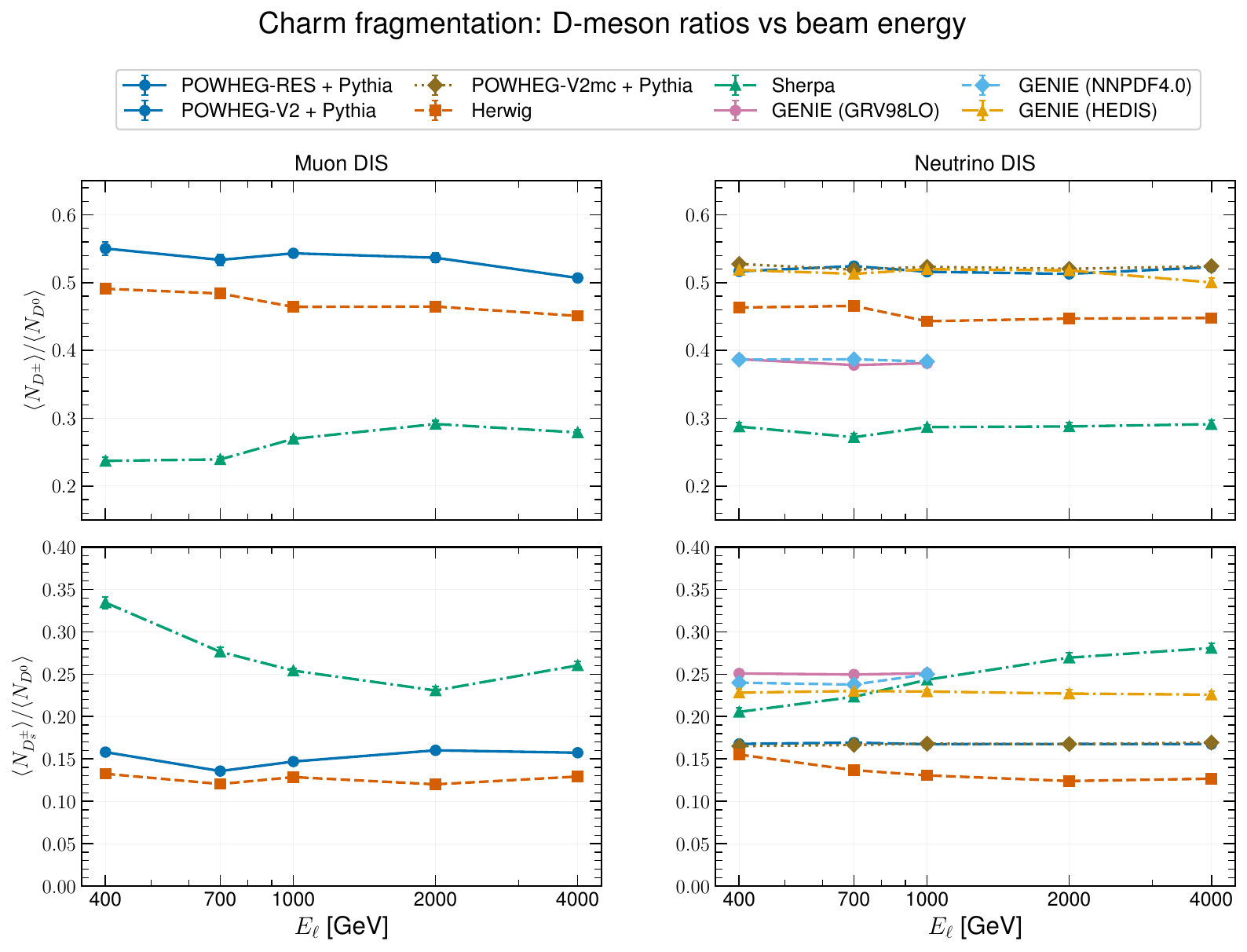}
  \caption{The ratios $\langle N_{D^\pm}\rangle/\langle N_{D^0}\rangle$ (top) and $\langle N_{D_s^\pm}\rangle/\langle N_{D^0}\rangle$ (bottom) between different $D$-meson species in charm-tagged events on tungsten as a function of the beam energy, for muon (left) and neutrino (right panels) DIS evaluated in the same fiducial region as Fig.~\ref{fig:charm-vs-energy}.
    $D$-mesons from the decays of excited states, such as $D^*\to D\pi$, are also included.
    The predictions for \genie are shown only for neutrino DIS. 
    }
  \label{fig:dmeson-ratios}
\end{figure}

\paragraph{Charm fragmentation.}
The fact that the various NLO event generators considered in Fig.~\ref{fig:charm-vs-energy} agree at the charm-tagged cross-section level when a common set of input settings is adopted does not necessarily imply that they all predict the same pattern of charm fragmentation into different $D$-meson species. 
To assess this feature, Fig.~\ref{fig:dmeson-ratios} displays the ratios $\langle N_{D^\pm}\rangle/\langle N_{D^0}\rangle$ and $\langle N_{D_s^\pm}\rangle/\langle N_{D^0}\rangle$ between different $D$-meson species in charm-tagged events  evaluated in the same fiducial region as Fig.~\ref{fig:charm-vs-energy}.
$D$-mesons arising from the decays of excited states, such as $D^*\to D\pi$, are also included.

From Fig.~\ref{fig:dmeson-ratios} we find that the ratio of the mean numbers of charged to neutral $D$ mesons, $\la N_{D^\pm}\ra/\la N_{D^0}\ra$, and the ratio $\la N_{D_s^\pm}\ra/\la N_{D^0}\ra$ in charm-tagged events differ between generators.
For instance, concerning the $\la N_{D^\pm}\ra/\la N_{D^0}\ra$ ratio, predictions based on the Lund string (the \powheg calculations showered with \pythia, and the {\sc\small HEDIS} tune of \genie) give $0.50$--$0.55$, while the cluster hadronisation model of \herwig predicts $0.44$--$0.49$, and the default \genie tune around $0.38$.
The difference between the two \genie configurations is due to their charm hadronisation: in the default tune the charm-hadron species is drawn from fixed fractions, which above $100$~GeV are $D^0:D^\pm:D_s^\pm=0.60:0.23:0.15$ (hence $D^\pm/D^0=0.38$), with its energy sampled from a Peterson fragmentation function and only the hadronic remnant hadronised by \pythia, whereas in the {\sc\small HEDIS} tune the final state is hadronised with the Lund string model of \pythia.

The main outlier in Fig.~\ref{fig:dmeson-ratios} is the \sherpa prediction, in the range $0.24$--$0.29$ and a factor of 2 smaller than the Lund string hadronisation model.
The $\la N_{D^\pm}\ra/\la N_{D^0}\ra$ ratio is nearly flat in energy for neutrino DIS, as expected since it depends only on physics taking place near the hadronisation scale. 
A similar spread is found for the $\la N_{D_s^\pm}\ra/\la N_{D^0}\ra$ ratio, from $0.12$--$0.17$ for the \powheg calculations showered with \pythia and for \herwig to $0.21$--$0.33$ for \sherpa, which also displays a marked variation with the energy. 

While it is beyond the scope of this work to understand the origin of these differences and whether they could be reconciled via dedicated hadronisation tunes or analytical calculations of charm production, it is clear that accounting for them is instrumental for FASER and SND@LHC analyses sensitive to the reconstruction of specific $D$-meson species.

\begin{figure}[!tbp]
  \centering
  \includegraphics[width=0.99\textwidth]{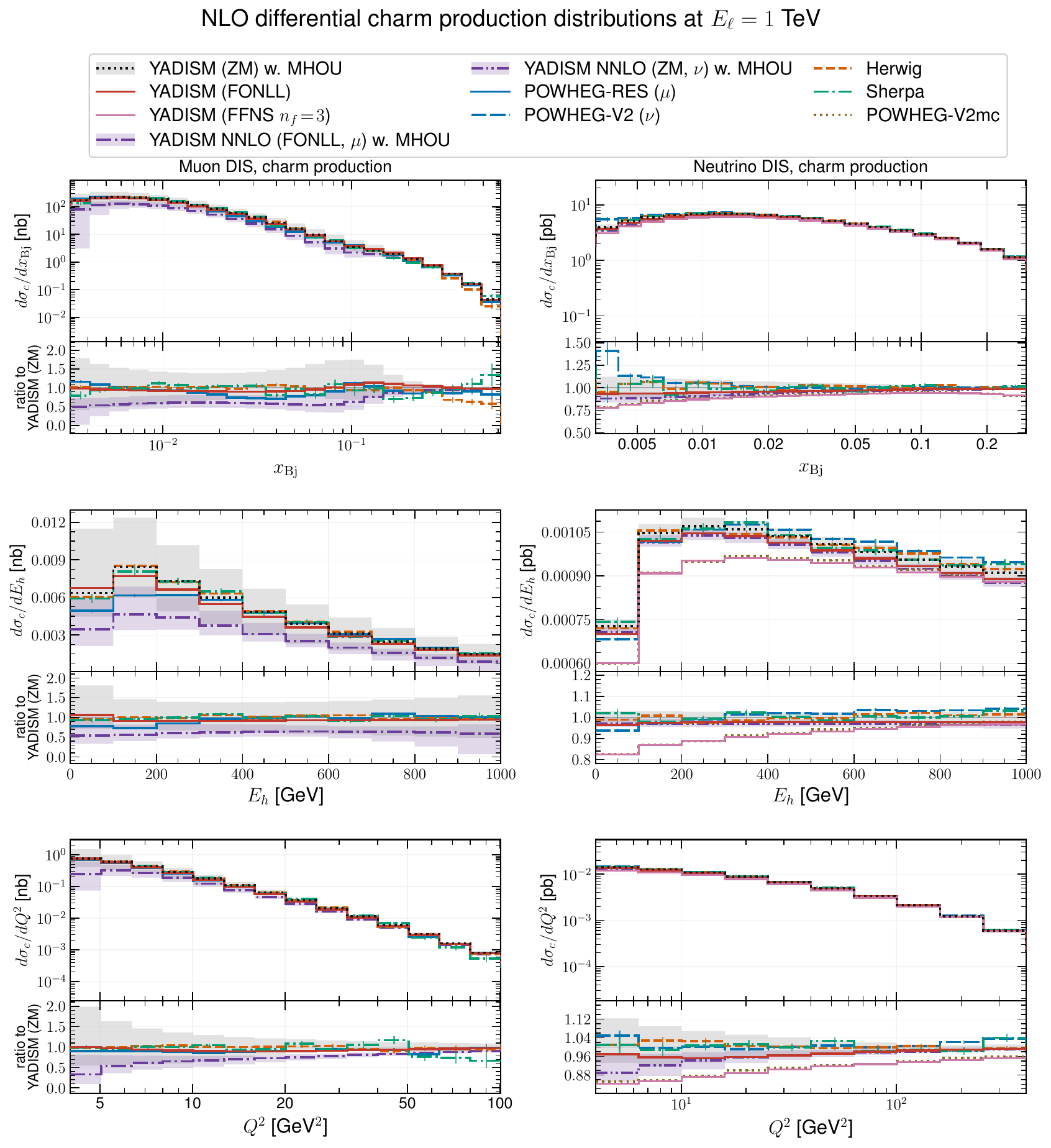}
  \caption{Same as Fig.~\ref{fig:dis-distributions}, now for the distributions in $x_{\rm Bj}$, $E_h$ and $Q^2$ in charm-tagged events for the same calculations displayed at the fiducial level in Fig.~\ref{fig:charm-vs-energy}.
  The \yadism NNLO predictions shown are obtained in the FONLL scheme for NC DIS and in the ZM-VFNS scheme for CC DIS.
  }
  \label{fig:charm-distributions}
\end{figure}

\paragraph{Differential distributions.}
Fig.~\ref{fig:charm-distributions} presents the DIS differential distributions in $x_{\rm Bj}$, $E_h$ and $Q^2$, in the same format as Fig.~\ref{fig:dis-distributions}, for charm-tagged events, for the same calculations displayed at the fiducial level in Fig.~\ref{fig:charm-vs-energy}.
The qualitative findings of this differential comparison are similar to those at the fiducial level: in general, there is good agreement between the NLO-matched event generators and the \yadism ZM reference for neutrino DIS, and agreement within about $10\%$ for muon DIS, with larger deviations near phase-space edges; MHOUs and the NNLO $K$-factors are larger for muon DIS than for neutrino DIS; 
the purely massive (FFNS) calculation of \powhegtwomc predicts a significant suppression of the differential cross-sections as compared to the FONLL-matched one, and is reproduced by its analytical \yadism counterpart. 
We conclude that this benchmark comparison is also successful for differential distributions in charm production. 

\begin{figure}[!tbp]
  \centering
  \includegraphics[width=0.99\textwidth]{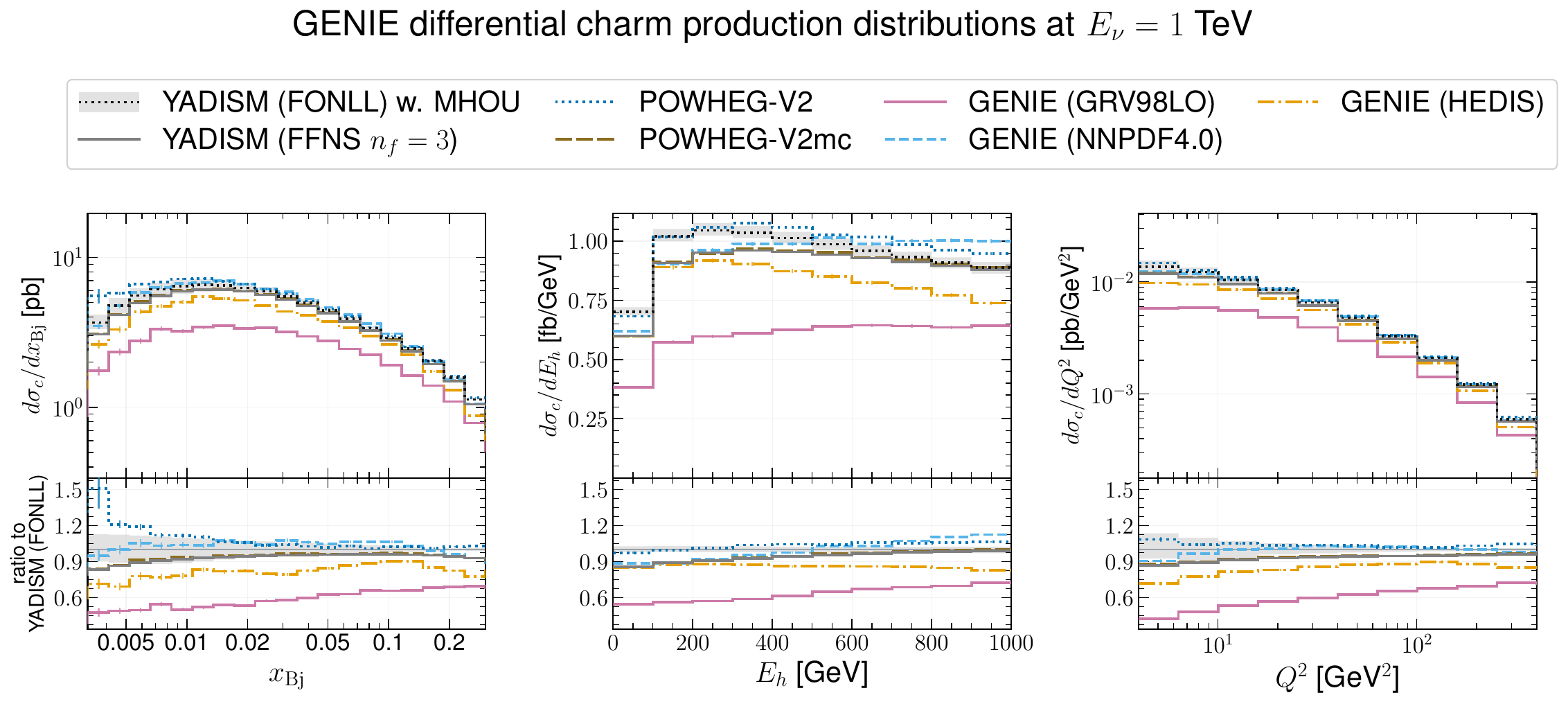}
  \caption{Same as Fig.~\ref{fig:genie-charm-vs-energy} for the differential distributions in $x_{\rm Bj}$, $E_h$ and $Q^2$ in charm-tagged neutrino DIS events at $E_\nu=1$ TeV.
    The bottom panels display the ratio to the \yadism FONLL prediction.
    }
  \label{fig:genie-charm-distributions}
\end{figure}

Finally, Fig.~\ref{fig:genie-charm-distributions} displays the same comparison for the DIS differential distributions as in Fig.~\ref{fig:charm-distributions} now including the \genie configurations, see Fig.~\ref{fig:genie-charm-vs-energy} for the corresponding predictions at the fiducial level.
From this comparison, we see that the deficit in the default \genie tune is not only a normalisation but also a shape effect, and is more marked at small $Q^2$ and small $x_{\rm Bj}$, where the GRV98LO strange sea PDF is most suppressed relative to \nnpdf.
Once \genie is run with \nnpdf, everything else unchanged,  the agreement improves but some differences remain, such as the tilt in the $E_h$ distribution, as a consequence of the charm production model adopted by \genie which predicts different yields as compared to the exact NLO QCD calculation. 
As was already the case for fiducial cross-sections, also at the differential level \powhegtwomc reproduces the \yadism FFNS $n_f=3$ calculation to within $3\%$.

\section{Predictions for hadronic final states}
\label{sec:hadronic}

The results presented in Sect.~\ref{sec:inclusive} focused on inclusive observables, where the only acceptance cuts imposed are those of Eq.~(\ref{eq:fiducial}), and where every observable is built from the initial- and final-state leptonic kinematics.
We established that the NLO-matched event generators agree with each other and with the analytic \yadism reference, while \genie displays larger variations for differential distributions together with an accidental agreement for inclusive cross-sections.
We then studied charm production in Sect.~\ref{sec:charm}, inclusive over the rest of the hadronic final state, for which similar qualitative conclusions apply, and where we also found that the various hadronisation models adopted predict rather different yields of the $D$-meson species.

Now we consider observables which depend on the modelling of the hadronic final state of the collision.
As opposed to the results of Sects.~\ref{sec:inclusive} and~\ref{sec:charm}, for these no analytic reference exists and the comparison is therefore purely between event generators.
First, we study the comparison between the different event generators considered for observables sensitive to the hadronic final state and which can be accessed at the FASER$\nu$ emulsion detector.
Then, we quantify the dependence of these results on the choice of parton shower and hadronisation algorithms within the same  generator.
Finally, we assess the impact of QED radiation on the stability of MC predictions.

\begin{figure}[!tbp]
  \centering
  \includegraphics[width=0.94\textwidth]{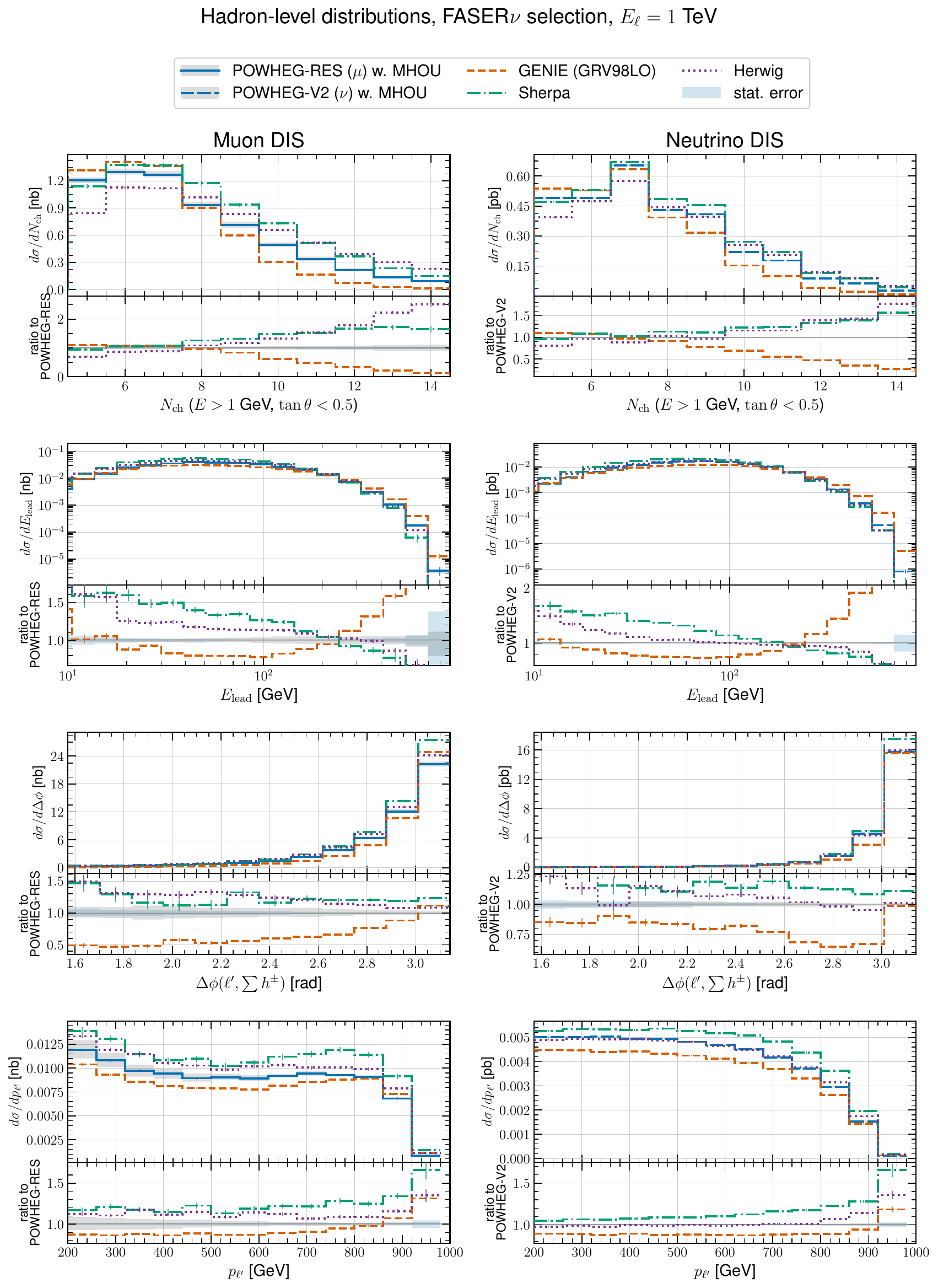}
  \caption{Hadron-level distributions per nucleon on tungsten at
    $E_\ell = 1$~TeV for the FASER$\nu$ selection,
    Eq.~(\ref{eq:tierE}), nested in the fiducial region
    of Eq.~(\ref{eq:fiducial}), for muon neutral-current (left) and neutrino charged-current
    (right panels) DIS.
    From top to bottom, we show the charged-hadron multiplicity $N_{\rm ch}$
    with energies above $1$~GeV within $\tan\theta < 0.5$ of the beam axis, the leading charged hadron's
    energy $E_{\rm lead}$, the azimuthal separation between the lepton and the summed
    charged-hadron system $\Delta\phi\lp \ell',\sum h^\pm\rp$, and the outgoing lepton's momentum $p_{\ell'}$.
    The \powheg, \sherpa and \herwig NLO-matched calculations are compared with \genie (GRV98LO), with 
    the bottom panels displaying the ratio to the \powheg predictions.
   }
  \label{fig:hadron-level}
\end{figure}

\paragraph{Hadronic final states for the FASER$\nu$ selection.}
Fig.~\ref{fig:hadron-level} displays hadron-level distributions at $E_\ell = 1$~TeV for the FASER$\nu$ selection, Eq.~(\ref{eq:tierE}), nested in the fiducial region Eq.~(\ref{eq:fiducial}), for muon neutral-current and neutrino charged-current DIS.
We show the charged-hadron multiplicity $N_{\rm ch}$, counting the final-state charged hadrons with energies above $1$~GeV within $\tan\theta < 0.5$ of the beam axis, the leading charged hadron's energy $E_{\rm lead}$, the azimuthal separation between the lepton and the summed charged-hadron system $\Delta\phi\lp \ell',\sum h^\pm\rp$, and the outgoing lepton's total momentum $p_{\ell'}$.
The latter can be evaluated purely from the lepton kinematics, but here it also (indirectly) depends on the hadronic final-state modelling via the FASER$\nu$ selection of Eq.~(\ref{eq:tierE}). 
The \powheg, \sherpa and \herwig NLO-matched calculations are compared with \genie (GRV98LO).
MHOUs from \powheg range between
$+4.3\%/-3.3\%$ for muon DIS and $+0.9\%/-0.7\%$ for neutrino DIS, a similar relative factor as found for the inclusive cross-sections in Sect.~\ref{sec:inclusive}, and are mostly flat in the considered kinematical region.

Fig.~\ref{fig:hadron-level} indicates that the agreement between NLO-matched generators found for DIS distributions (inclusive and charm production) is lost for observables sensitive to the modelling of the hadronic final state. 
For instance, \herwig and \sherpa predict a higher $N_{\rm ch}$ multiplicity distribution as compared to \powheg; the distributions in $E_{\rm lead}$ are likewise enhanced in \herwig and \sherpa relative to \powheg below 200 GeV and suppressed at higher energies; and also in $\Delta\phi$ and $p_{\ell'}$ the spread between the predictions of the three calculations at the same nominal perturbative accuracy is large, up to $50\%$ for $\Delta\phi$ and about $30\%$ for $p_{\ell'}$.
Significant differences are also found for the \genie predictions: if we compare with \powheg we find that for large values of $N_{\rm ch}$ \genie predicts significantly fewer charged hadrons; suppressed (enhanced) $E_{\rm lead}$ distributions at small (large) values of the leading charged-hadron energy; and a general suppression of the $\Delta \phi$ and $p_{\ell'}$ distributions, except possibly in the rightmost bins. 

The mean charged-hadron multiplicity within the FASER$\nu$ acceptance, which is a crucial parameter to select neutrino and muon DIS events, is $7.7$ for \powhegres, $7.0$ for \genie, $8.2$ for \sherpa and $8.8$ for \herwig in muon DIS, and $7.8$, $7.3$, $8.1$ and $8.2$ respectively in neutrino DIS: that is, as mentioned above, \genie lies below the matched calculations, \sherpa above \powheg and \herwig highest, a spread of up to about $25\%$.
Concerning the $p_{\ell'}$ predictions, their shape and normalisation agree for the three NLO-matched generators without the FASER$\nu$ cuts (Sect.~\ref{sec:inclusive}), and once the  FASER$\nu$ selection is applied the shapes remain mostly similar but the overall normalisation changes due to the effect of the $N_{\rm ch}$ vertex selection cut.

\paragraph{Parton shower and hadronisation dependence.}
The comparisons of Fig.~\ref{fig:hadron-level} highlight how differences in the way DIS event generators model the hadronic final state affect predictions for observables relevant for FASER$\nu$.
Since the three NLO-matched event generators considered agree for the DIS distributions both for inclusive and for charm production, the differences in Fig.~\ref{fig:hadron-level} must arise from the choice of parton shower algorithm and the treatment of hadronisation in each case.
Here we attempt to disentangle these effects, by comparing predictions based on the same generator where the parton shower algorithm, and possibly also the hadronisation model, is 
modified keeping all other settings unchanged.

\begin{figure}[!tbp]
  \centering
  \includegraphics[width=0.99\textwidth]{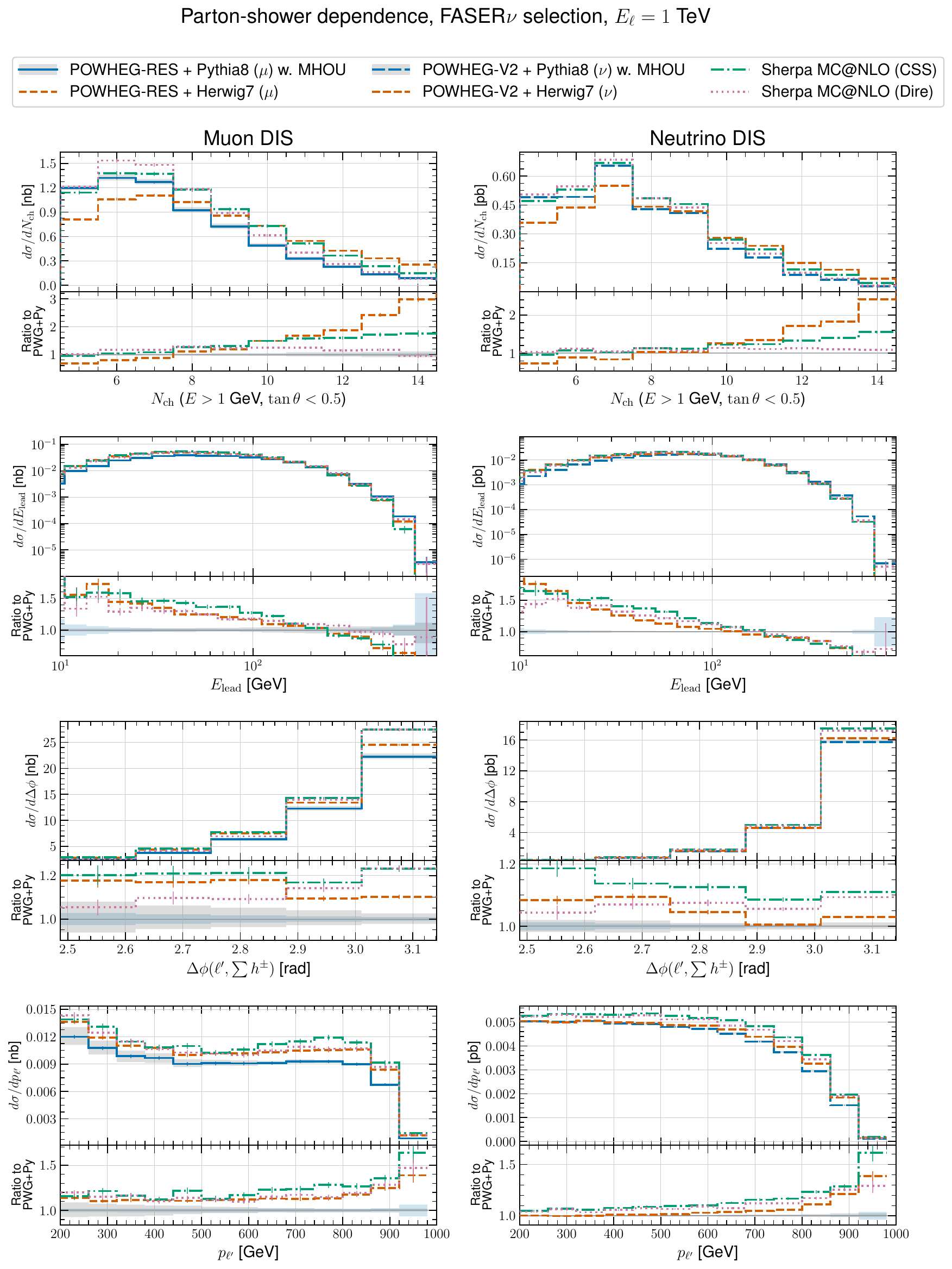}
  \caption{
  Same as Fig.~\ref{fig:hadron-level}, now with predictions for two of the NLO-matched generators being considered (\powheg and \sherpa) compared for alternative parton shower algorithms.
  For \powheg we compare the results of using the \pythia and \herwig shower and hadronisation algorithms, while for \sherpa we compare the default CSS with the Dire shower, keeping in this case the same hadronisation model.
  Bottom panels are normalised to the \powheg{}+\pythia prediction.
  }
  \label{fig:shower-dependence}
\end{figure}

Fig.~\ref{fig:shower-dependence} displays the same final-state hadronic observables as in Fig.~\ref{fig:hadron-level} now with predictions for two of the NLO-matched generators being considered (\powheg and \sherpa) compared for alternative parton shower algorithms.
Specifically, for \powheg we compare the results of using \pythia ($p_T$-ordered shower, Lund string hadronisation) and \herwig (angular-ordered shower, cluster model hadronisation), while for \sherpa we compare the default Catani--Seymour shower (CSS) with the Dire shower, keeping in this case the same hadronisation model.

The results of Fig.~\ref{fig:shower-dependence} indicate that $\la N_{\rm ch}\ra$, taken as representative hadronic observable,  moves by $6\%$ ($3\%$) between the two choices of parton shower algorithm available in \sherpa in muon (neutrino) DIS, and by $16\%$ ($10\%$) between the two shower algorithms (\pythia and \herwig) available for \powheg.
An uncertainty estimated by switching the shower algorithm within a single event generator is therefore not a complete estimate of the hadronic modelling spread between generators, in particular since parton fragmentation remains a major contribution.
Every variation of the  shower algorithm considered here raises the fiducial rates relative to the \powheg reference by up to $20\%$, since the FASER$\nu$ selection demands five charged tracks with four of them narrow, and hence is sensitive to the modelling of the $N_{\rm ch}$ distribution. 
As in Fig.~\ref{fig:hadron-level}, the $E_{\rm lead}$ distribution is where the variations separate most: all of them enhance the rate of soft leading hadrons relative to \powheg{}+\pythia, by up to about $80\%$ for \powheg{}+\herwig between $10$ and $100$~GeV, and suppress it at higher energies, by up to $40\%$ between $200$ and $600$~GeV.

Two main messages can be drawn from Figs.~\ref{fig:hadron-level} and~\ref{fig:shower-dependence}.
First, that even at the same formal NLO accuracy, differences for hadronic observables remain large, calling for higher-accuracy shower algorithms and tuned fragmentation models for the FASER kinematics.
Second, that FASER$\nu$ analyses should assess the stability of their vertex selection cuts with respect to the modelling of distributions such as $N_{\rm ch}$, to avoid biasing their neutrino reconstruction. 

\paragraph{Impact of QED radiation.}
We also consider the impact of QED radiation for differential distributions in muon and neutrino DIS, both for the DIS variables and for final-state hadronic observables.
So far, as indicated in Table~\ref{tab:settings}, predictions presented are pure-QCD calculations evaluated at fixed $\alpha_{\rm em}$ in the $G_\mu$ scheme.
We quantify the role of QED corrections by re-showering the same \powheg events with the QED algorithm of \pythia switched on in stages: final-state radiation (FSR) off the lepton line, initial-state radiation (ISR) and FSR off the lepton line (FSR + ISR), and the full QED shower which also includes radiation off the quark lines.
For neutrino DIS there is no ISR off the incoming lepton.

\begin{figure}[!tbp]
\centering
\includegraphics[width=0.90\textwidth]{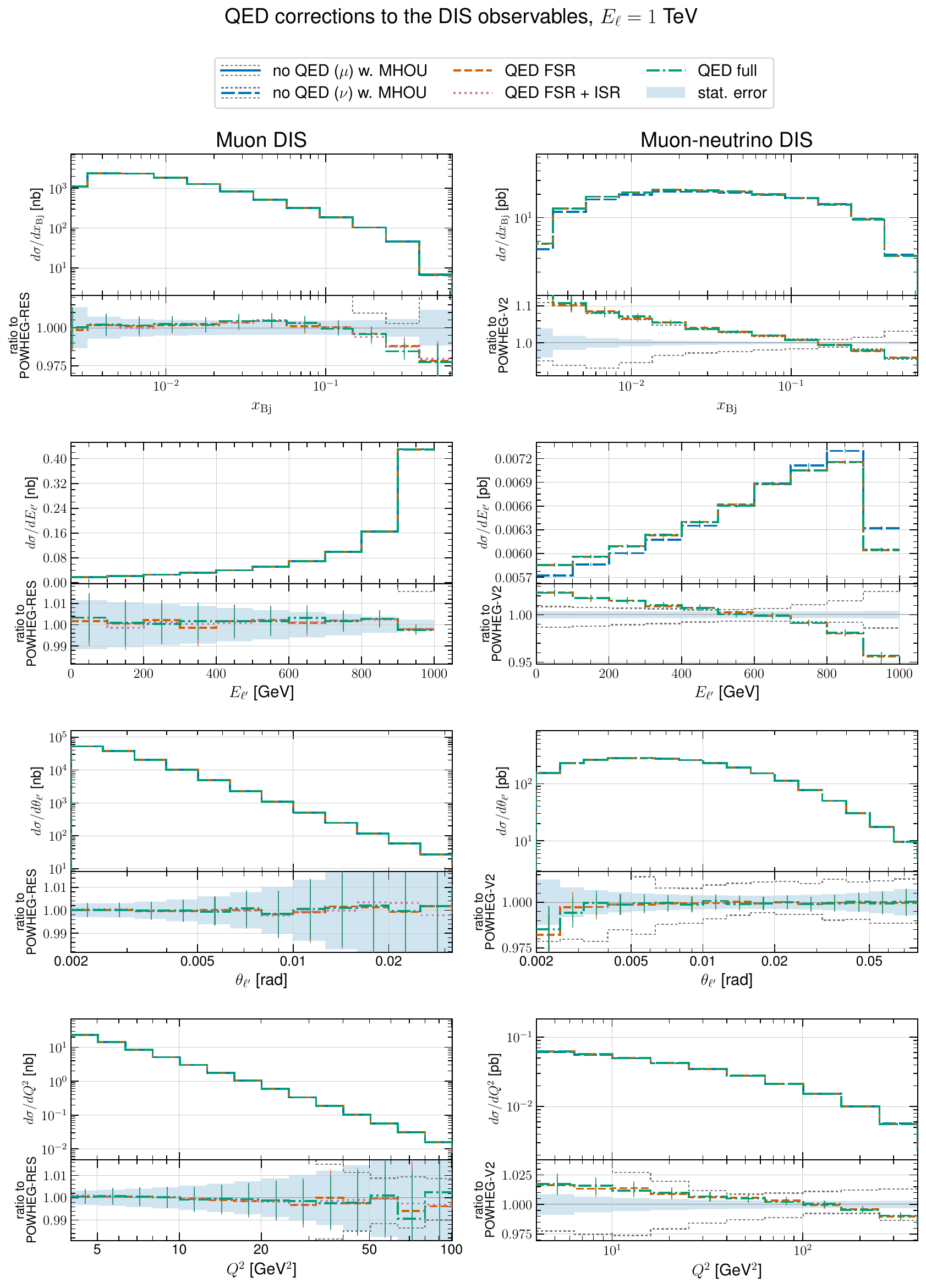}
\caption{The effect of QED radiation on the DIS observables per nucleon on tungsten at $E_\ell = 1$~TeV in the fiducial region Eq.~(\ref{eq:fiducial}) for muon neutral-current (left) and muon-neutrino charged-current (right panels) DIS. 
We compare the results of showering a common set of \powheg events with four different options (three for neutrino DIS, where there is no ISR off the incoming lepton): QED radiation off, with final-state radiation only, with initial- and final-state radiation, and then with the full QED shower; the lower panels are ratios to the prediction without QED corrections accounted for.
The blue band corresponds to the MC statistical uncertainties, and the dotted lines to the NLO MHOU of the prediction without QED corrections.
}
  \label{fig:qed-dis}
\end{figure}

\begin{figure}[!tbp]
  \centering
  \includegraphics[width=0.99\textwidth]{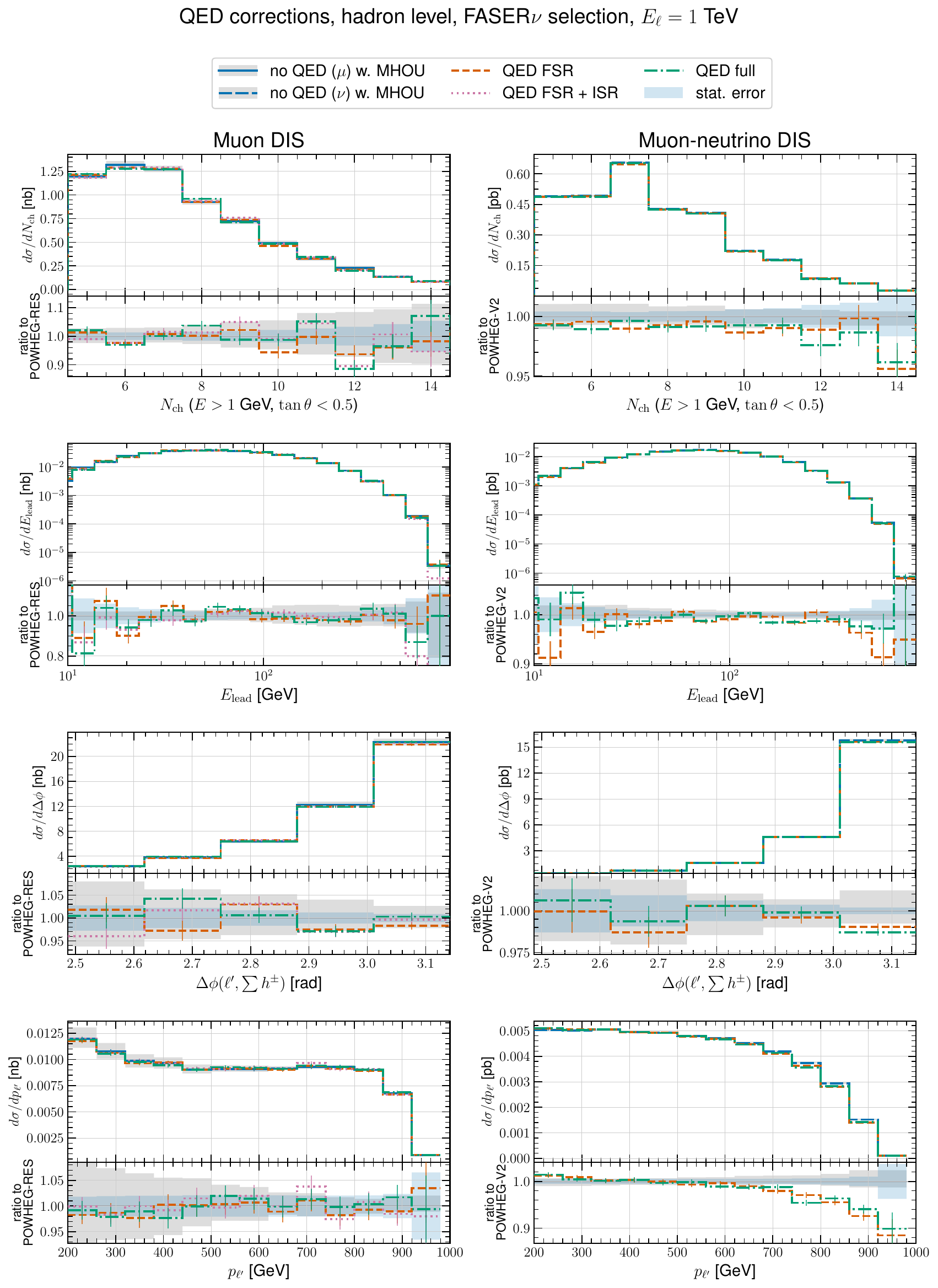}
  \caption{Same as Fig.~\ref{fig:qed-dis} for the
    hadron-level observables of Fig.~\ref{fig:hadron-level} corresponding to the FASER$\nu$ selection. }
  \label{fig:qed-hadron}
\end{figure}

Fig.~\ref{fig:qed-dis} quantifies the effect of QED radiation on DIS observables at $E_\ell = 1$~TeV in the fiducial region of Eq.~(\ref{eq:fiducial}) for muon neutral-current and muon-neutrino charged-current DIS.
In such an assessment it is important to identify the neutrino flavour, since QED radiation effects vary with the charged-lepton mass, and in particular they are larger for electron-neutrino than for muon-neutrino scattering, {\it e.g.}~\cite{vanBeekveld:2024ziz}.

Integrated over the fiducial region, QED radiation effects turn out to be negligible (below the per-mille level for both currents).  
Differentially, the NC and CC distributions behave differently due to the charge of the incoming lepton in each case.
In neutrino DIS, final-state radiation softens the scattered muon systematically: the $E_{\ell'}$ distribution increases by a few per cent below $E_{\ell'}\sim 600$ GeV, while at higher energies it decreases by up to about $4\%$.
Since $x_{\rm Bj}$ and $Q^2$ are reconstructed from the scattered lepton kinematics, they shift accordingly.\footnote{Should an electromagnetic calorimeter be available, one could define dressed leptons to integrate QED radiation both in the measurement and the associated observable.} 
These migrations take place inside the fiducial region, and hence the integrated effect vanishes.
In muon DIS the same distributions move by at most $2\%$: the muon is charged both in the initial and in the final state, so the radiation forms a dipole between the two lepton legs and its effect on the reconstructed kinematics largely cancels out.
Radiation off the quark line, measured from the difference between the full calculation and the partial FSR + ISR one, is not resolved in either current. 
For muon DIS, and for $\theta_{\ell'}$ and $Q^2$ in neutrino DIS, the effects of QED radiation are smaller than the NLO MHOU; for $x_{\rm Bj}$ and $E_{\ell'}$ in neutrino DIS they exceed it, reaching about $10\%$ at the smallest $x_{\rm Bj}$ and $4\%$ at the largest $E_{\ell'}$.

Fig.~\ref{fig:qed-hadron} presents the corresponding comparison for the hadronic observables of Fig.~\ref{fig:hadron-level} within the FASER$\nu$ selection.
In neutrino DIS, since the selection cuts on the lepton momentum, the softening of the scattered muon costs the selected rate around $1\%$ and removes about $5\%$ of the events with $p_{\ell'}>800$~GeV.
Beyond this overall selection cut, the hadronic observables remain stable.
As for the DIS observables, for muon scattering all QED effects are smaller than  other sources of theory uncertainty.

The comparisons presented in this section provide a shower-level estimate of QED effects in lepton-nucleon DIS but are not the complete picture.
A full calculation involves both the full electroweak virtual corrections, evaluated in a consistent electroweak scheme, and photon-initiated contributions to lepton-nucleon scattering~\cite{Manohar:2016nzj,Bertone:2017bme,Manohar:2017eqh,Cridge:2021pxm,NNPDF:2024djq}.
The conclusion of this study is nevertheless robust: for the FASER fiducial kinematics, QED radiation is a sub-per-cent effect on the inclusive rates, and its only visible signature is the softening of the scattered-lepton spectrum in charged-current scattering, which shifts the reconstructed $x_{\rm Bj}$ and $E_{\ell'}$ distributions by several per cent, beyond the NLO MHOU.
The latter finding is however restricted to muon-neutrino DIS, and larger effects arise for electron-neutrino scattering.

\section{Applications to FASER}
\label{sec:faser-app}
\label{sec:rates}

The comparisons presented so far are carried out at fixed $E_\ell$ values.
Here we use the same pipeline together with the LHC electron- and muon-neutrino fluxes~\cite{FASER:2024ykc} to provide predictions for representative observables at FASER.
We consider three specific applications: comparison of generator predictions with the latest FASER data from the emulsion and electronic sub-detectors, a study of the PDF dependence in the predictions for the dimuon signal at FASER's electronic detector, and predictions for charged-hadron production in SIDIS at FASER$\nu$. 
The expected number of events for a fiducial selection is 
\be
  N \, = \, \int dE_\nu \; \Phi(E_\nu) \, \sigma_{\rm fid}(E_\nu) \, T \, ,
  \label{eq:rate}
\ee
with $\Phi$ the neutrino flux through the detector for a given integrated luminosity~\cite{Kling:2021gos}, $T$ the column density of the tungsten target, and $\sigma_{\rm fid}$ the fiducial cross-section per nucleon, evaluated as explained in Sect.~\ref{sec:settings} from separate proton and neutron samples.
The cross-section in Eq.~(\ref{eq:rate}) is required as a continuous function of $E_\nu$, and
here we build a common fine-grained energy ladder from $10$~GeV to several TeV from the \yadism NLO ZM prediction and then rescale it, for each generator, by its ratio to \yadism interpolated in $\log E_\nu$ between the energies at which that generator has been run.
In the following, the FASER Run 3 flux model for $\Phi(E_\nu)$ in Eq.~(\ref{eq:rate}) is taken to be the same as in~\cite{FASER:2024ykc}, namely {\sc\small EPOS-LHC}~\cite{Pierog:2013ria} for light-hadron production and POWHEG-BOX heavy-quark production matched to \pythia~\cite{Buonocore:2023kna} for charm-hadron production, and no flux uncertainties are considered. 

\begin{figure}[!tbp]
  \centering
  \includegraphics[width=0.95\textwidth]{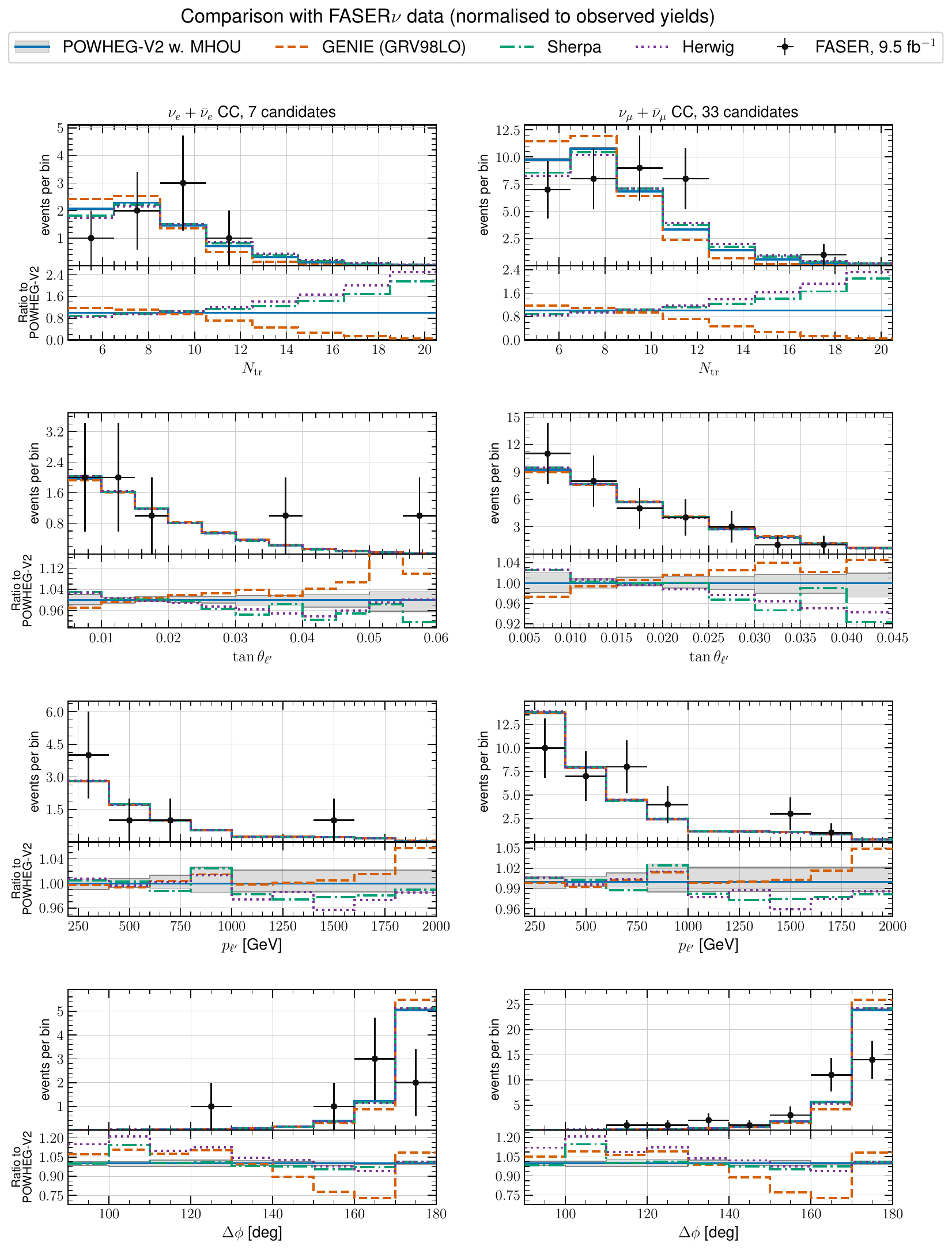}
  \caption{Differential distributions for $\nu_e+\bar{\nu}_e$ (left panels) and $\nu_\mu+\bar{\nu}_\mu$ (right panels) charged-current DIS candidate events at FASER$\nu$ for $\mathcal{L}=9.5$ fb$^{-1}$~\cite{FASERnu:2026conf}.
  From top to bottom, we show $N_{\tr}$, $\tan \theta_{\ell'}$, $p_{\ell'}$, and $\Delta \phi$ (defined as the azimuthal separation between the lepton and the summed remaining tracks).
   We display the corresponding predictions from \powhegtwo, \sherpa, and \herwig with the benchmark settings as well as from \genie (GRV98LO). 
  For each generator, distributions are normalised to the observed event yields.
  The bottom panels show the ratio of each generator to the \powhegtwo predictions. }
  \label{fig:faser-emulsion}
\end{figure}

\paragraph{Comparison with FASER data.}
The most recent FASER$\nu$ measurement~\cite{FASERnu:2026conf} reports $33$ muon-neutrino and $7$ electron-neutrino CC candidates collected with an integrated luminosity of $\mathcal{L}=9.5$~fb$^{-1}$.
Fig.~\ref{fig:faser-emulsion} compares four measured differential distributions ($N_{\tr}$, $\tan \theta_{\ell'}$, $p_{\ell'}$, $\Delta \phi$) for $\nu_e+\bar{\nu}_e$  and $\nu_\mu+\bar{\nu}_\mu$ from~\cite{FASERnu:2026conf} with the corresponding predictions from \powhegtwo, \sherpa, and \herwig as well as from \genie with the default tune (GRV98LO).
For each generator, distributions are normalised to the observed event yields to account for the experimental selection and acceptance efficiencies. 

The four predictions differ where the modelling of the  hadronic final state becomes relevant.
The outgoing lepton momentum and the scattering angle follow the agreement at the level of inclusive distributions found in Sect.~\ref{sec:inclusive}.
In particular, for $p_{\ell'}$ the four predictions agree to within $6\%$ in every bin.
On the track multiplicity distribution, \genie prefers lower values of $N_{\tr}$ as compared to \powhegtwo, \sherpa, and \herwig,  the same ordering found at fixed neutrino energy in Fig.~\ref{fig:hadron-level}.
For the azimuthal separation $\Delta\phi$, \genie lies around $7\%$ above \powhegtwo well below the back-to-back peak and a third below it just short of the peak, differences which are much larger than the estimated MHOUs.
For the bulk of the $\Delta\phi$ distribution, \powhegtwo, \sherpa, and \herwig display good agreement. 
While the limited statistics in the FASER$\nu$ measurement are not sufficient to arbitrate these differences, future data will constrain the modelling of the hadronic final state for this process, improving {\it e.g.} vertex selection. 

\begin{figure}[!tbp]
  \centering
  \includegraphics[width=0.99\textwidth]{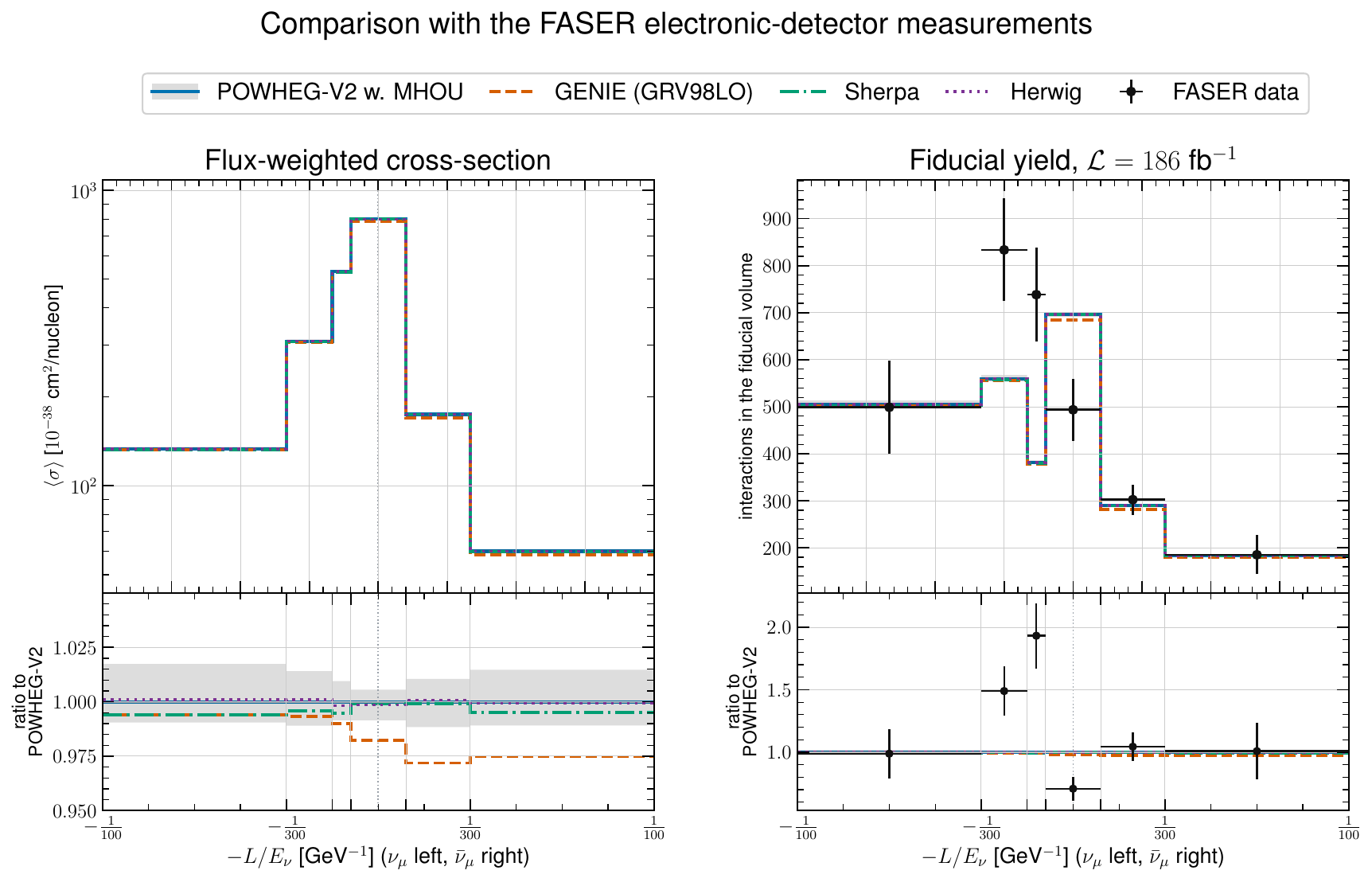}
  \caption{Left: predictions from \powhegtwo, \sherpa, \herwig, and \genie for the flux-weighted cross-section for CC DIS in the $-L/E_\nu$ kinematics of the FASER muon-neutrino electronic detector measurements of~\cite{FASER:2026enu2d}.
The central bin corresponds to the $\nu_{\mu}+\bar{\nu}_{\mu}$ sum, and the bottom panel displays the ratio to the \powhegtwo prediction.
The \powhegtwo, \sherpa, and \herwig predictions are complemented with the $Q^2< 4$~GeV$^2$ contribution taken from \genie to match the fiducial selection, see text.
  Right: the corresponding predictions for the number of muon-neutrino CC interactions expected for $\mathcal{L}=186$~fb$^{-1}$ compared to the unfolded FASER data~\cite{FASER:2026enu2d}. 
}
  \label{fig:faser-electronic}
\end{figure}

In addition to the neutrino emulsion data, FASER has also measured the $\nu_\mu$ and $\bar{\nu}_\mu$ CC interaction rate and fluxes with its electronic detector~\cite{FASER:2024ref,FASER:2026enu2d}, in bins of $-L/E_\nu$, where $L$ is the lepton number, using the spectrometer selection described in Sect.~\ref{sec:faser-selections}.
The latter measurement is also presented double-differentially in $E_\nu$ and in the neutrino rapidity, following an earlier measurement of the muon-neutrino flux as a function of rapidity~\cite{FASER:2025nurap}; here we compare only with its energy dependence, since the flux rapidity does not affect DIS modelling.
Fig.~\ref{fig:faser-electronic} displays the predictions from \powhegtwo, \sherpa, \herwig, and \genie for the flux-weighted cross-section for CC DIS for the kinematics of~\cite{FASER:2026enu2d}.
Since \powhegtwo, \sherpa, and \herwig predictions are generated for $Q^2>4$~GeV$^2$, while the FASER measurement corresponds to the total cross-section without a $Q^2$ cut, we extend these predictions with the $Q^2<4$~GeV$^2$ contribution estimated by \genie, which amounts to between $1\%$ and $23\%$ ($37\%$) of the $\nu_\mu$ ($\bar\nu_\mu$) cross-section from the highest to the lowest neutrino energies.
Then the right panel of Fig.~\ref{fig:faser-electronic} shows the corresponding predictions for the number of muon-neutrino CC interactions expected in the fiducial volume at $\mathcal{L}=186$~fb$^{-1}$ compared to the unfolded FASER data~\cite{FASER:2026enu2d}. 

From the comparisons of Fig.~\ref{fig:faser-electronic} one observes that the \sherpa and \herwig cross-sections agree with \powhegtwo to better than $1\%$ in every bin, and \genie to within $3\%$.
Without the extension below $Q^2=4$~GeV$^2$, \genie would instead lie up to $25\%$ above \powhegtwo in the two lowest $E_\nu$ bins, where this region is not negligible at the inclusive level. 
Concerning the comparison with the FASER measurements, the data agree with \powhegtwo within uncertainties in three of the six bins, but exceed it by $50\%$ and $90\%$ in the $300$--$1000$~GeV $\nu_\mu$ bins and fall $30\%$ below it above 1~TeV, differences which are much larger than the differences between theoretical predictions as well as the experimental uncertainties.
This discrepancy may indicate that the current predictions for the light-hadron fluxes that dominate muon-neutrino production need to be improved. 
In other words: for this specific observable, the data-vs-theory comparison is driven by the assumed neutrino fluxes and not by the modelling of the neutrino-nucleon scattering cross-section.

\paragraph{Charm production and the dimuon signal.}
\label{sec:faser-dimuon}

Charm production in neutrino DIS represents a powerful probe of the strangeness content of protons~\cite{Faura:2020oom,Gao:2017kkx} and nuclei.
Charm production can be measured at FASER through the emulsion detector, and a first search of this process was presented in~\cite{FASER:2026charm}.
Complementing these direct charm production measurements, FASER may be able to access this process by reconstructing the dimuon final state, Eq.~(\ref{eq:tierD}), via its tracking spectrometer.
In the dimuon process, the scattered muon is accompanied by an oppositely charged muon from the semileptonic decay of the charm hadron~\cite{Helenius:2024fow,Helenius:2025fpy,Helenius:2026uuz}.
While other neutrino experiments have measured the dimuon signal~\cite{NuTeV:2001dfo,NOMAD:2013hbk}, previous measurements involved much lower neutrino energies than those that FASER can reach.
Here we provide an initial estimate of the expected dimuon rates at the FASER electronic detector, and illustrate its PDF sensitivity by providing predictions for the dimuon yields for different PDF sets. 

\begin{figure}[!tbp]
  \centering
  \includegraphics[width=0.80\textwidth]{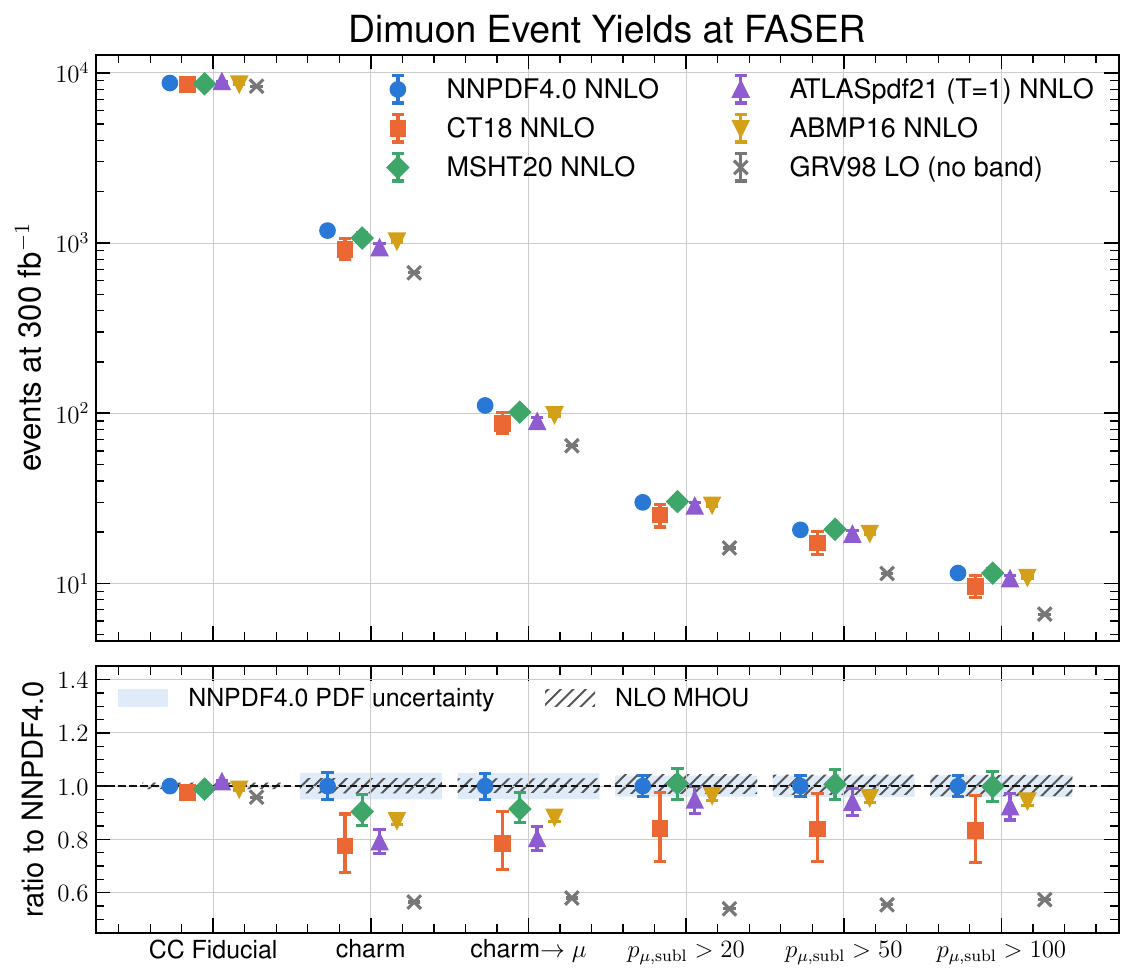}
  \caption{
  NLO predictions from \powhegtwo for the dimuon signal ($\nu_\mu + \bar\nu_\mu$) arising from CC muon-neutrino DIS recorded in FASER's electronic detector for an integrated luminosity of $\mathcal{L}=300$~fb$^{-1}$. 
 Predictions are provided for different PDF sets at the following analysis levels: inclusive CC scattering in the fiducial volume, after requiring a charm tag (information only accessible at the truth level), after requiring a semileptonic charm decay into a muon, and then requiring both muons inside the spectrometer acceptance and imposing three possible cuts on the total momentum of the subleading muon, $p_{\mu,\rm subl}$.   
 GRV98 does not provide a PDF uncertainty estimate.
The bottom panels display the ratio to NNPDF4.0 and also include the MHOUs from \powhegtwo. }
  \label{fig:dimuon-pdf}
\end{figure}

Fig.~\ref{fig:dimuon-pdf} presents NLO predictions from \powhegtwo for the dimuon signal arising from charged-current DIS recorded in FASER's electronic detector assuming an integrated luminosity of $\mathcal{L}=300$~fb$^{-1}$, using the dimuon selection defined in Sect.~\ref{sec:faser-selections}.
 Predictions are provided for different PDF sets at the following analysis levels: first, inclusive CC scattering within the fiducial volume, then, after requiring a charm tag (information only accessible at the truth level), then after requiring a semileptonic charm decay into a muon, and finally after requiring both muons inside the spectrometer acceptance and imposing three possible cuts on the total momentum of the subleading muon, $p_{\mu,\rm subl}$.   
 Note that GRV98 does not provide a PDF uncertainty estimate.
The bottom panels display the ratio to NNPDF4.0 and also include the MHOUs which are always subdominant as compared to the differences between PDF sets. 

From Fig.~\ref{fig:dimuon-pdf} one observes that FASER, for the assumed integrated luminosity, should record between 10 and 30 dimuon events, depending on the choice of PDF set and on how strict the cut on $p_{\mu,\rm subl}$ is.
This cut is necessary to reduce the backgrounds, for example charged pions produced near the end of the tungsten target, which can traverse the spectrometer and mimic a muon, but with a softer momentum spectrum. 
Hence, given the predicted event yields, and provided systematic errors and background subtraction can be kept under control, it is possible that FASER reports an observation of the dimuon signal with the full Run 3 dataset.

Furthermore, Fig.~\ref{fig:dimuon-pdf} also displays a significant dependence on the choice of PDF: as compared to the reference NNPDF4.0, CT18 predicts around 20\% fewer events and GRV98 40\% fewer events.
In addition, the predictions from different PDF sets do not agree in all cases within their nominal uncertainties. 
Hence a measurement of the dimuon signal, already based on the Run 3 dataset, should provide information to constrain the strange PDFs of the nucleon complementing existing dimuon measurements at lower energies which are already part of global PDF fits. 
As mentioned above, additional constraints will be provided by FASER$\nu$ measurements~\cite{FASER:2026charm} of charm production (corresponding to the second column of Fig.~\ref{fig:dimuon-pdf}), where despite a lower efficiency for charm reconstruction of $\epsilon_c\sim 50\%$, the separation between signal and background events is greatly facilitated as compared to the dimuon final state. 

\paragraph{Inclusive charged-hadron production in SIDIS.}
As a third representative application, we consider the possible measurement of the total charged-hadron production rate in semi-inclusive DIS.
Such a measurement would be enabled by the unique capabilities of FASER$\nu$, which reconstructs every charged track at the interaction vertex and hence can access single-inclusive hadron production, $\nu N \to \ell h X$, summing over all charged hadrons $h$ in the final state.
This is a process directly sensitive to the hadron fragmentation functions~\cite{Borsa:2021ran,Bertone:2017tyb,Bertone:2018ecm,Albino:2008gy}, and which can be evaluated up to NNLO in the QCD expansion both for charged-current~\cite{Bonino:2025ident,Bonino:2025qta} and for neutral-current scattering~\cite{Bonino:2024qbh,Goyal:2023zdi}, and for the polarised case in both currents~\cite{Goyal:2024tmo,Bonino:2024wgg,Bonino:2025sidis}.
In particular, this would be the first measurement of charged-hadron fragmentation in neutrino-initiated SIDIS at TeV energies.

\begin{figure}[!tbp]
  \centering
  \includegraphics[width=0.98\textwidth]{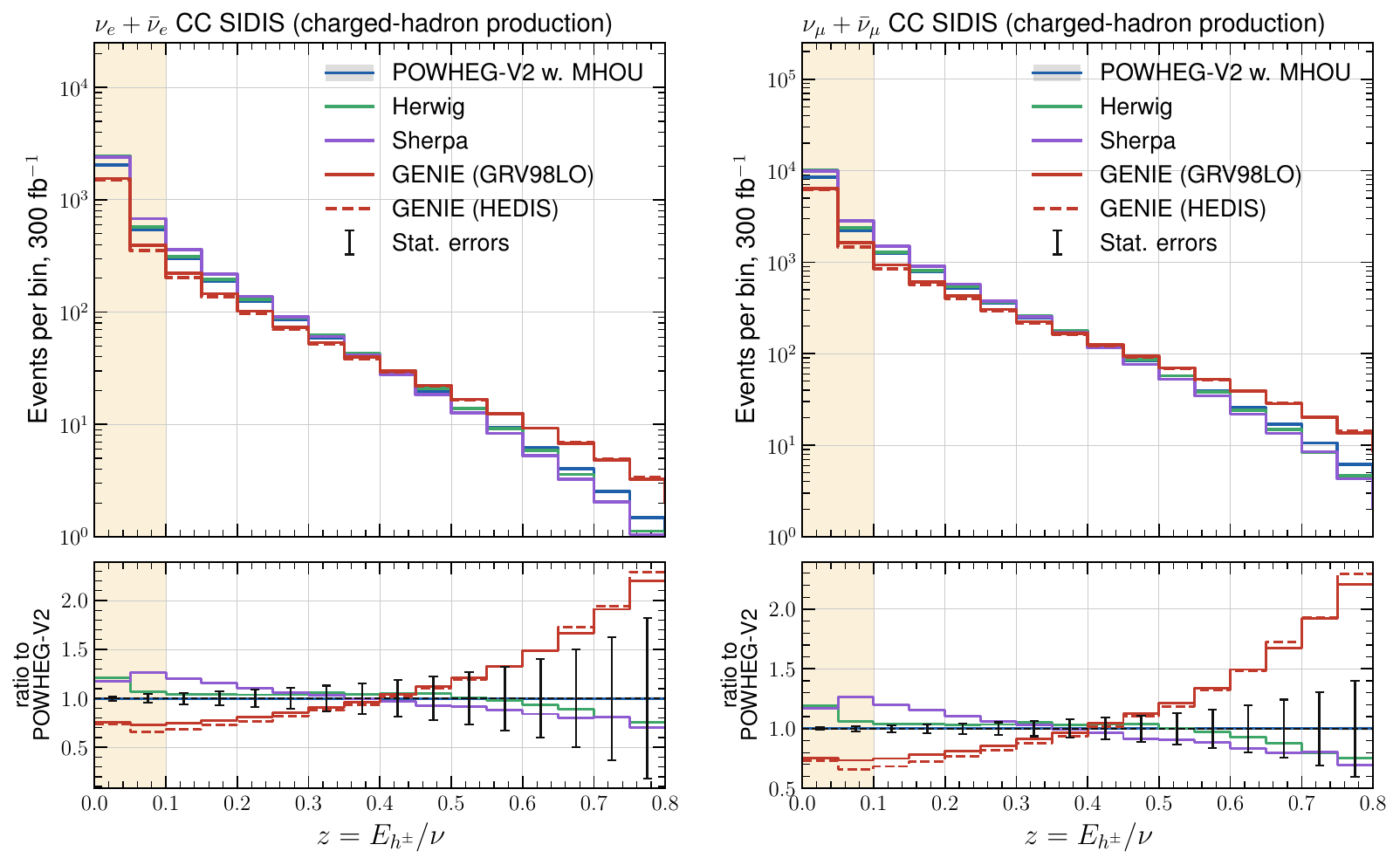}
  \caption{Predictions for the expected event yields in semi-inclusive charged-current DIS at FASER$\nu$ for $\mathcal{L}=300$ fb$^{-1}$ as a function of the SIDIS variable $z = E_{h^\pm}/\nu$.
  We display the yields for SIDIS with charged-hadron production summing over all stable charged hadrons (mostly pions and kaons) and assuming an $80\%$ hadron selection efficiency, for $\nu_e+\bar{\nu}_e$ (left panel) and $\nu_\mu+\bar{\nu}_\mu$ (right panel) scattering. 
The shaded band indicates the $z < 0.1$ region where fixed-order SIDIS calculations may not be reliable. 
The bottom panels display the ratios to \powhegtwo, with error bars indicating the expected statistical uncertainty in each bin.
    }
  \label{fig:sidis-yields}
\end{figure}

To estimate the expected event yields, Fig.~\ref{fig:sidis-yields} displays predictions for the expected event yields in semi-inclusive charged-current DIS at FASER$\nu$ for $\mathcal{L}=300$ fb$^{-1}$ as a function of the SIDIS variable $z = E_{h^\pm}/\nu$, where $z$ corresponds to the fraction of the energy transfer carried by the reconstructed charged hadron $h^\pm$.
  We evaluate total charged-hadron production summing over all stable charged hadrons (mostly pions and kaons, but also protons and hyperons) for events passing the FASER$\nu$ selection of Eq.~(\ref{eq:tierE}) and assuming an $80\%$ hadron selection efficiency, a conservative value given the single-film and tracking efficiencies of the FASER$\nu$ emulsion detector~\cite{FASER:2025qaf}, for $\nu_e+\bar{\nu}_e$  and $\nu_\mu+\bar{\nu}_\mu$ charged currents. 
The shaded band indicates the $z < 0.1$ region where fixed-order SIDIS calculations may not be reliable. 
While not shown here, charged-hadron production in muon DIS can also be evaluated using the same pipeline, and the event yields are substantially higher than those of the neutrino counterparts. 

From Fig.~\ref{fig:sidis-yields}, we expect that FASER$\nu$ should record around 900 (3700) charged hadrons in electron- (muon-)neutrino scattering, summing over neutrinos and antineutrinos, in the region $z>0.1$ where perturbative predictions are reliable.
Note that in Fig.~\ref{fig:sidis-yields} hadrons in a bin are not independent counts, since each event in general contains multiple charged hadrons within the FASER$\nu$ acceptance. 
Despite the steeply falling distribution in $z$, it is possible to cover up to $z\sim 0.4$--$0.5$ with sufficient statistics, especially in muon-neutrino SIDIS, hence accessing hadron fragmentation functions in kinematic regions which are currently poorly constrained. 
Concerning differences between generators,  \powhegtwo and \herwig agree at the few-per-cent level for most of the $z$ range (and never differ by more than 20\%), with \sherpa displaying somewhat larger differences both at small and large $z$.
Both \genie tunes predict a completely different hadron fragmentation distribution, underestimating the NLO rate by about 30\% at small $z$ and overestimating it by more than a factor of two at large $z$, highlighting that \genie, which fragments the hadronic system with the Lund string but without a preceding parton shower, predicts harder fragmentation than the three NLO-matched generators considered. 

The analysis of Fig.~\ref{fig:sidis-yields} strongly suggests that a first measurement of charged-hadron neutrino SIDIS at TeV energies should be possible with the full Run 3 dataset of FASER$\nu$, and the predicted statistical precision should be enough to distinguish between different calculations of hadron fragmentation, both those embedded in the hadronisation models used by the event generators as well as those based on fragmentation functions within the analytical QCD factorisation framework. 

We note again that in this study we do not account for (cold) nuclear final-state effects. 
The hard scattering process takes place inside a tungsten nucleus, which the outgoing quark must leave while undergoing parton shower and hadronisation. 
Interactions of both partons and hadrons with the cold nuclear medium will in general lead to the formation of more particles, thereby increasing the hadron multiplicity while decreasing the typical hadron energies and therefore the hadronic energy fraction $z$~\cite{Doradau:2024wli,deFlorian:2011fp}.
A complete theoretical model describing how parton showers evolve inside dense nuclei and how partons interact with cold nuclear matter is currently lacking. 
While these phenomena will be studied in detail at the Electron-Ion Collider (EIC)~\cite{Accardi:2012qut,AbdulKhalek:2021gbh}, the LHC neutrino and muon programme can perform the same measurements a decade in advance.
\FloatBarrier
\section{Summary and outlook}
\label{sec:conclusions}

We have presented a systematic benchmark of state-of-the-art NLO event generators modelling neutrino- and muon-initiated deep-inelastic scattering at the LHC far-forward experiments, comparing three independent general-purpose generators, \sherpa, \herwig and \powheg, against the analytic calculations of \yadism and the dedicated neutrino generator \genie, under common settings and assuming FASER's tungsten target.
The three NLO generators reproduce the \yadism (ZM-VFNS) inclusive fiducial cross-sections, and each other, to within about $2\%$ for both currents and over an order of magnitude in beam energy, with residual differences in the differential distributions at the per-cent level largely covered by MHOUs.

Our analysis shows that the default \genie configuration agrees with these NLO calculations at the inclusive level only through an accidental cancellation, and its differential distributions deviate from them by up to about $20\%$, well outside the MHOU band. 
NNLO corrections, estimated with \yadism (FONLL for NC and ZM-VFNS for CC), are moderate but not negligible, up to $-7\%$ for muon DIS and around $-1.5\%$ for neutrino DIS at the inclusive level and up to $-12\%$ in some regions of phase space for differential distributions, which highlights the urgency of extending available DIS event generators to NNLO accuracy.

Predictions from the event generators considered differ instead for observables sensitive to the modelling of the hadronic final state. 
For instance, the mean charged-hadron multiplicity $\la N_{\rm ch}\ra $ spreads by up to about $25\%$ between generators, with \genie below and \sherpa and \herwig above the \powheg predictions in both currents.
Since the FASER$\nu$ vertex selection algorithm relies on a cut on $N_{\rm ch}$, the event generator choice may influence the selection efficiency. 
Furthermore, changing the shower and the hadronisation model together moves this multiplicity by about three times more than changing the shower alone, indicating that a shower variation within one generator underestimates the overall modelling uncertainty.
These differences are well outside the MHOUs and motivate the need for a dedicated tune of non-perturbative QCD effects for TeV-energy lepton scattering using data from FASER and SND@LHC.

Of particular interest for phenomenology is charm production, which provides sensitive constraints on proton and nuclear structure and is closely related to the reconstruction of tau-neutrino events.
On charm production, the NLO-matched generators agree to $3\%$ for neutrino DIS, and charm mass effects reach $1$--$4\%$ ($3$--$7\%$) for neutrino (muon) DIS against a sub-per-cent effect at the inclusive level; the massive-charm \powhegtwo calculation agrees with the analytic \yadism counterpart in the same mass scheme (FFNS with $n_f=3$).
The default \genie tune produces about $40\%$ fewer charm events than the NLO calculations based on the benchmark settings, because of its choice of PDF (GRV98), whose strange sea is about half that of NNPDF4.0 in the relevant $x$ region. 

We also presented representative applications to neutrino DIS at FASER.
We find that differences between generator predictions are large for some distributions, such as the charged-track multiplicity in FASER$\nu$, though data statistics are limited and do not exhibit a clear preference for a specific prediction. 
Concerning the charm production rate, as well as the corresponding dimuon signal, at FASER, modern NNLO PDF sets differ in their predictions by up to 20\%, with the PDF of the default \genie tune (GRV98) suppressing it by $40\%$.
Finally, semi-inclusive charged-hadron production rates in SIDIS are large enough to enable the first measurements of light-hadron fragmentation functions in high-energy charged-current DIS. 

In summary, our analysis highlights the excellent agreement of modern NLO DIS event generators at the inclusive level and for differential DIS distributions, once common settings are adopted; that NNLO corrections are not negligible; that hadronic final states can markedly differ between generators; and that a careful treatment of heavy-quark mass effects is crucial for the accurate modelling of charm production in muon and neutrino DIS. 
We find that non-DIS effects can be neglected at FASER once the fiducial selection of Eq.~(\ref{eq:fiducial}) is applied, but also that contributions from outside the DIS region (the SIS and resonance regions) reach $20\%$ ($37\%$) of $\nu_\mu$ ($\bar\nu_\mu$) CC interactions at $E_\nu=100$~GeV, and about $10\%$ ($20\%$) of the total FASER$\nu$ event rate.
Furthermore, our study confirms that  nuclear PDF corrections should be accounted for in general.

Our work motivates progress in DIS event generators modelling lepton scattering at TeV energies towards increased perturbative accuracy (at least NNLO~\cite{Monni:2019whf,Hoche:2018gti}, eventually N$^3$LO~\cite{Vermaseren:2005qc,Moch:2004xu,Moch:2008fj,Currie:2018fgr}), improved shower logarithmic accuracy~\cite{vanBeekveld:2023chs,vanBeekveld:2022ukn}, and a dedicated treatment of non-perturbative effects including hadronisation, fragmentation, and nuclear effects related to rescattering and absorption in the tungsten target~\cite{Sjostrand:2020gyg}.
This theoretical progress must be aligned with experimental efforts, where measurements at FASER, such as the ones considered in this work, and at SND@LHC will constrain different aspects of the simulation pipeline on which the interpretation of the forward LHC neutrino programme rests.

Furthermore, the results of this work are directly applicable to the modelling of neutrino scattering signals at neutrino telescopes such as KM3NeT~\cite{KM3Net:2016zxf,KM3NeT:2021ozk} and IceCube~\cite{IceCube:2016zyt,IceCube:2013low}, both for their oscillation programme~\cite{IceCube:2017lak,IceCube:2023ewe,KM3NeT:2021ozk} and for their study of astrophysical neutrinos at ultra-high energies (UHE)~\cite{IceCube:2014stg,KM3NeT:2025npi,IceCube:2017roe,IceCube:2018pgc}.
Our pipeline can be straightforwardly applied to neutrino DIS in these experiments, replacing the FASER fluxes with the atmospheric and astrophysical ones and then applying the corresponding selection requirements.
For instance, while precision QCD calculations are available for UHE neutrinos~\cite{Gandhi:1998ri,Connolly:2011vc,Cooper-Sarkar:2011jtt,Gauld:2019pgt,Garcia:2020jwr,Xie:2023suk}, these are mostly limited to inclusive cross-sections, and final-state distributions with fiducial cuts have received less attention so far (see~\cite{FerrarioRavasio:2024kem} for an exception).

\bigskip

\rule{12cm}{0.4pt}\\

The complete benchmark presented in this work, including generator cards, analysis code and derived results, is publicly available from its {\tt GitHub} repository
\begin{center}
\url{https://github.com/LHCfitNikhef/evgen_dis_lhc}
\end{center}
as summarised in App.~\ref{app:implementation}.

\subsection*{Acknowledgements}

We are grateful to Luca Buonocore for assistance with \powhegtwo, Daniel Reichelt for support with the \sherpa event generation, and Alfonso Garcia for help with \genie. 
We are grateful to many FASER colleagues, in particular Akitaka Ariga and Tomoko Ariga, for information concerning the data selection strategy.
We thank Melissa van Beekveld for feedback on the manuscript.
We thank Giovanni Limatola, Rhorry Gauld, and Alfonso Garcia for discussions during the 8th FPF Workshop at CERN, which triggered this work. 

\FloatBarrier
\appendix

\section{Implementation and reproducibility}
\label{app:implementation}

The complete setup used in this study, including generator cards, analysis codes, and the derived results
in machine-readable form, is publicly available from {\tt GitHub}~\cite{benchmarkrepo}.  
In this appendix we provide information in order to reproduce and extend our results. 

\paragraph{Cross-section normalisation.}  
Generators that discard and regenerate events during showering report two totals: the integral over the generation region, and that integral scaled down by the fraction of events that survive. 
When the generation cuts coincide with the fiducial cuts, every discarded event was a fiducial event, so normalising with the survival-scaled number reports a fiducial cross-section low by exactly the discard rate.
In this work, we normalise with the
integrated cross-section throughout.

\paragraph{Sample closure.}  For weighted samples, it is common to fix the normalisation to the integrator's cross-section while taking the shapes from
the delivered events.  
This is self-consistent only if the events actually
reproduce that cross-section.
Here we compute the delivered cross-section for every weighted sample, compare it against the integrator, and record the ratio with the result.  
For the samples studied in this work, this closure holds at the level of a few per mille, and within $1.6\%$ for every sample at the energies of the benchmark.

\paragraph{Generation region wider than fiducial region.}
For some of the event generators considered in this work,  generation cuts are deliberately looser than the fiducial region Eq.~(\ref{eq:fiducial}).
Two examples are \powhegres and \powhegtwo, whose $Q^2$ cut acts on the underlying Born configuration and hence generation cuts must be looser than the fiducial selection. 
In such cases, the fiducial cross-section is obtained as the integrated cross-section times the fiducial weight fraction, so that migration across the boundary is simulated in both directions.

\paragraph{Reproducibility.} 
The repository of this project should be complemented with stand-alone installations of the various event generators considered:  \sherpa~3.0.5, {\sc\small Herwig}~7.3.0 (alongside ThePEG~2.3.0), {\sc\small Pythia}~8.311, \powhegres, \powhegtwo, \genie, and \yadism, together with HepMC3, \lhapdf and ROOT, at the versions listed in the repository's \texttt{README.md} and \texttt{environment.yml}.
During the course of this project, a small number of local modifications to these codes were found to be necessary and are documented in \texttt{patches/}.
These include the CKM treatment in the \powhegtwo neutrino
process and in the \sherpa virtual correction, the hadron remnant of the \sherpa neutron beam, the event-retry logic of ThePEG, the electroweak parameter ranges of \pythia, and the $Q^2$ floor of the electromagnetic DIS channel and the $W$-boson propagator of the charm-production cross-section in \genie.  

\paragraph{Production.} 
The event production is driven by one script per generator, described in \texttt{PRODUCTION.md}, each of which skips any sample that already exists and can therefore be interrupted and restarted.  The number of events, the number of parallel jobs and the energies are set by variables at the top of each driver.  
The event samples at $E_\ell=1$~TeV, which is taken as the representative energy throughout the project, occupy about $100$~GB at the considered statistics, and a single NLO integration takes of order an hour of CPU time per energy and nucleon.  
Most event files are deleted once analysed; the files
\texttt{REGENERATE\_*.md} list every deleted sample with its generator, energy, target and number of events, and the command that recreates it. 
Since the production scripts use fixed random seeds, a sample regenerated with the same settings reproduces the original results, up to differences well below their statistical errors. 
It is not necessary to regenerate the events to use the
results: every result presented in this work is stored in the repository in machine-readable form.

\paragraph{Agentic AI usage.}  This project was carried out with the support of Anthropic's Claude models (Opus 5.0/5.5 and Fable 5.0/5.1), used as software agents, under physicist supervision, with OpenAI Codex for external reviews.
A detailed collection of standing rules enforces physical decisions and every result records the files it was computed from.  
\section{Impact of non-DIS processes}
\label{app:nondis}

Among the event generators for muon- and neutrino-scattering considered in this work, only the dedicated neutrino generators (\genie, {\sc\small NuWro} and {\sc\small GiBUU}) account for non-DIS contributions: quasi-elastic, resonance, and diffractive scattering; here we use \genie to quantify these contributions.
Here we evaluate the size of these non-DIS effects for the observables relevant for this study, showing that they can be for all practical purposes neglected in FASER simulations. 

First, we consider total integrated cross-sections.
Table~\ref{tab:nondis} provides the percentage contribution to the total inclusive cross-section, evaluated with \genie, for NC muon DIS and CC neutrino DIS at three different values of $E_\ell$. 
  The non-DIS processes are quasi-elastic scattering (QEL), resonance production (RES), and diffractive scattering (DFR).
For muon DIS we apply a $Q^2 > 4$~GeV$^2$ cut, while for neutrino DIS no such cut is necessary since \genie allows extrapolation down to $Q^2\to 0$. 
No cut on $W$, the invariant mass of the hadronic final state, is applied for the numbers quoted in this table.
For this calculation, the target is assumed to be a free proton. 

We find that for energies above a few hundred GeV, as relevant for FASER, the non-DIS channels contribute at the per-cent level or below, $1.5\%$ at most (for $\bar\nu_\mu$ at 400~GeV), and hence can be safely neglected, at least at the inclusive level.
For example, for $\nu_\mu$ CC scattering, their sum is 0.8\% of the total at 400~GeV, 0.3\% at 1~TeV and 0.1\% at 4~TeV, dominated at all energies by resonance production, while for $\bar\nu_\mu$ scattering this non-DIS contribution is about twice as large because its quasi-elastic channel is open on the proton.  
For muon DIS, for which the condition $Q^2> 4$~GeV$^2$ is imposed, non-DIS corrections are always below the 1\% level and are dominated by resonance production.
As in the neutrino case, also for NC scattering it is clear that non-DIS contributions are negligible for fiducial cross-sections. 
Note that in Table~\ref{tab:nondis} ``DIS'' is understood in the \genie sense, so also includes the SIS region at low $Q^2$ which is phenomenologically relevant for inclusive neutrino rates, cf.\ the discussion around Fig.~\ref{fig:faser-electronic}.

\begin{table}[!tbp]
  \centering
    \renewcommand{\arraystretch}{1.60}
  \begin{tabularx}{\textwidth}{X | ccc | ccc | ccc}
    \toprule
    & \multicolumn{3}{c|}{$\mu^-$ NC ($Q^2 > 4$~GeV$^2$)} & \multicolumn{3}{c|}{$\nu_\mu$ CC ($Q^2 > 0$)}  & \multicolumn{3}{c}{$\bar\nu_\mu$ CC ($Q^2 > 0$)} \\
    channel & 400~GeV & 1~TeV & 4~TeV & 400~GeV & 1~TeV & 4~TeV & 400~GeV & 1~TeV & 4~TeV \\
    \midrule
    DIS  (\%)        & 99.20 & 99.42 & 99.62 & 99.25 & 99.68 & 99.91 & 98.48 & 99.37 & 99.83 \\
    \midrule
    RES   (\%)       &  0.75 &  0.54 &  0.36 &  0.58 &  0.24 &  0.07 &  0.78 &  0.32 &  0.09 \\
    QEL   (\%)       &  0.05 &  0.03 &  0.02 &  0.01 & 0.005 & 0.001 &  0.56 &  0.23 &  0.06 \\
    DFR   (\%)       & ---   & ---   & ---   &  0.16 &  0.08 &  0.03 &  0.18 &  0.09 &  0.03 \\
    \midrule
    non-DIS total (\%)    &  0.80 &  0.58 &  0.38 &  0.75 &  0.32 &  0.09 &  1.52 &  0.63 &  0.17 \\
    \bottomrule
  \end{tabularx}
  \vspace{0.3cm}
  \caption{Percentage contribution to the total inclusive cross-section, evaluated with \genie, for NC muon DIS and CC neutrino DIS at three different values of $E_\ell$.
  The non-DIS processes are QEL, RES and DFR.
For muon DIS we apply a $Q^2 > 4$~GeV$^2$ cut, while for neutrino DIS no such cut is necessary since \genie allows extrapolation down to $Q^2\to 0$. 
No cut on $W$, the invariant mass of the hadronic final state, is applied for the numbers quoted in this table.
Here DIS is understood in the \genie sense, and hence includes also the SIS region at low $Q^2$ (see text).
}
  \label{tab:nondis}
\end{table}

\begin{figure}[!tbp]
  \centering
  \includegraphics[width=0.94\textwidth]{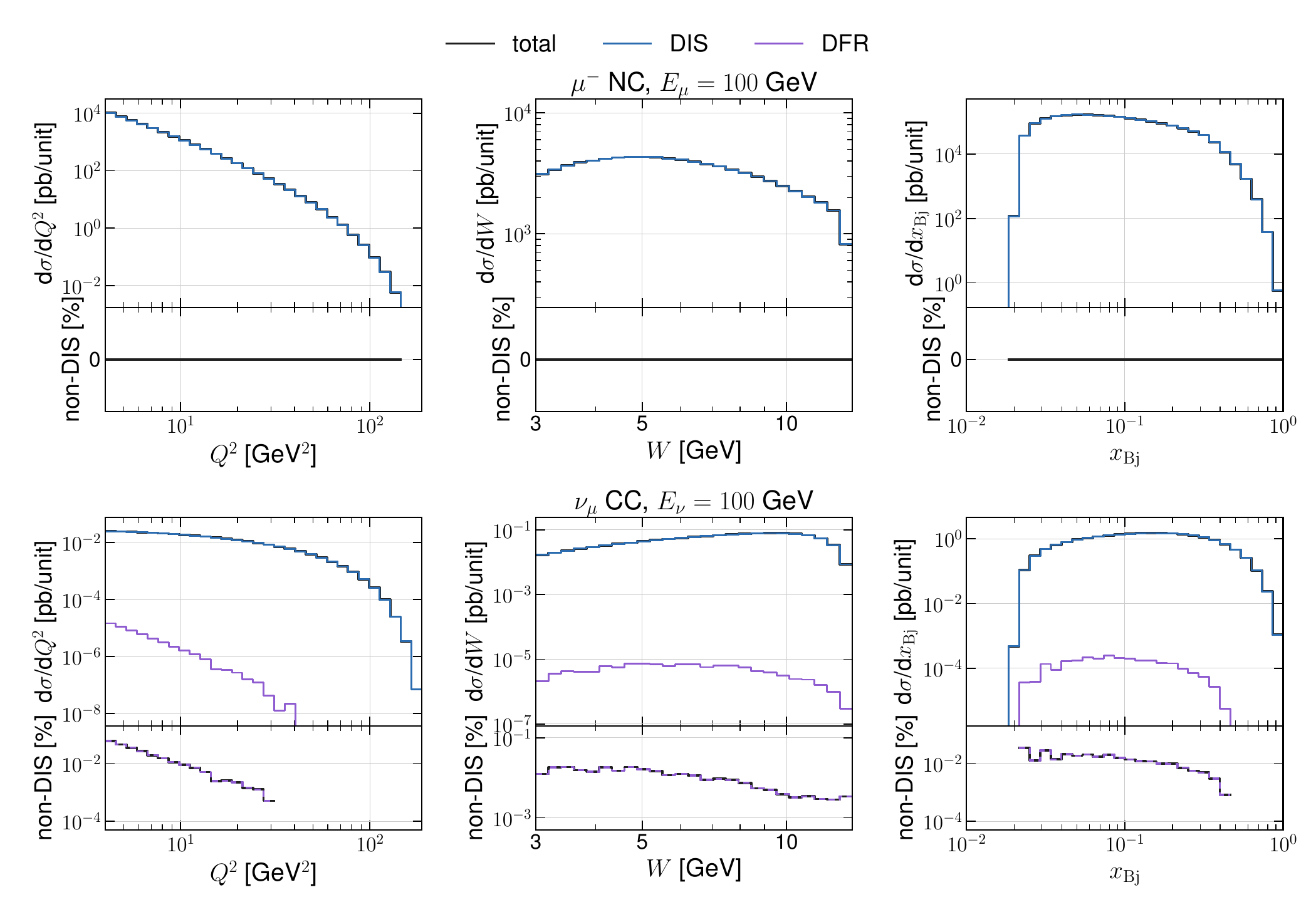}
  \includegraphics[width=0.94\textwidth]{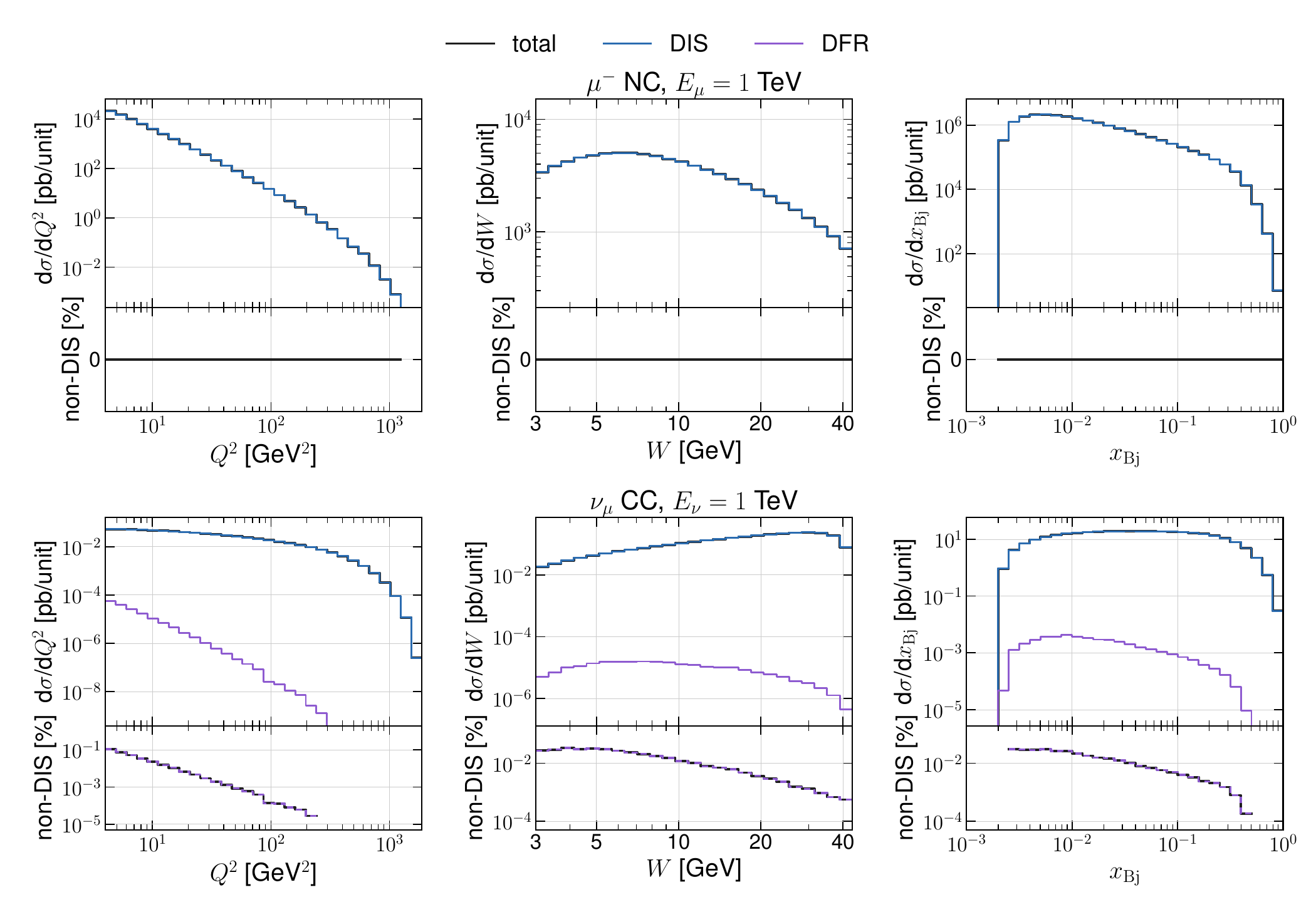}
  \caption{Differential distributions in $Q^2$, $W$, and $x_{\rm Bj}$ evaluated with \genie (GRV98LO) per nucleon on a tungsten target for muon NC and muon-neutrino CC scattering  for $E_\ell=100$ GeV (top panels) and $E_\ell=1$ TeV (bottom panels), including also non-DIS contributions. 
Only events that fall inside the fiducial region of  $Q^2 > 4$~GeV$^2$ and $W > 3$~GeV are retained. 
For each distribution, the bottom panels indicate the relative contribution of non-DIS processes to the predicted bin yields: in the case of muon DIS, this contribution is exactly zero within the statistics of the generated sample.
}
  \label{fig:nondis}
\end{figure}

Then Fig.~\ref{fig:nondis} shows the corresponding differential distributions in $Q^2$, $W$, and $x_{\rm Bj}$ evaluated with \genie (GRV98LO tune) for muon NC and muon-neutrino CC scattering  for $E_\ell=100$ GeV and $E_\ell=1$ TeV including non-DIS contributions, now per nucleon on a tungsten target rather than on the free proton of Table~\ref{tab:nondis}. 
In this case, and as opposed to Table~\ref{tab:nondis}, only events that fall inside the fiducial region of the benchmark, namely  $Q^2 > 4$~GeV$^2$ and $W > 3$~GeV, are retained. 
For each distribution, the bottom panels indicate the relative contribution of non-DIS processes to the predicted bin yields.
In the case of muon DIS, this contribution is exactly zero within the statistics of the generated sample, since all non-DIS contributions fall outside the considered fiducial region and \genie does not simulate diffractive scattering for charged leptons. 
For neutrino DIS, the only non-DIS contribution which survives is diffractive scattering, which is nevertheless below the per-mille level for all distributions at both values of $E_\nu$.
Hence from Fig.~\ref{fig:nondis} we conclude that also for differential distributions in the fiducial region Eq.~(\ref{eq:fiducial}) all non-DIS processes can be safely neglected when producing predictions for FASER observables.  

We also note that  \genie labels as DIS the whole inelastic continuum, including the SIS region at low $Q^2$ where perturbative QCD is not applicable, as well as the non-resonant background below the $W$ cut of the fiducial selection, Eq.~(\ref{eq:fiducial}). 
The contribution of this SIS region for a tungsten target is not negligible: between 3 and 6\% of the neutrino inelastic scattering channel (both DIS in the benchmark sense and SIS) sits below $Q^2 = 4$~GeV$^2$ at $E_\nu=400$~GeV and 1~TeV, while more than 99.5\% of it lies above $W = 2$~GeV; at $E_\nu=100$~GeV these fractions are 17\% and 98\% respectively.  
Hence, while the contributions from outside the DIS region in the sense of Eq.~(\ref{eq:fiducial}), namely the SIS region together with the non-DIS processes of Table~\ref{tab:nondis}, are not negligible for fully inclusive FASER neutrino cross-sections, they become negligible once cuts in $Q^2$ and $W$ ensuring the validity of the perturbative QCD calculations are applied.

However, the cuts of Eq.~(\ref{eq:fiducial}) can often not be applied, for instance in measurements in which only the outgoing muon is reconstructed, such as the electronic-detector analysis of Fig.~\ref{fig:faser-electronic}.
It is therefore useful to quantify how CC interactions are distributed among the different kinematic regions, as done in~\cite{Feng:2022inv} with \pythia alone.
Table~\ref{tab:kin-regions} reports, for $\nu_\mu$ and $\bar\nu_\mu$ CC scattering on tungsten evaluated with the default \genie tune and including all channels, the fraction of events in the DIS region ($Q^2>4$~GeV$^2$ and $W>2$~GeV), split into the fiducial region of the benchmark ($W>3$~GeV) and the strip $2<W<3$~GeV; in the SIS region ($Q^2<4$~GeV$^2$ and $W>2$~GeV); and in the resonance region ($W<2$~GeV), which contains the QEL and RES channels together with the non-resonant inelastic background.
These fractions are given at $E_\nu=100$~GeV and 1~TeV, as well as weighted with the FASER$\nu$ Run~3 flux of Sect.~\ref{sec:faser-app}, that is, as fractions of the expected number of CC interactions.

\begin{table}[!tbp]
  \centering
  \small
  \renewcommand{\arraystretch}{1.50}
  \begin{tabularx}{\textwidth}{X | ccc | ccc}
    \toprule
    & \multicolumn{3}{c|}{$\nu_\mu$ CC} & \multicolumn{3}{c}{$\bar\nu_\mu$ CC} \\
    region & 100~GeV & 1~TeV & FASER$\nu$ flux & 100~GeV & 1~TeV & FASER$\nu$ flux \\
    \midrule
    DIS: $Q^2>4$ GeV$^2$ \& $W>3$~GeV (\%)& 79.5 & 96.5 & 90.8 & 62.9 & 93.4 & 82.3 \\
    DIS: $Q^2>4$ GeV$^2$ \& $2<W<3$~GeV (\%)            &  1.3 & 0.14 & 0.56 &  2.3 & 0.28 & 0.97 \\
    SIS: $Q^2<4$ GeV$^2$ \& $W > 2$ GeV (\%)                         & 15.5 &  3.0 &  7.2 & 27.5 &  5.7 & 13.5 \\
    Resonance region: $W<2$~GeV (\%)       &  3.6 & 0.31 &  1.5 &  7.3 & 0.62 &  3.2 \\
    \bottomrule
  \end{tabularx}
  \vspace{0.3cm}
  \caption{Fraction of $\nu_\mu$ and $\bar\nu_\mu$ CC interactions on tungsten in each kinematic region, evaluated with the default \genie tune including all channels (that is, also with the RES, QEL and DFR processes of Table~\ref{tab:nondis}).
  We show results at two fixed energies, $E_\nu=100$~GeV and 1~TeV, and weighted with the FASER$\nu$ Run~3 flux used in Sect.~\ref{sec:faser-app}.
  In this table, DIS denotes $Q^2>4$~GeV$^2$ and $W>2$~GeV, split into the fiducial region of the benchmark, Eq.~(\ref{eq:fiducial}), and the strip $2<W<3$~GeV; SIS denotes the region defined by $Q^2<4$~GeV$^2$ and $W>2$~GeV; and the resonance region corresponds to $W<2$~GeV, which contains the QEL and RES channels together with the non-resonant inelastic background.
  The kinematic variables are reconstructed from the outgoing muon.
}
  \label{tab:kin-regions}
\end{table}

Weighted with the predicted flux, about $90\%$ ($80\%$) of the $\nu_\mu$ ($\bar\nu_\mu$) CC interactions fall inside the fiducial region of the benchmark and are hence described by the NLO-matched generators studied in this work, while the SIS region accounts for $7\%$ ($13\%$) and the resonance region for $1.5\%$ ($3\%$) of the events.
The fraction of events outside the fiducial region grows rapidly as the energy decreases, from $3\%$ ($7\%$) at $E_\nu=1$~TeV to $20\%$ ($37\%$) at 100~GeV and to about one half (three quarters) at 30~GeV, consistently with the size of the SIS ($Q^2<4$~GeV$^2$) contribution added to the NLO predictions in Fig.~\ref{fig:faser-electronic}.
The strip $2<W<3$~GeV between the conventional DIS boundary and the benchmark cut is at most $5\%$ at any energy and below $1\%$ after flux weighting.

\section{Impact of nuclear PDFs}
\label{app:npdf}

The predictions presented in this work neglect nuclear effects in the tungsten target of the detector, which we model as an isospin combination of free nucleons.
In this appendix we quantify the impact of nuclear PDF (nPDF) uncertainties~\cite{Klasen:2023uqj,Ethier:2020way} on DIS differential distributions in the fiducial region of Eq.~(\ref{eq:fiducial}).
For this purpose, we evaluate \yadism NLO structure functions in the ZM-VFNS scheme with target-mass corrections (TMCs) for three nPDF sets, namely nNNPDF3.0~\cite{AbdulKhalek:2022fyi}, EPPS21~\cite{Eskola:2021nhw} and nCTEQ15HQ~\cite{Kovarik:2015cma,Duwentaster:2022kpv}, in all cases at NLO accuracy, and compare with the corresponding free-nucleon baseline. 
See also~\cite{Francener:2025tyh} for a related study.

\begin{figure}[!tbp]
  \centering
  \includegraphics[width=0.99\textwidth]{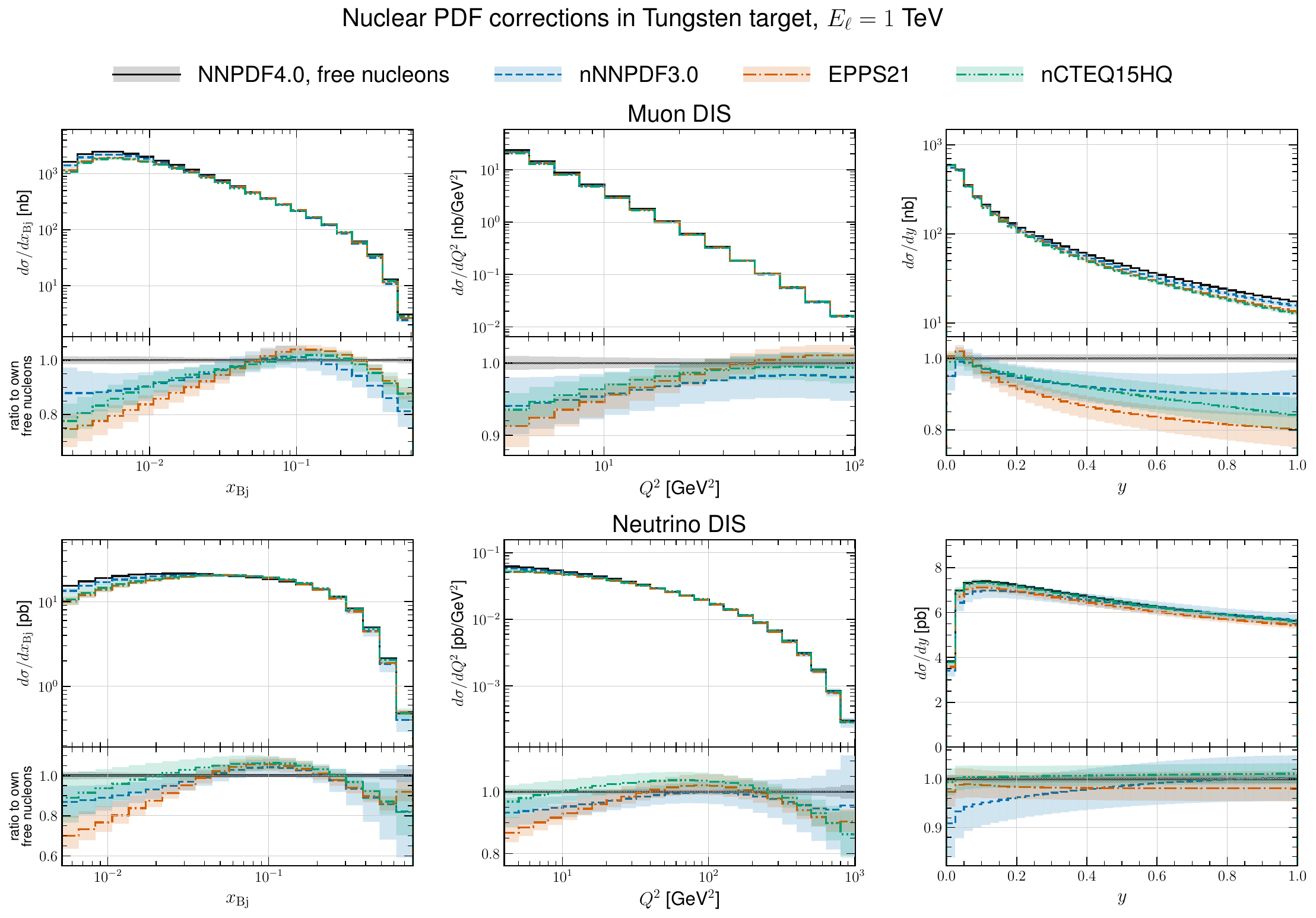}
  \caption{Differential inclusive distributions in $x_{\rm Bj}$, $Q^2$ and $y$ for muon (top) and neutrino (bottom) DIS on a tungsten target at $E_\ell=1$~TeV  in the fiducial region of Eq.~(\ref{eq:fiducial}), computed with \yadism\ at NLO in the ZM-VFNS and with TMCs.
The baseline predictions, evaluated with NNPDF4.0, are compared with three nPDF sets nNNPDF3.0~\cite{AbdulKhalek:2022fyi}, EPPS21~\cite{Eskola:2021nhw} and nCTEQ15HQ~\cite{Kovarik:2015cma,Duwentaster:2022kpv}. 
The bottom panels display the ratio of each prediction to the corresponding free-nucleon case, namely the $A=1$ variants of each nPDF determination.}
  \label{fig:npdf}
\end{figure}

Fig.~\ref{fig:npdf} shows differential distributions in $x_{\rm Bj}$, $Q^2$ and $y$ for muon and neutrino DIS on a tungsten target at the reference energy of $E_\ell=1$~TeV  in the fiducial region of the benchmark for these three nPDF sets as well as for NNPDF4.0.
The bottom panels display the ratio of each prediction to the corresponding free-nucleon case, namely the $A=1$ variants of each nPDF determination.
Note that the same nPDF dependence would be found should one use the NLO-matched event generators (following the findings of Sect.~\ref{sec:inclusive}), since the observables considered here do not depend on the modelling of the hadronic final state, and hence showing the \yadism predictions is sufficient. 

From Fig.~\ref{fig:npdf} one observes that nPDF effects are in general sizeable, though the associated uncertainties remain large.
Indeed, nPDF uncertainties are $2$ to $6$ times larger than the free-nucleon one.
In general, predictions from the three different nPDF sets are in qualitative agreement, mostly overlapping within their $1\sigma$ uncertainties.
nPDF-associated shifts tend to be smaller for neutrino DIS (where a nuclear ratio of unity is still allowed within uncertainties, except at small $x$ and low $Q^2$) than for muon DIS, where a non-unity value of the nuclear ratio is clearly preferred.
The latter is especially true for the $y$ distribution, the low-$Q^2$ region, and the large-$x$ (EMC effect) and small-$x$ (shadowing) regions.

\begin{figure}[!tbp]
  \centering
  \includegraphics[width=0.80\textwidth]{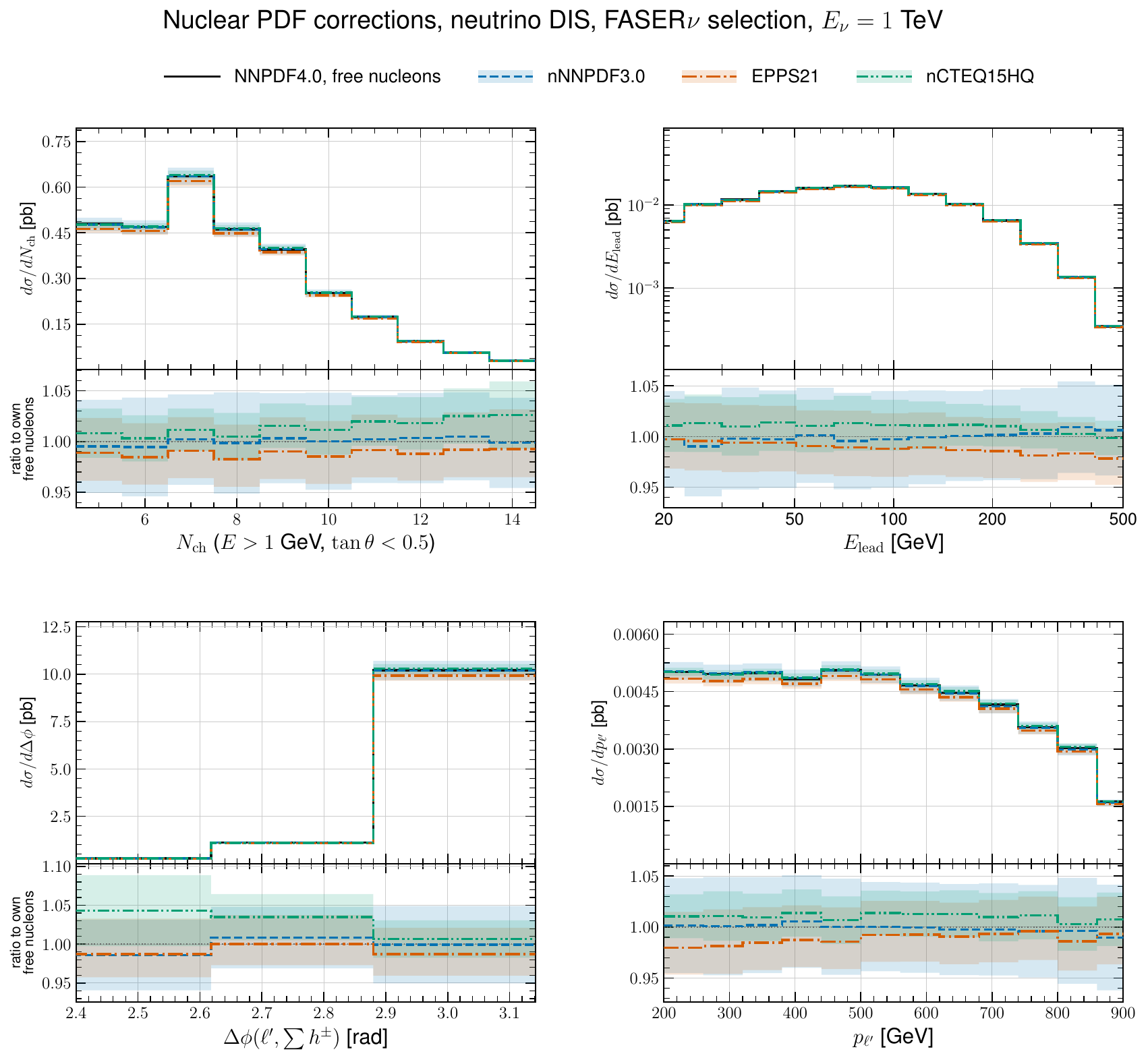}
  \caption{Same as Fig.~\ref{fig:npdf} for the hadron-level distributions of Fig.~\ref{fig:hadron-level} ($N_{\rm ch}$, $E_{\rm lead}$, $\Delta\phi$, $p_{\ell'}$) in neutrino DIS on tungsten at $E_\nu=1$~TeV under the FASER$\nu$ selection, from \powheg{}+\pythia evaluated with NNPDF4.0 (free nucleons), nNNPDF3.0, EPPS21 and nCTEQ15HQ.
    The bottom panels show the ratio of each set to its own free-nucleon baseline.}
  \label{fig:npdf-hadron}
\end{figure}

Fig.~\ref{fig:npdf-hadron} then shows how nPDF uncertainties propagate to the hadron-level observables of Fig.~\ref{fig:hadron-level}, for neutrino DIS at $E_\nu=1$~TeV under the FASER$\nu$ selection.
The nuclear modifications remain at the per-cent level, at most 4\% in $\Delta\phi$ and below 3\% elsewhere, are nearly flat, and lie well within the nPDF uncertainty bands of 2--5\%, as was the case for the inclusive DIS distributions of Fig.~\ref{fig:npdf}.
They are therefore much smaller than the differences between generators found in Sect.~\ref{sec:hadronic}, which are driven by the modelling of the parton shower and hadronisation.

All in all, Figs.~\ref{fig:npdf} and~\ref{fig:npdf-hadron} indicate that nPDF corrections must be accounted for when producing predictions for FASER observables and related simulations, since they are comparable to the NNLO corrections for muon DIS and larger than the spread between NLO-matched generators for inclusive NC and CC observables, though smaller than the generator differences on the hadronic final state.
Furthermore, this comparison confirms that FASER data itself should provide excellent opportunities to constrain nuclear PDFs alongside the free-nucleon ones, as already demonstrated in~\cite{Cruz-Martinez:2023sdv}.

\section{Comparison with NuWro and GiBUU}
\label{app:nugen}

In addition to \genie, other dedicated event generators are widely used by the neutrino community.
To highlight the flexibility of the pipeline assembled in this work, here we compare \genie with two other generators tailored to neutrino-nucleus scattering, namely {\sc\small NuWro}~\cite{Golan:2012rfa,Juszczak:2009qa} and {\sc\small GiBUU}~\cite{Buss:2011mx}, whose events are processed by the same analysis chain as those of the other generators of Table~\ref{tab:generators}.

\begin{figure}[!tbp]
  \centering
  \includegraphics[width=0.99\textwidth]{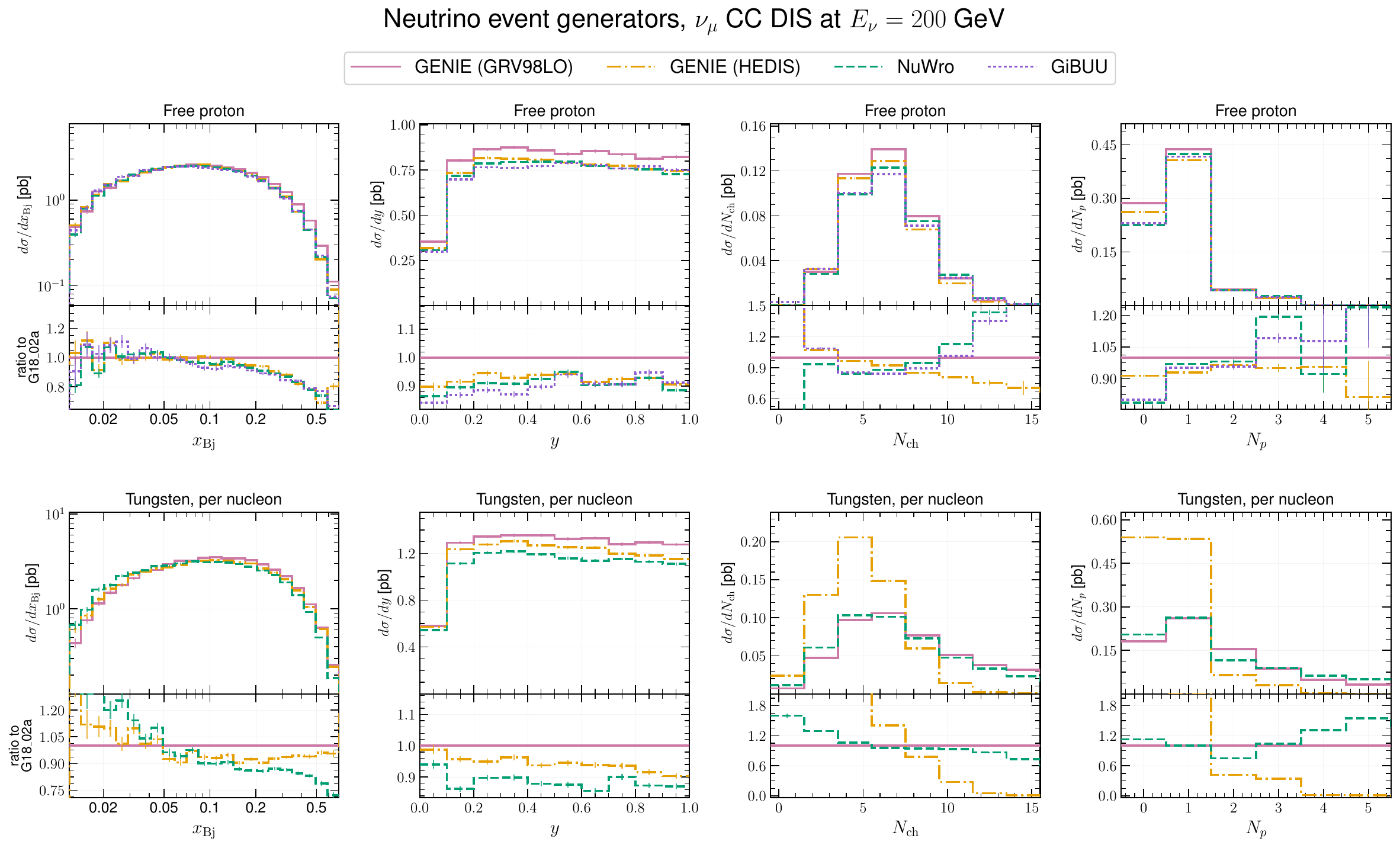}
  \caption{Distributions in $x_{\rm Bj}$, $y$, charged-hadron multiplicity $N_{\rm ch}$, and proton multiplicity $N_p$ evaluated in $\nu_\mu$ CC scattering at $E_\nu=200$~GeV in the fiducial region Eq.~(\ref{eq:fiducial}) both on a free proton  target (top) and on a tungsten target (bottom).
  Predictions from three different neutrino MC event generators are compared: \genie (both the default {\tt G18\_02a} tune and the {\sc\small HEDIS} tune), {\sc\small NuWro} and {\sc\small GiBUU}.
   {\sc\small GiBUU} is shown for the free proton only.
    The bottom panels display the ratio of the various predictions to \genie~{\tt G18\_02a}.}
  \label{fig:nugen}
\end{figure}

With this motivation, Fig.~\ref{fig:nugen} displays differential distributions in $x_{\rm Bj}$, $y$, charged-hadron multiplicity $N_{\rm ch}$, and proton multiplicity $N_p$ evaluated in muon-neutrino CC scattering at $E_\nu=200$~GeV in the fiducial region of this benchmark, both on a free proton  target and on a tungsten target.
Predictions from three different neutrino MC event generators are compared: \genie (both the default {\tt G18\_02a} tune and the {\sc\small HEDIS} tune), {\sc\small NuWro} and {\sc\small GiBUU}.
On tungsten, each generator applies its own nuclear model, and the DIS kinematic variables are defined with respect to the struck nucleon.
Note that {\sc\small GiBUU} is shown for the free proton only, since its tungsten transport does not complete at $E_\nu=200$ GeV, and its neutrino module is documented only up to $E_\nu=50$~GeV~\cite{Lalakulich:2012gm}, which limits its applicability to FASER physics.

On the free proton target, the generators mostly agree on the shapes of the distributions, but one observes large differences in the overall rate: relative to \genie~{\tt G18\_02a}, the fiducial cross-section is lower by $9\%$ in {\sc\small NuWro} and $10\%$ in {\sc\small GiBUU}.
Larger differences are found on the tungsten target: for example, the intranuclear cascade raises the mean charged multiplicity to $12.5$ in \genie~{\tt G18\_02a} and $10.6$ in {\sc\small NuWro}.
Note that the {\sc\small HEDIS} tune of \genie  does not simulate final-state interactions, so its hadron multiplicities remain at the free-nucleon level, see also~\cite{Feng:2022inv}.
More generally, we recall that the nuclear models of the various neutrino generators are tuned to data at much lower energies, and their extrapolation to the TeV energies of FASER neutrinos is in general not expected to work.

\bibliographystyle{utphys}
\bibliography{references}

\end{document}